\newif\ifconferencesubmission
\ifconferencesubmission
    \documentclass[format=acmsmall, review=false]{acmart}
    \usepackage{acm-ec-24-proc}
\else
    \documentclass[11pt, oneside]{article}
    \usepackage[margin=1in]{geometry}
    \usepackage[square,numbers]{natbib}
    \usepackage{hyperref}
    \usepackage{amssymb}
\fi

\usepackage{booktabs} 
\usepackage[ruled]{algorithm2e} 
\usepackage{natbib} 

\SetAlFnt{\small}
\SetAlCapFnt{\small}
\SetAlCapNameFnt{\small}
\SetAlCapHSkip{0pt}
\IncMargin{-\parindent}
\usepackage{amsfonts,amsthm,bbm,commath,amsmath,mathrsfs,enumitem,booktabs,xcolor, thm-restate,svg, subcaption}
\ifconferencesubmission
    \setcitestyle{authoryear}
\fi
\usepackage{cleveref}

\newcommand{\gm}{IPNG}
\newcommand{\game}{Informed Professional Networking Game}
\newcommand{\indiscgame}{Professional Networking Game}
\newcommand{\generalizedgame}{Generalized Professional Networking Game}
\newcommand{\generalizedgm}{GPNG}
\newcommand{\indiscgm}{PNG}
\newcommand{\dmin}{d_{\min}}
\newcommand{\dmax}{d_{\max}}
\newcommand{\umin}{u_{\min}}
\newcommand{\umax}{u_{\max}}
\newcommand{\informedy}{Y^{\mathrm{infd.}}}
\newcommand{\infd}{{{infd.}}}

\newcommand{\gini}{Gini}
\def\undertilde#1{\mathord{\vtop{\ialign{##\crcr
$\hfil\displaystyle{#1}\hfil$\crcr\noalign{\kern1.5pt\nointerlineskip}
$\hfil\tilde{}\hfil$\crcr\noalign{\kern1.5pt}}}}}

\crefname{ineq}{ineq.}{inequalities}
\creflabelformat{ineq}{#2{\upshape(#1)}#3} 

\newcommand{\yesnum}{\addtocounter{equation}{1}\tag{\theequation}}

\newcommand{\indic}[1]{\mathbbm{1}\left\{{#1} \right\}}

\newcommand{\R}{\mathbb{R}}
\newcommand{\Z}{\mathbb{Z}}
\newcommand{\N}{\mathbb{N}}

\newcommand{\defeq}{:=}

\newcommand{\prob}[1]{\mathbb{P}\left[ {#1} \right]}

\newcommand{\argmax}[1]{\operatorname{arg}\,\underset{#1}{\operatorname{max}}\;}

\newcommand{\ceil}[1]{\lceil {#1} \rceil}
\newcommand{\floor}[1]{\lfloor {#1} \rfloor}
\newcommand{\paren}[1]{\left( {#1} \right)}

\newcommand{\sqparen}[1]{\left[ {#1} \right]}
\newcommand{\ex}[1]{\mathbb{E}\sqparen{#1}}

\newcommand{\exgiv}[2]{\mathbb{E}\left[ {#1} \;\middle|\; {#2} \right]}

\newcommand{\curly}[1]{\left\{ {#1} \right\}}
\newcommand{\setcomp}[2]{\left\{ {#1} \;\middle|\; {#2} \right\}}
\newcommand{\probgiv}[2]{\mathbb{P}\left[ {#1} \;\middle|\; {#2} \right]}

\SetKwComment{Comment}{/* }{ */}

\newcommand{\cD}{\mathcal{D}}
\newcommand{\cE}{\mathcal{E}}

\newcommand{\cO}{\mathcal{O}}

\newcommand{\cS}{\mathcal{S}}
\newcommand{\cT}{\mathcal{T}}

\newcommand{\cV}{\mathcal{V}}

    \newtheorem{theorem}{Theorem}[section]
    
    \newtheorem{definition}[theorem]{Definition}

    \newtheorem{proposition}[theorem]{Proposition}

    \crefname{section}{Section}{Sections}
    \crefname{theorem}{Theorem}{Theorems}
    \crefname{observation}{Observation}{Observations}
    \crefname{proposition}{Proposition}{Propositions}
    \crefname{claim}{Claim}{Claims}
    \crefname{condition}{Condition}{Conditions}
    \crefname{example}{Example}{Examples}
    \crefname{fact}{Fact}{Facts}
    \crefname{lemma}{Lemma}{Lemmas}
    \crefname{corollary}{Corollary}{Corollaries}
    \crefname{definition}{Definition}{Definitions}
    \crefname{remark}{Remark}{Remarks}
    \crefname{algorithm}{Algorithm}{Algorithms}

\SetAlgorithmName{Algorithm}{Algorithm}{Algorithms}
\SetKwInput{kwInit}{Initialize}
\SetKwFor{For}{For}{do}{end}
\SetKwFor{ForEach}{For each}{do}{end}

\newcommand{\pzero}{q}

\newcommand{\ptwo}{p}
\newcommand{\g}[1]{\paren{1 - \frac{\ptwo}{\pzero} \cdot \frac{1-(1 - \pzero)^{{#1}}}{#1}}}

\newcommand{\h}[1]{h({#1})}

\newenvironment{proofof}[1]{\textit{Proof of #1.}}

 \renewcommand{\paragraph}[1]{
    \textbf{#1} 
 }

\newcommand{\ch}[1]{\textcolor{brown}{[CH: {#1}]}}
\newcommand{\cd}[1]{\textcolor{green}{[CD: {#1}]}}
\newcommand{\cynthia}[1]{\textcolor{green}{[CD: {#1}]}}
\newcommand{\mr}[1]{\textcolor{purple}{[MR: {#1}]}}
\newcommand{\jk}[1]{\textcolor{blue}{[JK: {#1}]}}

\renewcommand{\ch}[1]{}
\renewcommand{\cd}[1]{}
\renewcommand{\cynthia}[1]{}
\renewcommand{\mr}[1]{}
\renewcommand{\jk}[1]{}

\newcommand{\poa}{{PoA}}
\newcommand{\equilibrium}{defection-free pairwise Nash equilibrium}
\newcommand{\equilibria}{defection-free pairwise Nash equilibria}
\newcommand{\eqbm}{DFPNE}
\newcommand{\nashedgeset}{E_{\mathrm{DFPNE}}}

\newif\ifproofsinbody
\proofsinbodyfalse

\ifconferencesubmission
\else
    
\begin{document}
    \title{Equilibria, Efficiency, and Inequality in Network Formation for Hiring and Opportunity}
\fi
\date{\today}

\ifconferencesubmission
    \title[Network Formation for Hiring and Opportunity]{Equilibria, Efficiency, and Inequality in Network Formation for Hiring and Opportunity}
    \author{Cynthia Dwork} 
    \email{dwork@seas.harvard.edu}
    \affiliation{
        \institution{Harvard University}
        \city{Cambridge}
        \state{MA}
        \country{USA}
        }
    
    \author{Chris Hays}
    \email{jhays@mit.edu}
    \affiliation{
        \institution{Massachusetts Institute of Technology}
        \city{Cambridge}
        \state{MA}
        \country{USA}
        }
    
    \author{Jon Kleinberg}
    \email{kleinberg@cornell.edu}
    \affiliation{
        \institution{Cornell University}
        \city{Ithaca}
        \state{NY}
        \country{USA}
        }
    
    \author{Manish Raghavan} 
    \email{mragh@mit.edu}
    \affiliation{
        \institution{Massachusetts  Institute  of  Technology}
        \city{Cambridge}
        \state{MA}
        \country{USA}
        }
\else
    \author{Cynthia Dwork
    \thanks{ {Department of Computer Science, Harvard University} }
    \and
    {Chris Hays}
        \thanks{Institute for Data, Systems, and Society, Massachusetts Institute of Technology}
    \and
    {Jon Kleinberg}
        \thanks{Departments of Computer Science and Information Science, Cornell University}
    \and
    {Manish Raghavan} 
        \thanks{Sloan School of Management and Department of Electrical Engineering and Computer Science, Massachusetts  Institute  of  Technology}}
\fi

\ifconferencesubmission

    \copyrightyear{2024}
    \acmYear{2024}
    \setcopyright{rightsretained}
    \acmConference[EC '24]{The 25th ACM Conference on Economics and Computation}{July 8--11, 2024}{New Haven, CT, USA}
    \acmBooktitle{The 25th ACM Conference on Economics and Computation (EC '24), July 8--11, 2024, New Haven, CT, USA}
    \acmDOI{10.1145/3670865.3673451}
    \acmISBN{979-8-4007-0704-9/24/07}

       \makeatletter
        \gdef\@copyrightpermission{
          \begin{minipage}{0.1\columnwidth}
               \href{https://creativecommons.org/licenses/by/4.0/}{\includegraphics[width=0.90\textwidth]{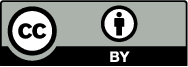}}
          \end{minipage}\hfill
          \begin{minipage}{0.9\columnwidth}
           \href{https://creativecommons.org/licenses/by/4.0/}{This work is licensed under a Creative Commons Attribution International 4.0 License.}
          \end{minipage}
          \vspace{5pt}
        }
        \makeatother

    
\fi

\ifconferencesubmission
\else
    \maketitle
\fi

\begin{abstract}
    Professional networks --- the social networks among people in a given line of work --- can serve as a conduit for job prospects and other opportunities. Here we propose a model for the formation of such networks and the transfer of opportunities within them. In our theoretical model, individuals strategically connect with others to maximize the probability that they receive opportunities from them. We explore how professional networks balance \textit{connectivity}, where connections facilitate opportunity transfers to those who did not get them from outside sources, and \textit{congestion}, where some individuals receive too many opportunities from their connections and waste some of them.
    
    We show that strategic individuals are over-connected at equilibrium relative to a social optimum, leading to a \textit{price of anarchy} for which we derive nearly tight asymptotic bounds. We also show that, at equilibrium, individuals form connections to those who provide similar benefit to them as they provide to others. Thus, our model provides a microfoundation in professional networking contexts for the fundamental sociological principle of \textit{homophily}, that ``similarity breeds connection'' \cite{mcpherson2001birds}, which in our setting is realized as a form of \textit{status homophily} based on alignment in individual benefit. We further explore how, even if individuals are \textit{a priori} equally likely to receive opportunities from outside sources, equilibria can be unequal, and we provide nearly tight bounds on how unequal they can be. Finally, we explore the ability for online platforms to intervene to improve social welfare and show that natural heuristics may result in adverse effects at equilibrium. Our simple model allows for a surprisingly rich analysis of coordination problems in professional networks and suggests many directions for further exploration.
\end{abstract}

\ifconferencesubmission
    \begin{document}
    
    \maketitle
\fi

\section{Introduction}

    Individuals' networks fundamentally shape their economic outcomes.
    Perhaps most importantly, they help people find employment: a large fraction of jobs across skill levels are found through personal contacts \cite{lester2021heterogeneous}, either by passing information about job openings or providing employee referrals.
    Indeed, networks have been shown to account for how about half of all jobs are acquired \cite{topa2011labor} and referred workers have been found to earn more across many contexts \cite{schmutte2015job, burks2015value}.
    Networks have also been shown to play a role in providing mentorship \cite{jamison2022mentorship}, fostering entrepreneurship \cite{burchardi2013economic}, facilitating access to different industries \cite{lin1986access}, leading to in-kind favors \cite{jackson2012social, ambrus2014consumption} or otherwise yielding desirable professional outcomes \cite{lin2017building}.
    In other words, networks are key to providing people with professional opportunities.
    
    The influence of networks on labor markets was motivation for and subject of seminal papers in network science \cite{granovetter1973strength}, and the role of networks is a central topic of study in labor economics \cite{jackson2017economic}.
    However, the equilibrium effects of how professional networks are \textit{formed} are still not well understood.
    Here, we propose and analyze a model of professional network formation motivated by the following two observations.

    First, modern professional networking increasingly takes place online and is increasingly mediated by algorithms.
    Individuals use platforms like LinkedIn to build and maintain large, dynamic networks, not necessarily constrained to their immediate local environment.
    Moreover, platforms play an active role in network formation, suggesting new links and controlling exposure to others on the platform through ranking algorithms.
    Our model will allow us to reason about how a platform's design choices can influence the characteristics of a professional network.
    
    Second, networking is strategic: individuals connect with others in the hopes that their connections will be beneficial to their careers.
    Individuals must choose who they connect with in order to best serve their professional goals.
    Indeed, LinkedIn's official blog publishes data-driven recommendations for building one's network for maximal career benefits \cite{Yu_SaintJacques_Matsiras_2021}, and experimental evidence has demonstrated how changes to who individuals form professional connections with lead to measurable impacts on their careers and professional lives \cite{rajkumar2022causal, Yin_2021}.
    In what follows, we propose and analyze a game-theoretic model of professional network formation, drawing on prior literature on network formation games and employee referrals.
    
    \paragraph{Our contributions.}
    We propose a model for the transmission of opportunities in professional networks, and we analyze efficiency and equity of game-theoretic equilibria in the formation of these networks. 
    Depending on the context, opportunities may represent jobs, talks, gigs, favors, referrals, interviews, co-authorship, advice or any other professional engagement which individuals seek and which can be transmitted person-to-person. 
    For concreteness, we will usually think of them as referrals.
    Our focus is the analysis of how jobs and referrals are passed within a social network in an \textit{ad hoc} manner; we capture situations where there is limited information about others’ availabilities and no central jobs clearinghouse. 
    The game we analyze is simple and symmetric: each individual receives a number of opportunities drawn from the same distribution.
    When an individual receives more opportunities than they can use on their own, they pass excess opportunities to their neighbors.
    Individuals are therefore incentivized to form connections with one another to increase their likelihood of receiving opportunities.
    
    Despite its simplicity, this model offers a surprisingly rich array of insights.
    In \Cref{sec:noedgecost}, we discuss and analytically explore pathways through which there can be inefficiency in the transmission of opportunities resulting from incoordination.
    In particular, we show how professional social networks balance the need for connectivity (if someone has an extra opportunity and they do not have many connections, there may be no person who needs an opportunity who it will get passed to) and congestion (if the network is too densely connected, individuals may inadvertently route too many opportunities to some people and not enough to others). 
    Surprisingly, our model exhibits losses in efficiency --- formulated as a price of anarchy --- \textit{even in the absence of costs associated with forming connections with others}, unlike canonical network formation games in the literature. 
    Our model also provides a microfoundation for a kind of \textit{status homophily} \citep{mcpherson2001birds}, where at equilibrium, individuals contribute similar benefits to their neighbors as their neighbors provide to them (and potentially different from those in other parts of the network).
    Intuitively, this happens because, in a large enough population, if some individual provides more benefits to someone they are connected to than this person does for them,  another more favorable connection can be substituted in their place.

    We study the inequality that can arise from our model in \Cref{sec:inequality}.
    Surprisingly, even when individuals are \textit{a priori} equal, we show that equilibrium network structures can result in inequality.
    Finally, in \Cref{sec:interventions}, we explore how a platform on which the members of the professional network interact, like LinkedIn, might try to induce better equilibria for users and prove that two natural policy levers --- changing the amount of friction to connect and providing information about individuals' bandwidths --- have nonmonotonic effects and may have either beneficial or adverse consequences depending on whether they are applied in a sophisticated or naive manner.
    \ifconferencesubmission
        We defer proofs of our results to the full version of this paper.\footnote{\href{https://arxiv.org/abs/2402.13841}{https://arxiv.org/abs/2402.13841}}
    \else
        We defer proofs of our results to \Cref{sec:deferredproofs}.
    \fi

    The game that we propose is an instance of a broader class of network formation games that we call \textit{mutual support games} where individuals benefit from direct connections they form (by, e.g., providing each other with opportunities) but possibly impose negative externalities on others as a result. 
    These games are qualitatively different from canonical network formation games in the literature, but several existing works, like that of \citet{blume2013network} and the co-authorship game in \citet{jackson2003strategic} can be described as mutual support games.
    When paired with the equilibrium concept we use, which is inspired by stable matching, our results suggest directions for more general equilibrium analyses of mutual support games.

\section{Related work} \label{sec:relatedwork}
    There are two key strands of literature that we bring together in our work: game theoretic analyses of network formation and theory models of referrals in labor markets.
    The first provides the analytical framework we use.
    The second informs our model of opportunity transfer.
    
    One line of work on network formation games for social and economic contexts has focused on communication networks and the flow of information \cite{jackson2003strategic, jackson2002evolution, WATTS2001331, bala2000noncooperative, bilo2021selfish}. 
    In these games, individuals benefit from short paths to others and pay for any direct connections they create.
    Our work diverges from these in that the source of externalities induced by individuals' behavior is different.
    Whereas canoncical network formation games exhibit prices of anarchy due to free riding, when some individuals benefit from short paths along edges they did not pay for, ours comes from situations where a lack of coordination leads some opportunities to be routed to those who do not need them.
    
    Other work has considered social and economic applications beyond communication, such as contagious risk \cite{blume2013network} or bridging between different parts of a network \cite{kleinberg2008holes, buskens2008dynamics}.
    Like our work, new connections in their models induce negative externalities, although the contexts they have in mind are very different from ours, and our conclusions are tailored to hiring and employment contexts.
    
    Finally, some prior work has considered inequality in network formation games as we do \cite{johnson2015inequality, harvath2018social}. 
    Like our work, theirs shows that inequality can emerge in equilibrium from networks of individuals who started out with identical characteristics.

    A second strand of literature has considered the effects of referrals in labor markets \cite{montgomery1991social, calvo2004effects, burks2015value, bolte2020role, okafor2020social}.
    These works focus on explaining how homophily influences labor market outcomes and how persistent inequality can emerge as a result of referrals.
    Our opportunity transfer model is inspired by \citet{calvo2004effects};
    in our model and theirs, individuals may receive opportunities from outside of the network, and they may use them and pass any extras along to their social contacts.
    Additionally, our work provides a microfoundation for a kind of homophily in referral networks (i.e., in equilibrium, individuals connect to others who provide similar benefits to them as they provide to others), which is usually taken as given in this literature.

\section{A model of opportunity transfer in networks} \label{sec:model}

    We model network formation in a population of individuals optimizing for their own opportunities.
    The population will be represented by a set of individuals $\{ 1,\dots,n \}$.
    We will represent the professional network as a (simple) undirected graph $G = ([n], E)$ where individuals are nodes and two nodes are connected by an edge if the corresponding individuals are connected to one another.
    
    \paragraph{Opportunity creation and transfer.}
    Individuals' utilities come from the opportunities that they make use of.
    They will have limited capacities for using opportunities. 
    We will call this their \textit{bandwidths}. 
    For simplicity, each individual in our model will have a bandwidth of one unit of opportunity (i.e., to hold one job). 
    Each individual receives a discrete number of opportunities exogenously from a known distribution.
    For much of this work, for simplicity and interpretability, we consider the setting where each individual gets zero opportunities with probability $q \in (0, 1)$, two opportunities with probability $p \in (0, q]$, and one opportunity with probability $1-p-q$.
    (We always assume $p, q > 0$.)
    In the event that an individual gets an extra opportunity, they pass it along to a random neighbor, who can derive utility from it if they do not already have an opportunity.
    We assume that extra opportunities are passed randomly among neighbors as a way to abstract away from the idiosyncratic factors influencing real life job and referral transfers.
    We can think of individuals participating in the network as agreeing to an implicit social contract: each participant in the network promises to pass on any extra opportunities they receive to someone in the network under the condition that everyone else does as well.
    Consistent with the prior literature on network-based referral models (e.g., \citet{calvo2004effects}) and our intuition that opportunities are often transient and non-transferable, an opportunity cannot be passed within the network more than once.
    Importantly, throughout this work, we assume there is limited coordination among individuals passing and receiving opportunities: sometimes, individuals may pass an opportunity to a social contact that is already bandwidth-constrained and that opportunity is wasted as a result.
    In \cref{sec:broadcastbandwidths} and \Cref{sec:informed}, we consider a variant of the model where individuals are made aware of others' bandwidths and pass opportunities only to people who need them. 

    To illustrate how this lack of coordination occurs in real-world contexts, suppose an academic researcher is organizing a panel for an academic conference and there is space on the panel for another panelist (i.e., they have an opportunity to pass to a professional connection). 
    The organizer contacts a suitable researcher they know, but they already have another commitment at the same time. 
    Thus, they decline the invitation, and the opportunity is wasted.
    Of course, it is possible that the researcher will reach out to other possible panelists after their first choice declined the invitation, but in many situations, the organizer may not be persistent in their search for someone to fill out the panel, or there is not enough time to find another person. 
    This kind of situation is not limited to panel invitations. One can imagine various and diverse situations captured by our basic model: an opportunity falls into a person's lap, they either take the opportunity for themselves or, if they can't, they pass it to one of their social contacts without following up to make sure the opportunity eventually gets to someone who can use it.
    
    In our model, we explore how individuals would strategically choose to form connections in order to maximize their own utility.
    An individual derives utility 1 from having an opportunity (either realized exogenously or transferred from a neighbor), but must pay a cost $\gamma \ge 0$ for each edge they participate in.
    Connection costs can be imagined to be as simple and small as friction in a user interface --- each individual must spend some amount of time navigating the UI in order to connect others --- but it could also include other factors like the burdens of cultivating the relationship.
    We remark that our model includes the case that $\gamma = 0$, in which network formation is \textit{frictionless}, and it has interesting and informative case on its own: for example, canonical network formation games only exhibit a price of anarchy because of free riding induced by nonzero edge costs, whereas the frictionless version of our model exhibits a price of anarchy even with no edge costs.
    
    \paragraph{Notation.}
    We will let $d_j$ be the degree of node $j$, and $N_j(E)$ will be the set of neighbors of $j$ in an edge set $E$. 
    %
    %
    We will let $X_i(q, p)$ represent the number of exogenous opportunities an individual receives (drawn from a distribution given by $p, q$ with support over 0, 1 and 2, independent of the other individuals' draws).
    For fixed $n \in \N$, $E \subseteq [n]^2, q$ and $p$, we will also use a random variable $R_{ji}(E; q, p)$ which is defined to be equal to 1 if individual $j$ passes an opportunity to $i$ and 0 otherwise.
    For fixed $\gamma \geq 0$, we will use $Y_i(E; q, p, \gamma) \defeq \min\{1, X_i(q, p) + \sum_{(i,j) \in E} R_{ji}(E; q, p)\} - \gamma d_i $ to represent $i$'s utility, the opportunities they use minus connection costs.
    Throughout, when it is clear from context, we will drop the $q, p$, and $\gamma$ from notation. 
    
    \paragraph{The network formation game.} The main focus of our analysis is users' strategic choices about forming new connections in the network.
    In the game we consider, before any $X_i$ is determined, individuals form connections amongst themselves.
    The strategy space for an individual $i$ is the set of edges $(i,j)$ for $j \in [n]$ that they propose, and a connection between individuals is formed if they both propose.
    In the game, we will assume individuals optimize for expected utility 
    \begin{align*}
       u_i(E; q, p , \gamma) \defeq \ex{Y_i(E; q, p , \gamma)} 
    \end{align*}
    where $E$ is the set of links that were formed, and the expectation is taken over draws of exogenous opportunities and opportunity transfers.
    We will call it the \indiscgame~(\indiscgm).
    Social welfare will be the sum of individuals' expected utilities, and networks maximizing social welfare will be interchangeably called \textit{socially optimal} or \textit{efficient}.
    For a set $\cS \subseteq [n]$, we will use the notation $u_{\cS}(E; q, p, \gamma) \defeq \sum_{i \in \cS} u_i(E; q, p, \gamma)$ to be the sum of individuals' expected utilities in $\cS$.
    
    \paragraph{Equilibrium concept.}
    Because edge formation requires two individuals to simultaneously choose to form a connection, standard equilibrium definitions that consider a single player's strategy in isolation (i.e., Nash equilibrium) do not yield intuitively stable arrangements.
    For example, the strategy profile where no one proposes connections with anyone else is a Nash equilibrium, even though any pair of individuals might benefit from connecting.
    Instead of Nash equilibria, we will consider a coalitional refinement of Nash equilibria.
    
    Our equilibrium notion, which is similar to stability in stable matching, will allow for (1) any pair of individuals to coordinate to form a connection between them and, \textit{simultaneously}, (2) a member of the pair to unilaterally drop some of their connections with others.
    Both of these conditions are optional: a pair of individuals can form a connection between them without dropping any connections, and each individual can drop connections without forming a connection.
    If no such moves are possible, an edge set is said to be an \textit{\equilibrium} (\eqbm).
    Like stable matching, our equilibrium notion is motivated by our intuition that, in equilibrium, there should exist no pair of individuals who would each rather be connected with the other than one of their existing connections.

    Formally, an edge set $E$ is defection-free pairwise Nash if
    \begin{enumerate}
        \item 
        for all $i,j$ such that $(i,j) \not\in E$ and for all $S_i \subseteq N_i(E)$, $S_j \subseteq N_j(E)$
        \begin{align}
            0 &\geq {u_i(E \cup \{ (i,j) \} \setminus \{ (i,\ell) \; : \; \ell \in S_i \})} - {u_i(E)}, \;\text{or} \nonumber \\
            0 &\geq {u_j(E \cup \{ (i,j) \} \setminus \{ (j,\ell) \; : \; \ell \in S_j \})} - {u_j(E)}, \label{eq:noadd}
        \end{align}
        \item and for each $i,j$ such that $(i,j) \in E$, it holds
        \begin{align}
            0 &\geq {u_i(E \setminus \{ (i,j) \} )} - {u_i(E)}, \;\text{and}  \nonumber \\
            0 &\geq {u_j(E \setminus \{ (i,j) \} )} - {u_j(E)}.\footnotemark\label{eq:nodelete}
        \end{align}
    \end{enumerate}
    \footnotetext{In \cref{prop:multipleedgeeq}, we show that this condition is equivalent to one in which an individuals can sever an arbitrary number of edges in one move (rather than just a single edge as we define it).}
    A minor variation of \eqbm~is a equilibrium concept proposed but not analyzed in the final section of \citet{jackson2003strategic} and used in \citet{WATTS2001331}, although neither of these works use our name for it.\footnote{\citet{jackson2003strategic} and \citet{corbo2005price} both primarily analyze \textit{pairwise Nash equilibria} (PNE), in which a strategy set is an equilibrium if no node could unilaterally sever an edge and be strictly better off and no pair of nodes could form a link between them where at least one node would be strictly better off. \citet{blume2013network} considers a definition admitting an even larger set of equilibria (a relaxation of PNE) in which, informally, a network is stable if no node could drop \textit{all} of its connections and be strictly better off and no pair of nodes could form a new link between them and be strictly better off.}
    In our game, \eqbm~yields interesting structural properties in equilibria, namely \textit{status homophily}, which we explore in \cref{sec:noedgecost}.  

    One of the primary reasons we use \eqbm~as our equilibrium concept is for some of the same reasons that stability in the stable matching problem is defined to be defection-free: pairs of individuals are imagined to easily coordinate with each other to simultaneously form a connection and sever other connections.
    Our model is a generalization of the stable matching problem where each individual can match with multiple people and the preferences of individuals may vary depending on the degree of the others they are matched with. 
    We use the terms \textit{stable} and \textit{equilibrium} networks to describe graphs satisfying \eqbm.
    When we are referring to an equilibrium with a particular network structure, for example, an equilibrium on a regular graph, we sometimes just refer to it as a {regular} equilibrium as a shorthand.
    
    \paragraph{Price of anarchy.} The price of anarchy, denoted $\mathrm{\poa}(n, q, p, \gamma)$ \cite{koutsoupias1999worst} is the ratio between the optimal social welfare and the worst-case equilibrium social welfare in a game. 
    Formally, for fixed parameters $q, p , \gamma$ and a population of size $n$, let $\cE_{\mathrm{\eqbm}}(n, q, p, \gamma) \subseteq 2^{([n]^2)}$ be the set of \eqbm~edge sets.
    Then the price of anarchy for our context is defined as 
    \begin{align*}
        \mathrm{\poa}(n, q, p, \gamma) \defeq \frac{ \max_{E \subseteq [n]^2} u_{[n]}(E; q, p, \gamma)}{ \min_{E \in \cE_{\mathrm{\eqbm}}(n, q, p, \gamma)} u_{[n]}(E; q, p, \gamma)}.
    \end{align*}
    The price of anarchy is bounded below by one (because the set of edge sets over which we are maximizing in the numerator is a superset of the set of edge sets in the denominator), and in some games can be unboundedly large.
    The higher the price of anarchy, the greater the difference in utilities between a centrally planned solution and one that forms organically as a result of uncoordinated individual behavior.

\section{Equilibrium and efficiency analysis} \label{sec:noedgecost}

    We split our analysis into the regime without connection costs ($\gamma = 0$) and that with connection costs ($\gamma > 0$).
    Understanding the case where $\gamma = 0$, which we call \textit{frictionless network formation}, is an informative building block towards the perhaps more realistic case where connection costs are non-zero, which we call \textit{costly network formation}.
    The frictionless context also allows us to develop many of the key qualitative insights we will present about our model: 
    We observe how individuals over-connect at equilibrium and induce more congestion in opportunity transfer.
    This allows us to show that there is a price of anarchy, and that efficient networks yield strict Pareto improvements (i.e., every individual would be strictly better off) relative to equilibrium networks.
    
    When network formation is costly, reasoning about equilibrium networks becomes more complicated, since individuals must now balance the benefits of forming connections against their costs.
    We reproduce the qualitative findings of the frictionless case and make several new observations.
    First, we develop structural insights about equilibrium networks: there may exist many different equilibrium networks, unlike in the frictionless case, but they all exhibit a form of \textit{status homophily}, where almost all individuals connect exclusively with others who have a similar probability of transferring them an opportunity to their own.
    Second, we observe the price of anarchy can be much larger in the case where connection costs are nonzero, compared to frictionless, as a result of the fact individuals still over-connect \textit{and} individuals' utilities decrease as a result of the costs. 
    Surprisingly, the price of anarchy is worse when connection costs are smaller!
    
    Before we dive into the results, we will give an explicit form for the expected utility of each individual for a given network in terms of its adjacencies, the deficiency probability $q$, the surplus probability $p$ and connection costs $\gamma$:
    \begin{align}
        {u_i(E; q, p, \gamma)} &= 1 - \prob{X_i = 0} \probgiv{\cap_{j \in N_i(E)} R_{ji} = 0}{X_i = 0} - \gamma d_i \nonumber \\
        &= 1 - q \prod_{j \in N_i(E)} \probgiv{R_{ji} = 0}{X_i = 0} - \gamma d_i \nonumber \\
        &= 1 - q \prod_{j \in N_i(E)} \paren{1 - \frac{p}{d_j}} - \gamma d_i \label{eq:pngutil}
    \end{align}

    \subsection{Frictionless network formation.}

    In the frictionless regime ($\gamma=0$), we first describe stable networks.
    In the game, there is a unique equilibrium in which individuals form as many connections as possible.
    This is because an edge between individuals always benefits them.
    Next, we describe socially optimal networks. 
    Despite the fact that there are no costs associated with forming edges, efficient networks are sparse:
    they consist of a matching between individuals.
    This is informative about the relative costs and benefits of connectivity versus congestion:
    when an individuals forms a second connection, the negative externalities to society of that connection outweigh their benefits to the two individuals forming the connection.
    Finally, we calculate the price of anarchy in the game and make several observations.

        \paragraph{Equilibrium networks.} 
        Equilibria in our basic model without connection costs are easy to calculate: each person should form as many connections as possible.
        %
        
        \begin{restatable}{proposition}{indiscselfish}\label{prop:indiscselfish}
            In the \indiscgame, if $\gamma = 0$, for all $E \subseteq [n]^2$ and $(i,j) \not\in E$, 
            \begin{align*}
                u_i(E \cup \{(i,j) \}) > u_i(E).
            \end{align*}
        \end{restatable}
        
        Intuitively, \cref{prop:indiscselfish} holds because each individual unconnected to another individual has no chance of receiving an opportunity from them, but if they connect, there is some chance; therefore it is better to connect when given the chance than to not.
        The unique equilibrium is the complete network.

        \paragraph{Efficient networks.} A social planner optimizing over connections in the network must balance {connectivity}, which says that forming some connections is better than none,  with {congestion}, which says that too many connections lead more opportunities to be routed to individuals that don't need them.
    \ifproofsinbody
            \proofof{\cref{prop:indiscselfish}} 
            Let $E$ be any edge set and suppose there exists some pair of individuals $(i,j) \not\in E$.
            %
            %
            Observe that
            \begin{align*}
                u_i(E) = 1 - q \prod_{\ell \in N_i(E)} \paren{1 - \frac{p}{d_\ell}}
            \end{align*}
            and 
            \begin{align*}
                u_i(E \cup \{ (i,j) \}) = 1 - q \paren{1 - \frac{p}{d_j+1}}\prod_{\ell \in N_i(E)} \paren{1 - \frac{p}{d_\ell}}
            \end{align*}
            which implies
            \begin{align*}
                u_i(E \cup \{ (i,j) \}) - u_i(E\} &= \frac{qp}{d_j+1} \prod_{\ell \in N_i(E)} \paren{1 - \frac{p}{d_\ell}} \\
                > 0.
            \end{align*}
        \qed
    \fi
%
%
        %
        In \cref{prop:obliviousoptimal}, we state that, unlike equilibrium networks, efficient networks are sparse.
        In particular, maximal matchings are asymptotically efficient. 
        %
        
        \begin{restatable}{proposition}{obliviousoptimal} \label{prop:obliviousoptimal}
            In the \indiscgame, if $\gamma = 0$, a network is efficient if it is a perfect matching. Maximal matchings (if $n$ is odd) achieve average utility within $o(1)$ of the social optimum.
        \end{restatable}

        The proposition provides a quantitative characterization of how the two factors can be balanced optimally. Without connection costs, individuals should each form one connection; any more and sometimes multiple opportunities will be randomly allocated to the same person, and some of the opportunities will be wasted.
        The result tells us that any edge that is formed as an individuals' second connection causes greater negative externalities than the benefits it brings to the newly-connected pair.
        \ifproofsinbody
                \proofof{\cref{prop:obliviousoptimal}} 
            A tight upper bound for the optimization problem can be derived as follows:
            \begin{align*}
                \max_{E \subset [n]^2} \;\; \frac{1}{n} \sum_{i=1}^n {u_i(E)} &= \max_{E \subset [n]^2} \frac{1}{n} \sum_{i = 1}^n 1 - {q}\prod_{j \in N_i(E)} \paren{1 - \frac{p}{d_j}}\\
                &= 1 - \frac{q}{n} \min_{E \subset [n]^2} \sum_{i = 1}^n \prod_{j \in N_i(E)} \paren{1 - \frac{p}{d_j}} 
            \end{align*}
            where the right-hand side product will be set to 1 if $d_i = 0$.
            Next, we lower bound and simplify the expression inside the minimization. Define $a_j \defeq (1 - p /d_j)$ for all $j$. Notice:
            \begin{align*}
                \frac{1}{n}\sum_{i = 1}^n \prod_{j \in N_i(E) } a_j &\geq  \paren{\prod_{i = 1}^n \prod_{j \in N_i(E)} a_j }^{1/n}  \yesnum \label{eq:amgm}\\ 
                &= \paren{\prod_{i = 1}^n \prod_{j = 1}^n a_j^{\indic{(i,j) \in E}}}^{1/n}  \\ 
                &= \paren{\prod_{j = 1}^n \prod_{i = 1}^n a_j^{\indic{(i,j) \in E}}}^{1/n}  \\ 
                &= \paren{\prod_{j = 1}^n a_j^{\sum_{i=1}^n \indic{(i,j) \in E}}}^{1/n}  \\ 
                &= \paren{\prod_{j = 1}^n a_j^{d_j}}^{1/n} \yesnum \label{eq:degseq}
            \end{align*}
            where the inequality in \cref{eq:amgm} comes from application of the AM-GM inequality.
            Next, notice that, since \cref{eq:degseq} depends only on the degree sequence of the nodes, we can write the minimization over the lower bound in \cref{eq:degseq} equivalently as
            \begin{align*}
                \min_{d_1, \dots, d_n \in \N} \prod_{j \in [n]} \paren{1-\frac{p}{d_j}}^{d_j / n} &= \prod_{j \in [n]} \min_{d_j \in \N} \paren{1-\frac{p}{d_j}}^{d_j/n} \\
                &=  \min_{d \in \N} \paren{1-\frac{p}{d}}^{d} ,
            \end{align*}
            where we use the notation $(1 - p/0)^0 = 1$ for the case when $d_j = 0$. Additionally, the application of the AM-GM inequality in \cref{eq:amgm} is tight for a $d$-regular graphs.
            Now, we just want to show that the optimal choice of $d$ is $1$ and that the bound is achieved by a perfect matching. This is shown by observing the fact that $1 - \ptwo < 1$ (so the expression is smaller at $d = 1$ than at $d = 0$) and that 
            $\paren{1 - {\ptwo / d}}^d$ is monotone increasing as a function of $d > 1$ from \cref{lem:monotoneincreasing}. 

            Now we deal with the case in which a perfect matching is infeasible because the number of individuals $n$ is odd. 
            The utility of a maximal matching is at least $(n-1)(1-q(1-p))/n$ since the utility of the matched individuals is $1-q(1-p)$ and the unmatched individual can have utility no less than 0.
            The proof above shows that average utility can be no more than $1-q(1-p)$, which implies that average utility in the maximal matching is within $(1-q(1-p))/n$ of optimal.
            \qed
        \fi
        The intuition is as follows: 
        An individual $i$ with just one connection to another person with just one connection derives utility $1 - q(1-p)$. 
        An individual with $n$ connections connected exclusively to others with $n$ connections derives utility $1 - q(1-p/n)^n \approx 1 - q e^{-p}$.
        Since $1-p < e^{-p}$, the degree-one person connected to another degree-one person is better off.
        
        \paragraph{The price of anarchy.} Now, using our characterizations of equilibrium and efficient networks above, we are prepared to compute the price of anarchy under frictionless network formation.
        \begin{figure}[t]
            \centering
            \begin{subfigure}[b]{0.45\textwidth}
                \includegraphics[width=\textwidth]{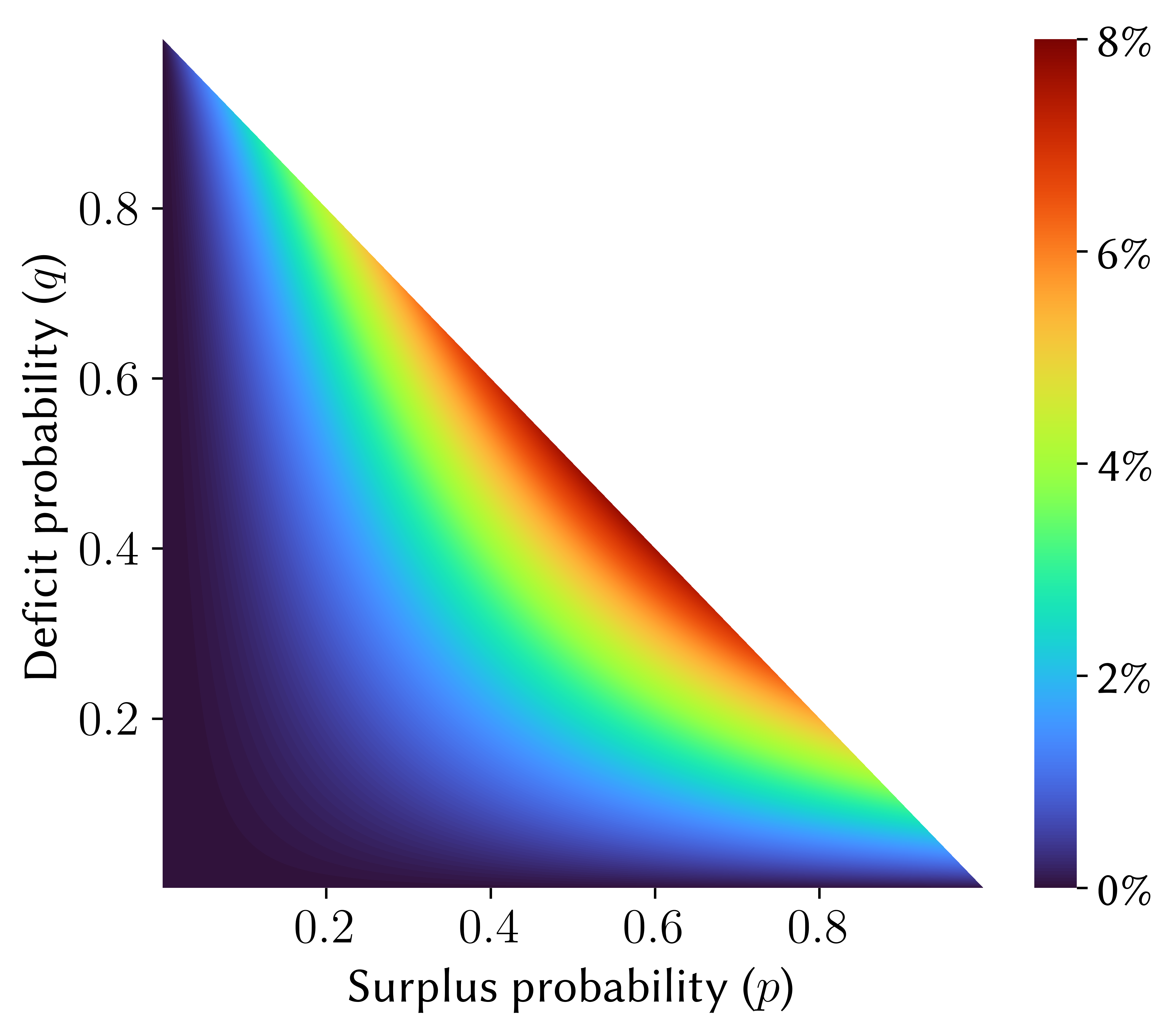}
                \caption{Frictionless network formation}
                \label{fig:oblivious}
            \end{subfigure}
            \hfill 
            \begin{subfigure}[b]{0.45\textwidth}
                \includegraphics[width=\textwidth]{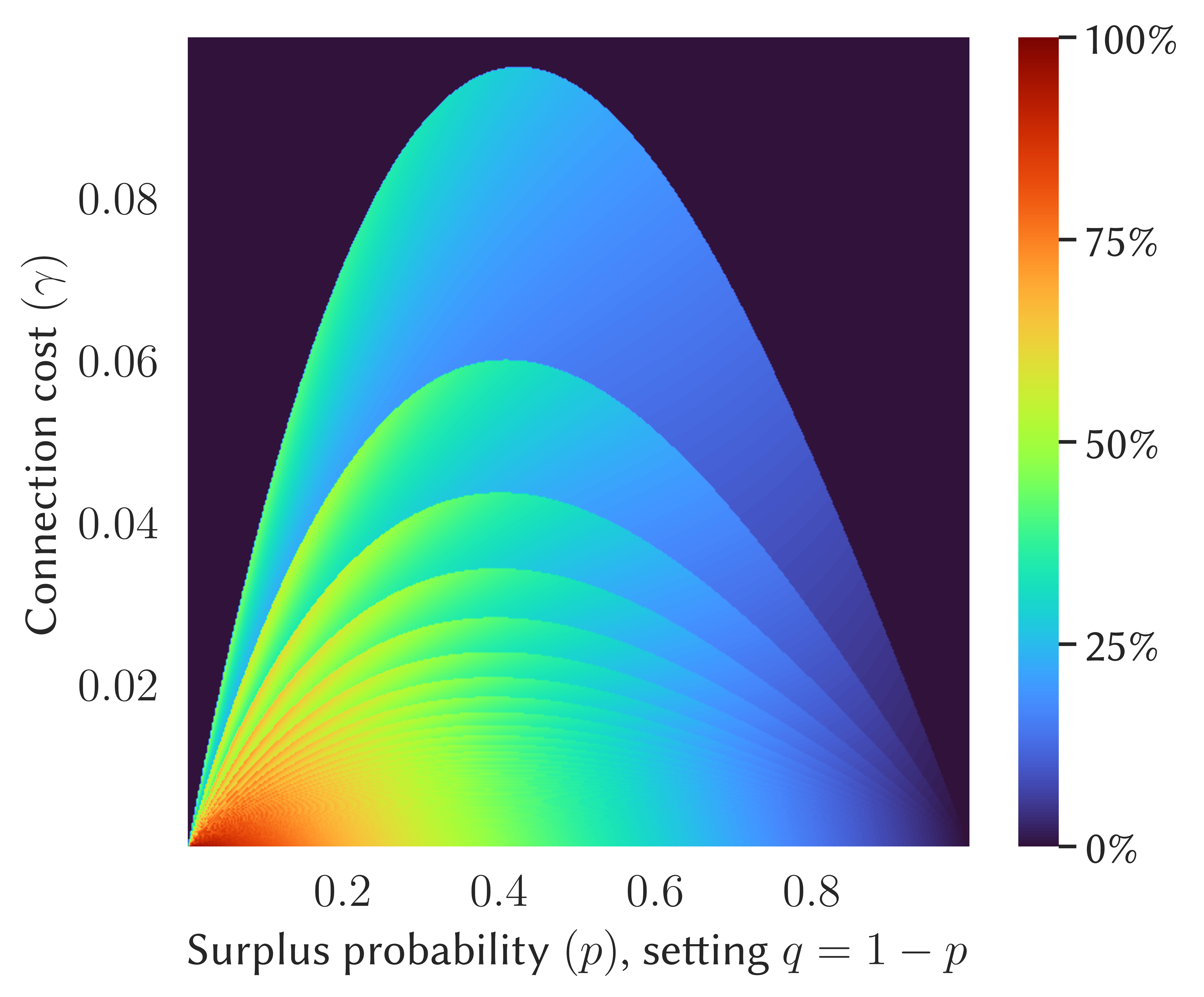}
                \caption{Costly network formation}
                \label{fig:obliviousedgecosts}
            \end{subfigure}
            \caption{A visualization of the price of anarchy in professional networking games without (left) and with (right) connection costs. Color represents the price of anarchy, where we normalize the scale as percentage above 1 (i.e., $100(\mathrm{\poa} - 1)$). Left: We vary over $q$ and $p$, plotting the result in \cref{prop:indiscpoa}. Since $p + q \leq 1$ the price of anarchy is given over the 2-dimensional simplex. Right: We vary over $p$ and $\gamma$, setting $q = 1-p$. The color displayed is the lower bound in \cref{prop:poaedgecosts}. }
            \label{fig:comparison}
        \end{figure}
    
        \begin{restatable}{theorem}{indiscpoa} \label{prop:indiscpoa}
            The asymptotic price of anarchy in the \indiscgame, if $\gamma = 0$, is 
            \begin{align*}
                \mathrm{\poa}(n, q, p, 0) &= \frac{1-\pzero(1-\ptwo)}{1-\pzero e^{-\ptwo}} \pm o(1)
            \end{align*}
            and efficient networks are a strict Pareto improvement over equilibrium networks.
        \end{restatable}

        The price of anarchy as a function of $q$ and $p$ is plotted in \cref{fig:oblivious}.
        One of our key and possibly counter-intuitive findings is that a price of anarchy exists at all!
        Canonical network formation games in the literature (see, e.g., \citet{fabrikant2003network}) exhibit a price of anarchy resulting from the fact that some nodes have to pay costs associated with edges while other nodes, which benefit from short paths to other nodes along the edges, do not.
        In our game, no individual benefits from a connection they are not incident to, so there is no possibility of free riding.
        Instead, the price of anarchy comes from congestion (in the transfer of opportunities) where multiple opportunities are routed to the same individual who cannot use them all.
    
        Notice that at extreme values of $p$ and $q$, the price of anarchy disappears.
        Intuitively, this is because the model becomes trivial and equilibrium effects become a non-factor: for example, when $q \to 0$, the probability that any individual needs an opportunity becomes negligible and connections become unnecessary for an individual to reach their bandwidth.
        The price of anarchy increases at intermediate values where individuals are more likely to need opportunities and also more likely to receive them if they connect with others.
        The maximum price of anarchy occurs at $\ptwo\approx0.55$ and $\pzero = 1 - \ptwo$.
        That it happens when individuals are slightly more likely than not to receive an extra opportunity is a result of the fact that the probability than an individual needs an opportunity but does not receive one from their network decreases faster in equilibrium as $p$ increases than it does at the social optimum, for fixed $q$.
    \ifproofsinbody
            \proofof{\cref{prop:indiscpoa}} 
            All that we need in addition to \cref{prop:indiscselfish} and \cref{prop:obliviousoptimal} is to calculate social welfare for each of the efficient and equilibrium networks.
            
            \cref{prop:indiscselfish} states that the complete graph is the unique (and therefore worst-case) equilibrium.
            Thus, the average utility is
            \begin{align*}
             \frac{1}{n}\sum_{i=1}^n {u_i(\nashedgeset)}  &= 1 - q \paren{1 - \frac{p}{n-1}}^{n-1} \\
             &= 1 - q e^{-p} \pm o(1).
            \end{align*}

            \cref{prop:obliviousoptimal} tells us that maximal matchings are efficient (up to a additive constant of $O(n^{-1}$).
            Thus, average utility is
            \begin{align*}
             \max_{E \in [n]^2} \;\; \frac{1}{n}\sum_{i=1}^n {u_i(E)}   = 1 - q (1 - {p}) \pm o(1).
            \end{align*}
            Dividing the social optimum by the unique equilibrium, we have
            \begin{align*}
                \frac{1 - q (1-p) \pm o(1)}{1 - q e^{-p} \pm o(1)} &= \frac{1 - q (1-p)}{1 - q e^{-p} \pm o(1)} \pm \frac{o(1)}{1 - q e^{-p} \pm o(1)} \\
                &= \frac{1 - q (1-p)}{1 - q e^{-p} \pm o(1)} \pm o(1) \\
                &= \paren{1 - q (1-p)} \sum_{k=1}^\infty \paren{qe^{-p} \pm o(1)}^k \pm o(1) \\
                &= \paren{1 - q (1-p)} \sum_{k=1}^\infty\paren{qe^{-p}}^k \pm o(1) \\
                &= \frac{1 - q (1-p) }{1 - q e^{-p} }\pm o(1)
            \end{align*}
        \qed
    \fi

    \subsection{Costly network formation.} \label{sec:edgecost}
        One of the reasons to consider a model with connection costs is the basic fact that individuals in real-world settings exert effort to connect with others: they must navigate the user interface of an online platform, message the potential connection, or otherwise spend effort making their acquaintance.
        In addition to the cost of creating the connection, individuals may incur costs for maintaining professional contacts from the time necessary to reach out, keep each other informed about each of their activities or otherwise regularly communicate.
        These facts are consistent with our intuition that the equilibria derived in the previous section seem unrealistic: real world platforms are just too big for everyone to connect with everyone else.

        In this section, we first develop structural insights about equilibrium graphs.
        The first insight is that individuals form only at most a constant number (depending on $p, q, \gamma$) of connections at equilibrium, unlike in the frictionless case.
        The second is that they exhibit a form of {status homophily} where almost all individuals have connections who provide an equal probability of transferring an opportunity to them as they do to others.
        In our basic model, the marginal value an individual contributes to a connection is solely determined by their degree, so status homophily manifests as individuals connecting exclusively to others with similar degree, which we call \textit{degree homophily}.
        At the end of this section, we generalize our model to individuals with heterogeneous exogenous opportunity distributions and provide a status homophily result for that more general model.
        We also show that the price of anarchy is much greater in the regime with connection costs, and that the price of anarchy grows larger as costs decrease to zero.
        \ifconferencesubmission
        \else 
        This reveals a discontinuity in the price of anarchy at $\gamma = 0$: when $\gamma$ is small but nonzero, the price of anarchy can be large, but when $\gamma$ is exactly zero it is smaller.
        \fi

        Before we state the results, we will give explicit expressions for the equilibrium conditions in \cref{eq:noadd} and \cref{eq:nodelete} in terms of the model parameters and equilibrium edge set. 
        These conditions play a central role in both the statements of certain results and extensively throughout the proofs. 
        Let $E$ be an equilibrium edge set for the \indiscgm~with parameters $q, p$ and $\gamma$.
        Then, for all $(i,j) \not \in E$ and for all $S_i \subseteq N_i(E)$, $S_j \subseteq N_j(E)$, it holds
        \begin{align}
            &0 \geq q \paren{\prod_{k \in S_i} \paren{1 - \frac{p}{d_k}} - \paren{1 - \frac{p}{d_j+1}}} \prod_{\ell \in N_j(E) \setminus S_j} \paren{1 - \frac{p}{d_\ell}}- \gamma \paren{1 - \abs{S_i}}, \; \text{or} \nonumber \\
            &0 \geq q \paren{\prod_{k \in S_j} \paren{1 - \frac{p}{d_k}} - \paren{1 - \frac{p}{d_i+1}}} \prod_{\ell \in N_j(E) \setminus S_j} \paren{1 - \frac{p}{d_\ell}}- \gamma \paren{1 - \abs{S_j}} \label{eq:noaddexplicit}
        \end{align}
        and for each $i,j$ such that $(i,j) \in E$, it holds 
        \begin{align}
            &0 \geq \gamma - \frac{q p}{d_j} \prod_{\ell \in N_i(E) \setminus \{ j \}} \paren{1 - \frac{p}{d_\ell}}, \;\text{and} \nonumber \\
            &0 \geq \gamma - \frac{q p}{d_i} \prod_{\ell \in N_i(E) \setminus \{ j \}} \paren{1 - \frac{p}{d_\ell}}. \label{eq:nodeleteexplicit}
        \end{align}
        The first set of conditions says that the costs of any new connections must outweigh the benefits of reducing the probability that they receive no opportunities (in addition to the corresponding costs and benefits of possibly dropping any number of other connections).
        The second set says that any existing connection must bring a greater marginal increase in the probability of receiving an opportunity than the cost of the connection.
        
        \paragraph{Equilibrium networks.} 
        First, we can use the conditions above to show that the degrees of individuals in equilibrium are bounded by a constant (depending on $p, q$, and $\gamma$), unlike in the frictionless case.
        (We prove that equilibria exist in \cref{prop:eqexists} in the appendix.)
        To see that individuals' degrees are bounded, fix parameters $q, p$ and $\gamma$ and an equilibrium network $E$.
        Notice from \cref{eq:nodeleteexplicit} that for any individual $j$ with nonzero degree, each $i \in N_j(E)$ must satisfy
        \begin{align*}
            &0 \geq \gamma - \frac{q p}{d_i} \prod_{\ell \in N_i(E) \setminus \{ j \}} \paren{1 - \frac{p}{d_\ell}}, \\
            \implies & 0 \geq \gamma - \frac{q p}{d_i}, \\
            \implies & d_i \leq \frac{qp}{\gamma}.
        \end{align*}
        In other words, because $i$ must be worth connecting to, they cannot have degree so high as to always provide negative marginal utility.
        In addition to bounded degree, equilibrium networks in the regime where $\gamma > 0$ exhibit a structural property where, for all but at most a constant number of individuals, each individual is connected exclusively to individuals of degree within one of their own.
        Since we use this property at various points later, we formally define $k$-degree homophily in \cref{def:degreehomophily} and then state the result in \cref{prop:layeredgraph}.

        \begin{restatable}{definition}{degreehomophily} \label{def:degreehomophily}
            An individual $i$ is said to be \textit{$k$-degree homophilous} on an edge set $E$ if for all $j \in N_i(E)$, it holds that $d_i - k \leq d_j \leq d_i + k$.
            A set of individuals is said to be $k$-degree homophilous on $E$ if all of its members are. 
        \end{restatable}

        The definition says that a set having $k$-degree homophily means that individuals have connections with degrees within $k$ of their own.
        Next, we state the result about equilibrium edge sets.
        
            \begin{restatable}{proposition}{layeredgraph}\label{prop:layeredgraph}
                For the \indiscgm~on a population of size $n$ with parameters $q, p, \gamma > 0$, in any \eqbm~edge set $\nashedgeset$,  there exists a set $\cS \subseteq [n]$ where ~$\;\abs{\cS} \geq n - (qp/\gamma)^{2}(1 + qp/\gamma)$ and $\cS$ is 1-degree homophilous on $\nashedgeset$.
            \end{restatable}
        
            Informally, \cref{prop:layeredgraph} states that if $n$ is large enough, in any equilibrium, almost all individuals are only connected to other individuals with degree within one of their own.
            Intuitively, this holds because relationships in our model are reciprocal: each participant in an edge expects to receive roughly as much as they are contributing.
            If there are individuals in asymmetric relationships (where one individual has a higher probability of passing an opportunity than the other), then they could sever that edge and form a new one with someone contributing the same amount.
            Because individuals all draw opportunities from the same exogenous distribution, an individual's contribution to an edge is dictated solely by their degree.
    
        \ifproofsinbody
                    \proofof{\cref{prop:layeredgraph}} 
                First, suppose for contradiction some pair $i,j \in [n]$ are unconnected in $\nashedgeset$, have $d_i = d_j$, and each have a neighbor $k \in N_i(\nashedgeset)$ and $\ell \in N_j(\nashedgeset)$ such that $d_k > d_i + 1$ and $d_\ell > d_j + 1$. 
                Then, $i, j$ would each rather defect and connect to each other than keep their existing connections to $k, \ell$, respectively, contradicting that $\nashedgeset$ is a \eqbm.
                Thus, in any equilibrium, any $i,j \in [n]$ such that $d_i = d_j$ where each has a neighbor $k \in N_i(\nashedgeset)$ and $\ell \in N_j(\nashedgeset)$ such that $d_k > d_i + 1$ and $d_\ell > d_j + 1$ must also be connected to each other.
                  
                However, at most $d$ nodes of a given degree $d$ can have this property: Each individual has $d$ connections; at least 1 edge goes to a node of degree more than $d+1$ by assumption; this implies that they are connected to no more than $d-1$ others with a connection of degree $> d+1$, forming at most a clique of $d$ nodes.
                Thus, for all $d$ such that there exists $i \in [n]$ such that $d_i = d$, it must hold that all but at most $d$ of $i \in [n]$ such that $d_i = d$ have the property that for all $j \in N_i(\nashedgeset)$ that $d_j \leq d_i + 1$. 
                 
                Define $\cV$ to consist of all individuals $i \in [n]$ such that for all $j \in N_i(\nashedgeset)$ that $d_j \leq d_i + 1$. (This is the right-hand inequality in the 1-degree homophily condition.)
                Notice that $\abs{\cV}$ is no smaller than $n - (qp/\gamma)^2$, since the number of distinct degrees of individuals in $\cV$ is no more than $qp/\gamma$ by \cref{prop:edgecostseqdegree} and at most $d$ individuals of degree $d$ can have connections of degree greater than $d+1$ in any equilibrium.
                Let $\cS \defeq \{ i \in \cV \; : \; \forall j \in N_i(\nashedgeset), j \in \cV \}$ consist of individuals whose neighbors are each also in $\cV$.
                Notice that for all $i \in \cS$ and $j \in N_i(\nashedgeset)$, it must also be that $d_i \leq d_j + 1$. (This is the left inequality in the 1-degree homophily condition.)
                 
                All that remains is to prove that $[n] \setminus \cS$ consists of no more than $(qp/\gamma)^{2}(1 + qp/\gamma)$ individuals. 
                But since $[n] \setminus V_n$ consists of no more than $(qp/\gamma)^{2}$ individuals and all individuals must have degree upper bounded by $qp/\gamma$, the number of individuals connected to individuals in $[n] \setminus \cV$ must also be no more than $(qp/\gamma)^{3}$. 
                Finally, noticing $\cS$ satisfies 1-degree homophily, we have proved the result.
            \qed

        \fi

        \paragraph{Efficient networks.} As in the case without connection costs, maximal matchings are asymptotically optimal in the \indiscgm~with connection costs as long as the costs are not too large.
        %
        
        \begin{restatable}{proposition}{obliviousoptimaledgecosts} \label{prop:obliviousoptimaledgecosts}
            In the \indiscgame, if $0 < \gamma \leq qp$, a maximal matching achieves average utility within $o(1)$ of the social optimum (and is exactly optimal if $n$ is even). 
            If $\gamma > qp$, the unique social optimum is the empty graph.
        \end{restatable}

        If connection costs $\gamma$ are too high, the benefit of even forming one connection is outweighed by the cost: The utility of someone with no connection is $1-q$ and with one connection of degree one is $1 - q (1-p) - \gamma$.
        If $\gamma > qp$, the former is greater than the latter.
        Otherwise, if $\gamma$ is not too large, the reason that a maximal matching is still socially optimal in the costly network formation case is that the negative externalities associated with forming more than one connection outweigh the benefits even when $\gamma = 0$, so reducing utility further with connection costs only makes it less socially desirable to create denser networks. 

        \ifproofsinbody
        
        \proofof{\cref{prop:obliviousoptimaledgecosts}} 
        It is straightforward to generalize the proof of \cref{prop:obliviousoptimal}.
        All that is needed is to notice that 
            \begin{align*}
                \min_{d_1, \dots, d_n \in \N} \prod_{j \in [n]} \paren{1-\frac{p}{d_j}}^{d_j / n} - \gamma \sum_{j=1}^n d_j &= 
                \min_{d \in \N} \paren{1-\frac{p}{d}}^{d} - \gamma n d,
            \end{align*}
            which can be used to see that, if $\gamma \leq qp$, the optimal choice of $d$ is again 1 and otherwise achieved at $d = 0$, the empty network.
        \qed
        \fi
        \paragraph{The price of anarchy.}
        Before we state \cref{prop:poaedgecosts}, we need to define notation we will use throughout the rest of this paper.
        Many of our results depend on upper and lower bounds on the degrees of individuals in equilibrium.
        Sometimes, there will be a gap of one or two between these bounds (due to the looseness of the 1-degree homophily condition), which will propagate through the results.
        \newcommand{\equnder}[1]{\underset{#1}{=}}
        To ease presentation, we give the following definition.
        
        \begin{definition}
            For a constant $a \in \R$ and function $f \; : \; [0,1]^k \to \R$ we will define
            \begin{align*}
                a \equnder{\delta_1,\dots,\delta_k} f(\delta_1,\dots,\delta_k)
            \end{align*}
            to mean
            \begin{align*}
                \min_{\delta_1, \dots, \delta_k \in [0, 1]} f(\delta_1, \dots, \delta_k) \leq a \leq \max_{\delta_1, \dots, \delta_k \in [0, 1]}f(\delta_1, \dots, \delta_k).
            \end{align*}
        \end{definition}
        Throughout this paper, we will use the symbols $\delta_1, \dots, \delta_k$ exclusively for this purpose. We will usually not define $f$ explicitly; instead it will be an expression with $\delta_1, \dots, \delta_k$ in it on the right-hand side of $=_{\delta_1, \dots, \delta_k}$.
        Using this notation, we state our price of anarchy result.
    
        \begin{restatable}{theorem}{poaedgecosts} \label{prop:poaedgecosts}
                The price of anarchy in the \indiscgame~with $0 < \gamma \leq qp$ is
                \begin{align*}
                    \mathrm{\poa}(n, p, q, \gamma) \equnder{\delta_1, \delta_2} \frac{1 - q(1-p) - \gamma}{1 - q \paren{1 - \frac{p }{d+\delta_1 + \delta_2}}^{d+\delta_1}- \gamma (d + \delta_1)} \pm o(1)
                \end{align*}
                and $d$ is the largest integer satisfying 
                \begin{align}
                    \frac{\gamma}{qp} \leq \frac{1}{d} \paren{1 - \frac{p}{d}}^{d-1}. \label{eq:dregval}
                \end{align}
                There is no price of anarchy if $\gamma > qp$.
            \end{restatable}

            Our lower bound on the price of anarchy (setting $\delta_1 = \delta_2 = 0$) is shown in \cref{fig:obliviousedgecosts}.
            Notice that \cref{eq:dregval} is a rearrangement of \cref{eq:nodeleteexplicit} for a $d$-regular network.
            This is not an accident: the denominator in our price of anarchy depends on showing that mean worst-case equilibrium utility cannot be much lower than that of the largest $d$ for which there is a $d$-regular equilibrium.
            Generally, throughout our work, the degree homophily result allows us to work with equilibria that are nearly regular, which simplifies things considerably.
            It also suggests that generalizations of degree homophily, which we turn to in \cref{sec:statushomophily}, are the key to understanding \equilibria~across a range of network formation games similar to ours.
            \ifconferencesubmission
            \else
            
            \Cref{eq:dregval} is also the reason that there are discontinuities in \cref{fig:obliviousedgecosts} as we move through the parameter space: as the parameters change, different values of $d$ become feasible, leading to different worst-case equilibrium utilities.
            That there is only a price of anarchy when $0 < \gamma \leq qp$ comes from the fact that, as we indicated in the discussion of \cref{prop:obliviousoptimaledgecosts}, if $\gamma > qp$, both equilibrium and efficient networks are the empty network.
            We can also notice that our upper and lower bounds (depending on $\delta_1, \delta_2$) get tighter for smaller $\gamma$, since smaller $\gamma$ leads to larger $d$ so that eventually, $d \approx qpe^{-p}/\gamma$ and the $\delta_1, \delta_2$ terms change the function infinitesimally.

            Additionally, notice that the price of anarchy increases to nearly 2 (or, in the scale of \cref{fig:obliviousedgecosts}, 100\%) when $p$ is close to 0, $q = 1-p$ and $\gamma$ is close to zero. 
            We can informally (and without precisely quantifying rates of convergence) understand why this is the case by making the following observations: When $\gamma$ is small, \cref{eq:dregval} tells us $d \approx qpe^{-p}/\gamma$, so that the denominator of \cref{prop:poaedgecosts} is close to $1 - q(1-p)e^{-p} \approx 1 - (1-p)^3$ for $p$ close to 0 and $q = 1-p$.
            Since the numerator approaches $1 - (1-p)^2$ for $\gamma \to 0$ and $q = 1-p$, we can see that $(1 - (1-p)^2)/(1 - (1-p)^3)$ becomes large for $p$ close to 0.
            On the other hand, for $\gamma = 0$, $q = 1-p$ and $p$ small, we have from \cref{prop:indiscpoa} that the numerator of the price of anarchy is $1-(1-p)^2$ and the denominator is $1 - (1-p)e^{-p} \approx 1 - (1-p)^2$.
            Thus, the numerator and denominator approach 1 at approximately the same rate, and the price of anarchy is close to 1.
            The combination of the observations in this paragraph tell us something fundamental about the difference between network formation when $\gamma = 0$ and when $\gamma$ is small but positive.
            When $\gamma = 0$, the fact that the network is densely connected at equilibrium only incurs a modest loss of social welfare due to congestion: more opportunities are wasted due to the network being over-connected.
            When $\gamma > 0$, there is also a loss of social welfare due to nontrivial connection costs (of an additive factor of about $-pe^{-p}$ in the denominator).
            In both cases, the social welfare of efficient networks is about the same.
            Thus, equilibrium networks may have much lower worst-case utility when $\gamma > 0$ is small compared to when $\gamma = 0$.
            \fi
    \ifproofsinbody
    
            \proofof{\cref{prop:poaedgecosts}}
                First, we calculate the social optimum utility. 
                From \cref{prop:obliviousoptimaledgecosts}, we know that a maximal matching is approximately optimal (within an additive factor of $o(1)$), which yields average utility $1 - q (1-p) - \gamma \pm 0(1)$, if $\gamma < qp$, and average utility $1-q$ otherwise.

                Next, we calculate the equilibrium utility.
                If $\gamma > qp$, for the same reason that it is not socially optimal to add edges, the unique equilibrium will be the empty graph.
                To see this, for contradiction suppose some individual $i \in [n]$ has at least one connection.
                Then their utility is upper bounded by
                $1-q(1-p)^{d_i} - \gamma d_i$, which would occur if $i$ was each of their connections' only connection.
                But it is easy to show that for all ${d_i} \geq 1$, $i$'s utility would be higher if they had no connections:
                \begin{align*}
                    &1 - q > 1 - q(1-p)^{d_i} - \gamma d_i, \\
                    \impliedby &1 - q > 1 - q (1-p) - \gamma, \\
                    \impliedby &\gamma > qp, \\
                \end{align*}
                where the second line comes from the fact that the right-hand side is monotonic decreasing in $d_i$ for $d_i \geq 1$. This can be seen by the following:
                \begin{align*}
                    &\frac{\dif }{\dif d_i}1 - q(1-p)^{d_i} - \gamma d_i < 0 \\
                    \iff &\frac{\dif }{\dif d_i} q(1-p)^{d_i} + \gamma d_i > 0 \\
                    \impliedby &(1-p)^{d_i} \log(1-p) + p > 0 \\
                    \impliedby &(1-p) \log(1-p) + p > 0 \\
                \end{align*}
                and the last line can be verified by noticing that when $p=0$, the left-hand expression is 0 and that the derivative of the expression is positive for $p \in (0, 1)$
                \begin{align*}
                    \frac{\dif }{\dif p} (1-p) \log(1-p) + p &= - \log (1-p) > 0.
                \end{align*}
                Thus, when $\gamma > qp$, equilibrium average utility will be $1-q$, and there is no asymptotic price of anarchy when $\gamma > qp$.
                
                Next, we deal with the nontrivial case where $\gamma \leq qp$.
                We will first prove that the worst-case average utility in an equilibrium must not be more than 
                \begin{align}
                    1 - q \paren{1 - \frac{p}{d}}^d - \gamma d \pm o(1)\label{eq:equtilub}
                \end{align}
                for $d$ the largest integer satisfying \cref{eq:dregval}.
                First, we prove that there always exists a  equilibrium.

                \begin{restatable}{proposition}{eqexists} \label{prop:eqexists}
    In the \indiscgm, for all $q, p, \gamma \geq 0$, there exists a \eqbm~edge set.
\end{restatable}

\proofof{\cref{prop:eqexists}}
\ifconferencesubmission
    We split the analysis in to the cases that $\gamma > qp$ and $\gamma \leq qp$. If $\gamma > qp$, the unique equilibrium will be the empty graph.
    To see this, for contradiction suppose some individual $i \in [n]$ has at least one connection.
    Then their utility is upper bounded by
    $1-q(1-p)^{d_i} - \gamma d_i$, which would occur if $i$ was each of their connections' only connection.
    But it is easy to show that for all ${d_i} \geq 1$, $i$'s utility would be higher if they had no connections:
    \begin{align*}
        &1 - q > 1 - q(1-p)^{d_i} - \gamma d_i, \\
        \impliedby &1 - q > 1 - q (1-p) - \gamma, \\
        \impliedby &\gamma > qp, \\
    \end{align*}
    where the second line comes from the fact that the right-hand side is monotonic decreasing in $d_i$ for $d_i \geq 1$. This can be seen by the following:
    \begin{align*}
        &\frac{\dif }{\dif d_i}1 - q(1-p)^{d_i} - \gamma d_i < 0 \\
        \iff &\frac{\dif }{\dif d_i} q(1-p)^{d_i} + \gamma d_i > 0 \\
        \impliedby &(1-p)^{d_i} \log(1-p) + p > 0 \\
        \impliedby &(1-p) \log(1-p) + p > 0 \\
    \end{align*}
    and the last line can be verified by noticing that when $p=0$, the left-hand expression is 0 and that the derivative of the expression is positive for $p \in (0, 1)$
    \begin{align*}
        \frac{\dif }{\dif p} (1-p) \log(1-p) + p &= - \log (1-p) > 0.
    \end{align*}
\else
    We have already argued that if $\gamma > qp$, the unique equilibrium will be the empty graph (which is $0$-regular).
\fi
Next, the case that $\gamma \leq qp$.
If there exists a $d$-regular equilibrium edge set $\nashedgeset$, it must satisfy the equilibrium conditions: For all $i,j$ such that $(i,j) \not \in \nashedgeset$, 
\begin{align*}
    0 &\geq {u_i(\nashedgeset \cup \{ (i,j) \})} - {u_i(\nashedgeset)},
\end{align*}
and for all $i,j$ such that $(i,j) \in \nashedgeset$
\begin{align*}
    0 &\geq {u_i(\nashedgeset \setminus \{ (i,j) \})} - {u_i(\nashedgeset)}.
\end{align*}
We do not need to consider the defection-free condition since it is satisfied automatically: all nodes are the same degree and thus no node has an incentive to defect.
For the $d$-regular graph, these expressions reduce to
\begin{align}
    \frac{\gamma}{qp} \geq \frac{1}{d+1} \paren{1 - \frac{p}{d}}^d
\;\;\;\;\;\;\;\; \text{ and } \;\;\;\;\;\;\;\; 
    \frac{\gamma}{qp} \leq \frac{1}{d} \paren{1 - \frac{p}{d}}^{d-1}, \label{eq:dregeqconds}
\end{align}
by substituting the closed form for each expected utility and simplifying.
Intuitively, the first condition says that no two unconnected individuals would want to connect and the second says that no individual would want to sever any of their existing connections.
Next, we will verify that there always exists a $d$ satisfying these equations when $\gamma \leq qp(1-p)$ and thus there exists a $d$-regular equilibrium.
Our argument will be that
\begin{align}
    \bigcup_{d=1}^\infty \sqparen{\frac{1}{d+1} \paren{1 - \frac{p}{d}}^d, \frac{1}{d} \paren{1 - \frac{p}{d}}^{d-1}} = (0, 1] \label{eq:dregeqexists}
\end{align}
and since $0 < \gamma / qp \leq 1$ in this regime, for any $\gamma / qp$ there must exist a $d$ such that $\gamma / qp$ falls in the interval inside the union of the left-hand side.
To prove \cref{eq:dregeqexists}, first notice that for $d=1$, the right endpoint of the interval is 1. 
Next, it is easy to see that each interval overlaps on the left with the next interval in the sequence: for any $d \in \N$, the left endpoint of the interval corresponding to $d$ is less than the right endpoint of the interval corresponding to $d+1$:
\begin{align*}
    \frac{1}{d+1} \paren{1 - \frac{p}{d}}^d < \frac{1}{d+1} \paren{1 - \frac{p}{d+1}}^d
\end{align*}
Finally, the sequence of left-endpoints of the intervals converges to zero. Formally,
\begin{align*}
    \lim_{d \to \infty} \frac{1}{d+1} \paren{1 - \frac{p}{d}}^d = 0,
\end{align*}
which completes the proof of \cref{eq:dregeqexists}. 

Recall that if $n$ or $d$ is even and $n > d$, a $d$-regular graph always exists, so the analysis above is exact. 
If $n$ and $d$ are odd, a $d$-regular graph cannot exist. 
However, consider the following network that we will claim will have almost all individuals with degree $d$. 
Start with a $d$-regular graph on $n-1$ nodes with $i$ the remaining individual.
We will argue that we can pick either $\floor{d/2}$ or $\ceil{d/2}$ pairs of connected individuals arbitrarily and drop the connection between each pair and form a connection between individual $i$ and each member of the pair, and that an equilibrium is formed for one of these two choices.
Individual $i$ will have $2 \floor{d/2}$ or $2 \ceil{d/2}$ connections, each of $i$'s connections will have $d$ connections and everyone else will have $d$ connections. 
It is easy to see that one of these is an equilibrium by checking the equilibrium conditions: 
Since the degrees of all will be within one of each other, there is no incentive to defect. 
If it is $\floor{d/2}$, the equilibrium conditions say
\begin{align*}
   \frac{1}{d}\paren{1 - \frac{p}{d}}^d \leq \frac{\gamma}{qp}, \; \text{or} \\
   \frac{1}{d+1}\paren{1 - \frac{p}{d}}^{d-1} \leq \frac{\gamma}{qp},
\end{align*}
i.e., that no individual will want to form a connection with $i$ or $i$ will not want to form a connection with them. If it is $\ceil{d/2}$ then
\begin{align*}
   \frac{1}{d}\paren{1 - \frac{p}{d}}^d \geq \frac{\gamma}{qp}, \; \text{and} \\
   \frac{1}{d+1}\paren{1 - \frac{p}{d}}^{d-1} \geq \frac{\gamma}{qp},
\end{align*}
i.e., that no individual will want to break a connection with $i$ and $i$ will not want to break their connection with $i$.
Since these conditions are complementary, one or the other set must hold.
\qed
                
                The largest $d$ satisfying the equilibrium conditions is thus given by the largest $d$ satisfying \cref{eq:dregval} since the right condition in \cref{eq:dregeqconds} gives an upper bound on $d$. 
                Since the utility in \cref{eq:equtilub} is decreasing in $d$, this gives us the worst-case $d$-regular equilibrium, which is an upper bound on the worst-case equilibrium overall.
                For the rest of this proof, we will use $d$ (without subscripts) to mean the $d$ achieving the worst-case $d$-regular equilibrium utility.
                
                Next, we will prove that the worst-case average utility in an equilibrium must not be less than 
                \begin{align*}
                    1 - q \paren{1 - \frac{p}{d+2}}^{d+1} - \gamma (d + 1)
                \end{align*}
                which will complete our proof of the $\gamma \leq qp(1-p)$ case.
                To do so, we will argue that all networks satisfying the equilibrium conditions must have utility no worse than this lower bound.
                We will leverage \cref{prop:layeredgraph}. 
                From here on, for any equilibrium edge set $E_n$, we will reason only about the utility of individuals in a 1-degree homophilous set $\cS \subseteq [n]$ satisfying 1-degree homophily and no smaller than $n$ minus a constant.
                Since for any $\varepsilon > 0$ there exists some $m$ such that for all $n > m$ it holds $\abs{\cS} > (1-\varepsilon) n$, we can just compute the average utility of individuals in $\cS$ and, since each individuals' utility is bounded between 0 and 1, the average utility of all nodes will approach this as $n \to \infty$.

                Consider a node $i \in \cS$ with the largest degree of all nodes in $\cS$. 
                Define $\dmax \defeq d_i$.
                %
                %
                Recall from the equilibrium conditions that for all $j \in N_i(E_{\mathrm{\eqbm}})$
                \begin{align*}
                    \frac{\gamma}{qp} \leq \frac{1}{d_j} \prod_{\ell \in N_i(E_{\mathrm{\eqbm}}) \setminus \{ j \}} \paren{1 - \frac{p}{d_\ell}}
                \end{align*}
                which implies
                \begin{align}
                    \frac{\gamma}{qp} 
                    &\leq \max_{u, v \in \{ \dmax -1, \dmax \}} \frac{1}{v} \paren{1 - \frac{p}{u}}^{\dmax-1} \nonumber \\
                    &\leq \frac{1}{\dmax - 1} \paren{1 - \frac{p}{\dmax}}^{\dmax-1} \label{eq:dmaxconstraint}
                \end{align}
                where the first line is a relaxation of the equilibrium condition with 1-degree homophily.
                This gives us an upper bound on the degree of nodes in $\cS$ which is tighter than the simple bound we give in \cref{prop:edgecostseqdegree} that we used to prove \cref{prop:layeredgraph}.
                Next, we will argue that a lower bound on the utility of individuals in $\cS$ is that of a (hypothetical) individual of degree $\dmax$ connected exclusively to nodes of degree $\dmax + 1$.
                For any $k \leq \dmax$ such that there exists $j \in \cS$ with $d_j = k$, utility of such an individual is lower-bounded by
                \begin{align}
                    1 - q \paren{1 - \frac{p}{k + 1}}^k - \gamma k. \label{eq:klessthandmax}
                \end{align}
                As a direction implication of \cref{prop:concave}, \cref{eq:klessthandmax} is concave in $k$. 
                Thus, the minimum over $k$ in \cref{eq:klessthandmax} is achieved by either $k=1$ or $k = \dmax$. 
                The following sequence proves that the minimum is achieved at $k=\dmax$:
                \begin{align*}
                    &1 - q \paren{1 - \frac{p}{2}} - \gamma \geq 1 - q\paren{1 - \frac{p}{\dmax + 1}}^{\dmax} - \gamma \dmax \\
                    \iff & \dmax \geq \frac{q}{\gamma} \paren{\frac{p}{2} - 1 - \paren{1 - \frac{p}{\dmax + 1}}^{\dmax}} + 1 \\
                    \impliedby & \dmax \geq \frac{q}{\gamma} \paren{\frac{p}{2} - 1 - e^{-p}} + 1 \\
                    \impliedby & \dmax \geq 1
                \end{align*}
                where the third line comes from the fact that $(1-p/\dmax)^{\dmax-1}$ is decreasing in $\dmax$ from \cref{prop:decreasing} and the last line comes from the fact that $p/2 - 1 - e^{-p} < 0$.
                Thus, utility of any individual in $\cS$ is lower-bounded by \cref{eq:klessthandmax}, plugging in $\dmax$ for $k$.
                
                Our last step in the proof is to relate $d_{\max}$ to the $d$ of the worst-case $d$-regular equilibrium network we analyzed earlier.
                The following sequence of inequalities shows $d \geq d_{\max} - 1$:
                \begin{align*}
                    d &= \max \curly{k \in \N \; : \; \frac{\gamma}{qp} \leq \frac{1}{k} \paren{1 - \frac{p}{k}}^{k-1}} \\
                    &= \max \curly{k \in \N\; : \; \frac{\gamma}{qp} \leq \frac{1}{k - 1} \paren{1 - \frac{p}{k - 1}}^{k - 2}}  - 1\\
                    &\geq \max \curly{k \in \N \; : \; \frac{\gamma}{qp} \leq \frac{1}{k - 1} \paren{1 - \frac{p}{k}}^{k - 1}} - 1 \\
                    &\geq \dmax - 1
                \end{align*}
                The first line comes from the definition of $d$; the second line comes from a change of variable from $k$ to $k-1$; the third line comes \cref{prop:decreasing} which says that $(1-p/k)^{k-1}$ is monotonic decreasing; the fourth line comes from the fact that $\dmax$ must satisfy \cref{eq:dmaxconstraint}.
                \qed
    \fi
            Before we conclude this section, we will return to the degree homophily results of \cref{prop:layeredgraph} and show that a kind of status homophily holds in a much more general model.
    
        \subsection{Generalizing degree-homophily results.} \label{sec:statushomophily} 
        The results we presented in \cref{prop:layeredgraph} can be extended beyond our basic model. 
        In this section, we consider generalizations of our model with a larger class of exogenous opportunity distributions that may be heterogeneous across individuals. Formally, suppose each individual $i$ is assigned some $p_{ik} \defeq \prob{X_i = k}$ for all $k \in \Z_{\geq 0}$. 
        Notice that this allows for an individual to receive more than 2 exogenous opportunities and for different individuals to have different exogenous opportunity distributions.
        We will assume that each individual receiving $k$ more opportunities than they need chooses $k$ distinct social contacts uniformly at random to transfer them to.
        We will assume that if $X_i > d_i + 1$, all of $i$'s social contacts will receive opportunities and the rest of the extra opportunities will be wasted.
        To make things easier later, we work with a truncated function of the random variable $\widetilde X_i(d) \defeq \max\{X_i, d + 1\}$.
        As before, each individual will have bandwidth for one opportunity.
        We will call this the \generalizedgame~(\generalizedgm).
        For our equilibrium concept, we consider \equilibrium~as before.
        
        This generalization of our model can yield a significantly richer set of equilibrium and efficient network structures: for example, consider the case that for some $i$ it holds $p_{in} = 1$, $p_{jk} = 0$ for all $j \neq i$ and all $k \in \N$, and $0 < \gamma < 1$.
        Then it is easy to see that the unique social optimum is the star graph with individual $i$ at the center.

        Before we state \cref{prop:genhomophily}, we will establish an explicit expression for the expected utility of an individual:
        \begin{align}
            {u_i(E)} &= 1 - \prob{X_i = 0} \probgiv{\cap_{j \in N_i(E)} R_{ji} = 0}{X_i = 0} - \gamma d_i \nonumber \\
            &= 1 - p_{i0} \prod_{j \in N_i(E)} \paren{p_{j0} + \sum_{k=1}^\infty p_{jk} \paren{1 - \frac{\min \{ k-1, d_j \}}{d_j}}}  - \gamma d_i \nonumber \\
            &=  1 - p_{i0} \prod_{j \in N_i(E)} \paren{1 - \frac{\ex{\widetilde X_j (d_j)} - 1 + p_{j0}}{d_j} } - \gamma d_i. \label{eq:genutilexpr}
        \end{align}
        For convenience, we will define $\mu_j(d_j) = {\ex{\widetilde X_j (d_j)} - 1 + p_{j0}}$.
        The explicit expression says that each individual's utility depends on their own parameters and a single summary statistic $\mu_j(d_j) / d_j$ about each of their neighbors.

        \begin{restatable}{proposition}{genhomophily} \label{prop:genhomophily}
            In the \generalizedgame~with $\gamma > 0$, any \eqbm~edge set $\nashedgeset$ corresponding to a population of size $n$ has the following property. For all $\varepsilon > 0$, there exists a set $\cS \subseteq [n]$ where~ $\;\abs{\cS} \geq n - \gamma^{-3}(1 + \gamma^{-1}) \varepsilon^{-1}$ and for all $i \in \cS$ and $j \in N_i(\nashedgeset)$
            \begin{align}
                \frac{1}{d_i + 1} \left\lfloor\frac{\mu_i (d_i + 1)}{\varepsilon}\right\rfloor \varepsilon \leq \frac{\mu_j(d_j)}{d_j}  \;\;\;\;\;\;\;\; \text{and}  \;\;\;\;\;\;\;\; \frac{1}{d_j + 1}\left\lfloor\frac{\mu_j(d_j+1) }{\varepsilon}\right\rfloor \varepsilon \leq \frac{\mu_i(d_i)}{d_i}. \label{eq:marginalcontribution}
            \end{align}
        \end{restatable}
        The proposition says that $\mu_j(d_j)/d_j$ is close to $\mu_i(d_i) / d_i$ for all but a constant number of individuals (not depending on $n$), up to an arbitrarily fine $\varepsilon$-grid and an additive factor of one in each individual's degree.
        The statistic $1- \mu_j(d_j)/d_j$ is the probability that $j$ will pass a given one of their connections an opportunity.
        Thus, our result provides a microfoundation for status homophily: in sufficiently large networks, almost everyone can expect to connect with others who provide a similar probability of passing them an opportunity as they do. 
        Connections can be expected to be similarly beneficial because, if they are too asymmetric, there probably exists someone else who would be willing to connect who provides better value.

        A more complete analysis of the \generalizedgm~would be a valuable direction for future research.
        In particular, it may be worth considering a larger set of interventions to platforms: While social optima are harder to characterize in the \generalizedgm~than in the \indiscgm, it is easy to see that there may be circumstances where it is welfare-maximizing to connect high-status people to low-status people (for example, see our star graph example above).
        This suggests that, when considering interventions society can make to a platform to increase social welfare or reduce inequality, it may also be worth considering subsidizing certain edge formation costs.
        
        \ifproofsinbody
        
        \proofof{\cref{prop:genhomophily}} Our proof will follow the same structure as in \cref{prop:layeredgraph}, and the main idea is the same: almost all individuals will have the option of connecting to someone with a similar marginal contribution to their utility, which ensures that not many people will be connected to others who provide lower marginal utility. 
        Formally, fix a population of size $n$ with parameters $\{ p_{ik} \}_{i \in [n],k \in \N}$, and let $\nashedgeset$ be any \eqbm~edge set. 
        We will prove that at most a constant number of individuals not depending on $n$ violate \cref{eq:marginalcontribution}.
        First, we need to state the explicit form for the marginal contribution of a (potential) connection $i'$ to an individual $i$ where $(i,i') \not \in \nashedgeset$, since we will use it in this proof:
        \begin{align*}
            {u_i(\nashedgeset \cup \{ (i,i') \}} - {u_i(\nashedgeset)} = p_{i0} \paren{\frac{\mu_{i'}(d_{i'}+1)}{d_{i'} + 1}} \prod_{j \in N_i(E)} \paren{1 - \frac{\mu_j(d_j)}{d_j} } - \gamma.
        \end{align*}
        The expression can be derived directly from \cref{eq:genutilexpr} by plugging in $\nashedgeset \cup \{ (i,i') \}$ and $\nashedgeset$ and subtracting.
        Similarly, if $i'$ is already connected to $i$ in $\nashedgeset$, then its marginal utility is
        \begin{align*}
            {u_i(\nashedgeset \setminus \{ (i,i') \}} - {u_i(\nashedgeset)} = p_{i0} \paren{\frac{\mu_{i'}(d_{i'})}{d_{i'}}} \prod_{j \in N_i(E) } \paren{1 - \frac{\mu_j(d_j)}{d_j} } - \gamma.
        \end{align*}
        Now, suppose for contradiction some pair $i, i' \in [n]$ have the following properties
        \begin{itemize}
            \item $(i, i') \not \in \nashedgeset$,
            \item $d_i = d_{i'}$,
            \item $\floor{\mu_i(d_{i}) / \varepsilon} = \floor{\mu_{i'}(d_{i'}) / \varepsilon}$
            \item there exists $j, j'$ such that $(i,j), (i',j') \in \nashedgeset$ where
            \begin{align*}
                \frac{\mu_j(d_{j})}{d_j} < \frac{1}{d_i + 1}\left\lfloor\frac{\mu_i(d_{i} + 1) }{\varepsilon}\right\rfloor \varepsilon \;\;\;\;\;\;\;\; \text{and} \;\;\;\;\;\;\;\;
                \frac{\mu_{j'}(d_{j'})}{d_{j'}} < \frac{1}{d_{i'} + 1}\left\lfloor\frac{\mu_{i'}(d_{i'} + 1) }{\varepsilon}\right\rfloor \varepsilon.
            \end{align*}
        \end{itemize}
        In other words $i$ and $i'$ are unconnected, have the same degree, fall into the same cell of the $\varepsilon$-grid and are connected to others who are connected to others who provide lower marginal utility than any member of their cell with the same degree would.
        Then, $i, i'$ would rather defect and connect to each other while severing their connections to $j, j'$, since we assumed (in the fourth bullet) that the marginal utility of severing the connections to $j, j'$ are less than that of $i,i'$ connecting to each other.
        But this contradicts the fact that $\nashedgeset$ is an equilibrium.
        Thus, if the last three conditions are to hold, then $i, i'$ must be connected to each other.
        But this means that, in an equilibrium, at most $d_i$ individuals can exist where the last three conditions hold (for given $d_i$ and $\floor{\mu_i(d_i)/\varepsilon}$), forming $d_i$-clique, since they must all be connected to each other.

        Next, we will argue that there are only finitely many $\varepsilon$-grid cell-degree combinations. 
        Since $\mu_i(d_{i}) \leq d_i$ for all $i$, there can be only $\max_{i \in [n]} d_i/\varepsilon$ cells, so we can just prove a bound on the degree of all individuals.
        Let $i$ be the individual with largest degree.
        But since utility must be nonnegative, we have
        \begin{align*}
            0 &\leq {u_i(\nashedgeset)} \\
            &= 1- p_{i0} \prod_{j \in N_i(E) } \paren{1 - \frac{\mu_j(d_{j})}{d_j} } - \gamma d_i \\
            &\leq 1 - \gamma d_i
        \end{align*}
        which implies $d_i \leq \gamma^{-1}$. 
        Thus, there can be no more than $(\gamma \varepsilon)^{-1}$ cells and no more than $\gamma^{-2} \varepsilon^{-1}$ cell, degree combinations.

        Combining the two previous paragraphs, notice that all but at most $\gamma^{-3} \varepsilon^{-1}$ individuals satisfy the left-hand inequality in \cref{eq:marginalcontribution}: there are no more than $\gamma^{-2} \varepsilon^{-1}$ cell, degree combinations in the $\varepsilon$-grid and the number of individuals in each cell violating the inequality for each cell, degree combination is no more than $\gamma^{-1}$ since at most $d$ individuals of degree $d$ can violate the inequality and each individual's degree is no more than $\gamma^{-1}$.
        Define $\cV$ to consist of all individuals $i \in [n]$ such that for all $j \in N_i(\nashedgeset)$ it holds
        \begin{align*}
            \frac{1}{d_i + 1}\left\lfloor\frac{\mu_i(d_{i} + 1) }{\varepsilon}\right\rfloor \varepsilon \leq \frac{\mu_j (d_{j})}{d_j},
        \end{align*}
        i.e., all individuals satisfying the left-hand inequality in \cref{eq:marginalcontribution}.
        Define $\cS \defeq \{ i \in \cV \; : \; \forall j \in N_i(\nashedgeset), j \in \cV \}$ consist of all individuals in $\cV$ whose social contacts are all in $\cV$.
        Notice for all for all $i \in \cS$ and $j \in N_i(\nashedgeset)$ it must hold that
        \begin{align*}
            \frac{1}{d_j + 1}\left\lfloor\frac{\mu_j(d_{j}) }{\varepsilon}\right\rfloor \varepsilon \leq \frac{\mu_i(d_{i})}{d_i},
        \end{align*}
        which is the second inequality in \cref{eq:marginalcontribution}. 
        Since the number of individuals in $[n] \setminus \cV$ is upper bounded by $\gamma^{-3} \varepsilon^{-1}$ and each individual's degree is upper bounded by $\gamma^{-1}$, the number of individuals connected to individuals in $\cV$ must be upper bounded by $\gamma^{-3} (1 + \gamma^{-1})\varepsilon^{-1}$, so $[n] \setminus \cS$ must also be upper bounded by $\gamma^{-3}(1 + \gamma^{-1}) \varepsilon^{-1}$.
        Finally, since all $i \in \cS$ satisfy \cref{eq:marginalcontribution} we have the desired result.
        \qed
        \fi
        
\section{Inequality at equilibrium} \label{sec:inequality}

    \begin{figure}
        \centering
        \includegraphics[width=0.5\textwidth]{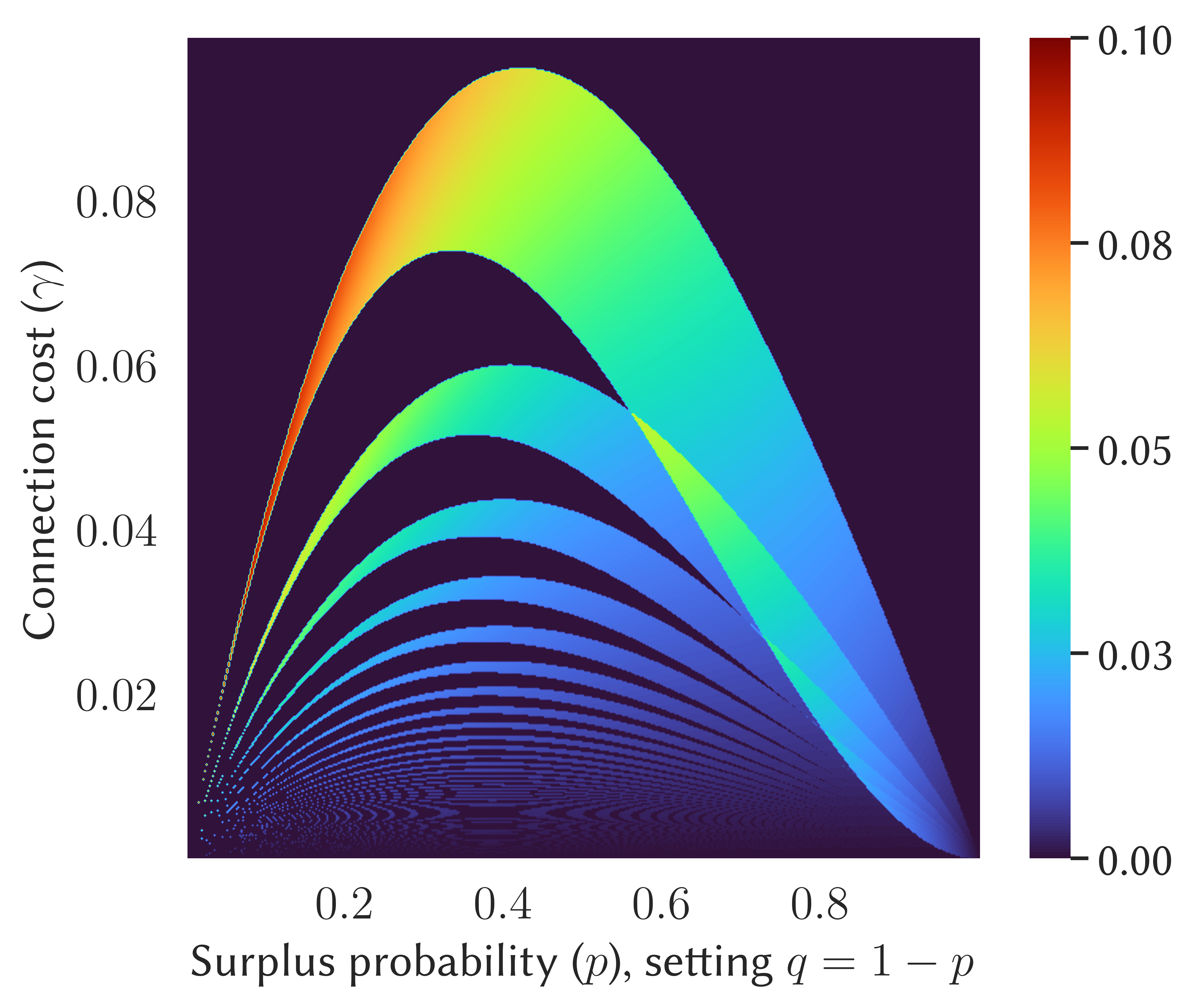}
        \caption{Worst-case Gini coefficient in equilibrium networks in the \indiscgame\ with connection costs $\gamma$. Color represents the a lower bound on the Gini coefficient in \cref{prop:inequality} achieved by setting $\delta_1 = \delta_2 = \delta_3 = 0$.}
        \label{fig:inequality}
    \end{figure}

    At equilibrium, some individuals may have higher or lower utility than others.
    In this section, we return to our analysis of the \indiscgm~and characterize how unequal the distribution of opportunities within the network can be, \textit{even when the exogenous opportunity distributions of individuals are identical}.
    To quantify dispersion, we use the canonical Gini coefficient and consider worst-case inequality for an equilibrium on a population.
    Gini coefficients vary between 0 and 1, where 0 is attained when utility is equally distributed among all individuals, and 1 is attained when one individual derives all utility and all others derive zero.
    The Gini coefficients of most OECD countries vary between 0.25 and 0.4 \cite{WB:2024}.
    Formally, for a population of size $n$, let $\cE_{\mathrm{\eqbm}}(n, q, p, \gamma)$ be the set of equilibria edge sets for fixed $n, q, p, \gamma$, and define the Gini coefficient for our context as follows.
    \begin{align}
        \mathrm{\gini}(n, q, p, \gamma) \defeq  \max_{E \in \cE_{\mathrm{\eqbm}}(n, q, p, \gamma)} \frac{\sum_{i,j \in [n]} \abs{{u_i(E)} - {u_j(E)}}}{ 2n \cdot u_{[n]}(E)}. \label{eq:gini}
    \end{align}
    We note that our results in this section can be directly extended to other dispersion measures.
    In the next theorem, we compute the worst-case Gini coefficient in an equilibrium. 
    The worst-case is attained by a network composed of two types of people, one with maximum utility and the other minimum utility (denoted $\umax$ and $\umin$) where the proportions of $\umax$ is determined by some carefully chosen $\lambda$.
    The values of $\umax$ and $\umin$ depend themselves on the least and greatest degrees (denoted $\dmin$ and $\dmax$), respectively, of individuals in an equilibrium.

    \begin{restatable}{theorem}{inequality} \label{prop:inequality}
        The worst-case equilibrium Gini coefficient in the \indiscgm~with $\gamma > 0$ is 
        \begin{align*}
            \mathrm{\gini}(n, q, p, \gamma) = \frac{\lambda (1 - \lambda) (u_{\max} - u_{\min})}{2 \paren{\lambda u_{\max} + (1-\lambda) u_{\min}}} \pm o(1)
        \end{align*}
        where
        \begin{align*}
            &u_{\max} \equnder{\delta_1} 1 - q \paren{1 - \frac{p}{d_{\min} + \delta_1}}^{\dmin} - \gamma d_{\min}, \\
            &u_{\min} \equnder{\delta_2, \delta_3} 1 - q \paren{1 - \frac{p}{d_{\max} + \delta_2 + \delta_3}}^{\dmax + \delta_3} - \gamma (d_{\max} + \delta_3),\;\; \text{and}\\
            &\lambda = \frac{\sqrt{\umax \umin} - \umin}{\umax - \umin}
        \end{align*}
        and where $d_{\max}, d_{\min}$ are the maximum and minimum $d$, respectively, satisfying
        \begin{align*}
            \frac{1}{d+1} \paren{1 - \frac{p}{d}}^{d} \leq \frac{\gamma}{qp} \leq \frac{1}{d} \paren{1 - \frac{p}{d}}^{d-1}.
        \end{align*}
    \end{restatable}

    The Gini coefficient for different values of $p$ and $\gamma$ (setting $q = 1-p$ and $\delta_1 = \delta_2 = \delta_3 = 0$) is displayed in \cref{fig:inequality}.
    \ifconferencesubmission
    \else
        To parse the figure, consider how equilibria change for fixed $p$ and $q = 1-p$ and varied $\gamma$.
        As $\gamma$ sweeps down to 0, there are points at which each of $\dmin$ and $\dmax$ increment.
        When $\dmax$ increments but $\dmin$ stays the same, inequality increases (these are the points at the top of the colored ribbons) and when $\dmin$ increases but $\dmax$ stays the same, inequality decreases (these are the points at the bottom of the colored ribbons).
        On the left-hand part of the figure, where the ribbons do not overlap with each other, $\dmax - \dmin$ alternates between 1 and 0 as $\gamma$ sweeps down: when $\dmax - \dmin > 0$ there can be inequality and when $\dmax - \dmin = 0$ there is no inequality.
        On the right-hand part, the ribbons overlap with each other, so the gap between $\dmax$ and $\dmin$ can be larger than 1.

    \fi
    \ifproofsinbody
    
    \proofof{\cref{prop:inequality}} 
    First, we will establish the worst-case Gini coefficient for individuals' utilities bounded between constants $a, b$ where $0 < a \leq b$.
    Suppose user utilities are $u_1 \leq u_2 \leq \dots \leq u_n$ and $a \leq u_i \leq b$ for all $i \in [n]$.
    Recall that the Gini coefficient is defined as
    \begin{align}
        \mathrm{\gini}(u_1,\dots,u_n) = \frac{\sum_{i,j \in [n]} \abs{u_i - u_j}}{ 2n \sum_{i \in [n]} u_i} \label{eq:giniu}
    \end{align}
    We will first argue that that the Gini coefficient is largest when each $u_i \in \{ a, b \}$.
    To see this, we will use an iterative argument starting from arbitrary $u_1, \dots, u_n$ and changing their values in such a way that the Gini coefficient never decreases and such that at the end, all utilities are either $a$ or $b$.
    Suppose at least one $u_i \not\in \{ a, b \}$. Consider the smallest index $i$ such that $u_i > a$ and the largest index $i'$ such that $u_{i'} < b$.
    Replacing $u_i$ with $a$ will change the sum in the numerator of the Gini coefficient in \cref{eq:giniu} by $2(u_i - a)(n + 1 - 2i)$ (which is positive for $i \leq \floor{n/2}$) and decrease the sum in the denominator by $u_i - a$.
    Replacing $u_{i'}$ with $b$ will increase the sum in the numerator by $2(b-u_{i'})(2i'-1-n)$.
    It will \textit{increase} the sum in the denominator by $b - u_i$.
    Let
    \begin{align*}
        c_1 &\defeq \sum_{j, k \in [n]} \abs{u_j - u_k}, \; \text{and} \\
        c_2 &\defeq \sum_{j \in [n]} u_j.
    \end{align*}
    Then it must hold either that
    \begin{align*}
        &\mathrm{\gini}(u_1,\dots, a, \dots, u_n) \geq \mathrm{\gini}(u_1,\dots, u_i, \dots, u_n) \\
        \iff &\frac{c_1 + 2 \paren{ u_i - a}(n + 1 - 2i)}{ 2n \paren{a - u_i + c_2 }} \geq \frac{c_1}{ 2n {c_2}} \\
        \iff &2(2i - 1 -n) \leq \frac{c_1}{c_2},
    \end{align*}
    or 
    \begin{align*}
        &\mathrm{\gini}(u_1,\dots, b, \dots, u_n) \geq \mathrm{\gini}(u_1,\dots, u_{i'}, \dots, u_n) \\
        \iff &\frac{c_1 + 2 \paren{b - u_{i'}}(2i' - 1 -n)}{ 2n \paren{b - u_{i'} + c_2 }} \geq \frac{c_1}{ 2n {c_2}} \\
        \iff &2(2i' - 1 -n) \geq \frac{c_1}{c_2},
    \end{align*}
    since $i' \geq i$. Repeating this process iteratively, replacing either $u_i$ with $a$ or $u_{i'}$ with $b$ proves the result.
    
    
    Thus, an upper bound on the Gini coefficient for individuals with utilities $a \leq u_1 \leq \dots \leq u_n \leq b$ is 
    \begin{align*}
        \frac{\lambda (1-\lambda) (b-a)}{2 \paren{\lambda b+ (1-\lambda) a}}
    \end{align*}
    for some $\lambda \in (0,1)$.
    If we maximize over $\lambda$, we can notice that the Gini coefficient is 0 when $\lambda = 0$ or $\lambda = 1$ and nonzero in between if $a < b$, so $\lambda \in (0, 1)$.
    Next, we take the derivative of the expression with respect to $\lambda$ and set it equal to 0.
    \begin{align*}
        \frac{\dif }{\dif \lambda}  \frac{\lambda (1-\lambda) (b-a)}{\lambda b+ (1-\lambda) a} &= \frac{(a + (b-a) \lambda) (1- 2\lambda) - \lambda (1-\lambda) (b-a)}{(a + (b-1) \lambda)^2}
    \end{align*}
    and solving for the zeros the expression, we get
    \begin{align*}
        \lambda = \frac{\sqrt{ab} - a}{b - a}.
    \end{align*}

    Next, we will derive upper and lower bounds on the worst-case inequality at equilibrium.
    We will derive the lower bounds first, constructively.
    Notice that the utilities of individuals (and therefore equilibrium conditions) in separate components do not depend on each other. 
    So if we have two equilibrium networks, we can combine them into the same network (with two components) and the resulting network is an equilibrium.
    Our plan will be to construct equilibria this way by combining two regular graphs, one with low degree and one with high degree.
    To do so, for a given $p, q, \gamma$, we define a set of the $d$ such that a $d$-regular graph is an equilibrium.
    Formally, define
    \begin{align*}
        \cD_{\text{reg}} \defeq {\setcomp{d \in \N }{\frac{1}{d+1} \paren{1 - \frac{p}{d}}^d \leq \frac{\gamma}{qp} \leq \frac{1}{d} \paren{1 - \frac{p}{d}}^{d-1}} }
    \end{align*}
    and as in the statement of the result, let $\dmax \defeq \max \cD_{\text{reg}}$ and $\dmin \defeq \min \cD_{\text{reg}}$.
    Since the utility of individuals is monotonic decreasing in their degree, $u_{\min} \defeq 1 - q(1 - p/\dmax)^\dmax - \gamma \dmax$ is indeed the minimum utility achieved by any individual in a $d$-regular component of a network for $d \in \cD_{\text{reg}}$.
    Similarly, $u_{\max} \defeq 1 - q(1 - p/\dmin)^\dmin - \gamma \dmin$ is the maximum utility achieved by any individual in a $d$-regular component of a network for $d \in \cD_{\text{reg}}$.
    Thus, the Gini coefficient when there is a $\dmax$-regular component of size $\floor{\lambda n}$ where
    \begin{align*}
        \lambda \defeq \frac{\sqrt{\umax \umin} - \umin}{\umax - \umin}
    \end{align*}
    and the remaining individuals constitute a $\dmin$-regular component of size $\ceil{(1-\lambda)n}$, is
    \begin{align*}
        \frac{\lambda (1-\lambda)(\umax - \umin)}{\lambda \umax + (1-\lambda) \umin}
    \end{align*}
    which is equal to the statement of the result when $\delta = 0$.
    Since our result is asymptotic, we do not need to worry about whether there exists a network of size $n$ with two components, one that is $\dmin$-regular and the other that is $\dmax$-regular.
    This is because we can just consider a subgraph with this property, and for large enough $n$, there exists a subgraph with this property that is an arbitrarily large fraction of $n$.
    Since all utilities are lower-bounded by 0, this means that the worst-case Gini coefficient for a sequence of equilibria associated with populations of size $n=1,2\dots$ will approach the value we derived.
    This is our lower bound.
    
    For our upper bound, we will reason about the maximum Gini coefficient for all equilibria.
    Consider any equilibrium edge set $E$.
    Recall from \cref{prop:layeredgraph} that all but a finite number of individuals exclusively have connections with degrees within one of their own.
    Call the set of individuals with this property $\cS$.
    Define a new set $\cT$ which contains each individual $i \in \cS$ such that there exists some $j \in S$ such that $d_i = d_j$ and $j \not\in N_i(E)$.
    In other words, for each individual in $\cT$, there exists someone else of the same degree in $\cT$ such that they are not connected and they share the same degree.
    Notice that $\abs{\cT}$ no smaller than the size of $\cS$ minus a constant, since there are no more than a constant number of distinct degrees of nodes in $\cS$ and any degree $d$ that contains more than $d + 1$ individuals satisfies this property for all individuals of that degree in $\cS$.
    Next, we will upper and lower bound the utilities of individuals in $\cT$.
    To do so, we will find the largest and smallest degrees possible in $\cT$ in an equilibrium.
    Then, we will use these to derive an upper bound on the Gini coefficient for any equilibrium.
    Since $\cT$ contains no fewer than $n$ individuals minus a constant, we can disregard the utilities of all other individuals (since they are bounded between 0 and 1) and our results for average utilities (or average mean absolute differences) in $\cT$ will approach their values for all $[n]$ in the limit as $n \to \infty$.
    \newcommand{\dmaxeq}{d_{\max}^{\mathrm{eq.}}}
    \newcommand{\dmineq}{d_{\min}^{\mathrm{eq.}}}
    Recall from the equilibrium conditions in \cref{eq:dmaxconstraint} that the largest degree $\dmaxeq$ in $\cT$ must satisfy
    \begin{align*}
        \frac{\gamma}{qp} \leq \frac{1}{\dmaxeq - 1} \paren{1 - \frac{p}{\dmaxeq}}^{\dmaxeq-1}. 
    \end{align*}
    Similarly, by the assumption that for each individual in $\cT$ there must be someone unconnected to them with the same degree (which, by the equilibrium conditions, is not a mutually beneficial connection), it must hold that the smallest degree $\dmineq$ in $\cT$ satisfies
    \begin{align*}
        \frac{\gamma}{qp} \geq \frac{1}{\dmineq + 1} \paren{1 - \frac{p}{\dmineq}}^{\dmineq}. 
    \end{align*}
    Next, notice that for any $k$ such that there exists $i \in \cT$ where $d_i = k$, the utility of individual $i$ is upper-bounded by
    \begin{align}
        1 - q \paren{1 - \frac{p}{k-1}}^k - \gamma k \label{eq:upperboundutil}
    \end{align}
    and lower-bounded by
    \begin{align}
        1 - q \paren{1 - \frac{p}{k+1}}^k - \gamma k. \label{eq:lowerboundutil}
    \end{align}
    We showed in the proof of \cref{prop:poaedgecosts} that \cref{eq:lowerboundutil} is minimized at $k = \dmaxeq$.
    We will next show that \cref{eq:upperboundutil} is maximized at $k = \dmineq$.
    From \cref{prop:increasing} we have $(1-p/(k-1))^k$ is increasing in $k$.
    Thus, the whole expression in \cref{eq:upperboundutil} is decreasing in $k$, implying that the expression is maximized at $k = \dmineq$.

    These results give us upper and lower bounds on the utility of individuals in $\cT$. Define
    \newcommand{\umaxeq}{u_{\max}^{\mathrm{eq.}}}
    \newcommand{\umineq}{u_{\min}^{\mathrm{eq.}}}
    \begin{align*}
        \umaxeq &\defeq 1 - q \paren{1 - \frac{p}{\dmineq - 1}}^{\dmineq} - \gamma \dmineq \\
        \umineq &\defeq 1 - q \paren{1 - \frac{p}{\dmaxeq + 1}}^{\dmaxeq} - \gamma \dmaxeq
    \end{align*}
    Recall from the beginning of this proof that the worst-case Gini coefficient for individuals' utilities would occur when all utilities are either $\umineq$ or $\umaxeq$, with the proportion of those who are $\umaxeq$ equal to 
    \begin{align*}
        \lambda^{\mathrm{eq.}} = \frac{\sqrt{\umaxeq \umineq} - \umineq}{\umaxeq - \umineq}.
    \end{align*}
    Our last step is to relate $\dmaxeq$ to $\dmax$ and $\dmineq$ to $\dmin$.
    Notice that the equilibrium conditions on $\dmin$ are the same as $\dmineq$ so $\dmin = \dmineq$.
    From the proof of \cref{prop:poaedgecosts}, we also have that $\dmax \geq \dmaxeq - 1$.
    \qed

    \fi

    One might wonder just how large the difference between $\dmin$ and $\dmax$ can be: where the ribbons overlap in \cref{fig:inequality}, how large can the gap between $\dmax$ and $\dmin$ be?
    Our next result states that the range of individuals' degrees in equilibrium can be arbitrarily large for parts of the parameter space.

    \begin{restatable}{proposition}{multiplicityofdreg} \label{prop:multiplicityofdreg}
        For all $K > 0$, there exists some \indiscgame~with parameters $q, p, \gamma$ such that there exists an equilibrium edge set $\nashedgeset$ such that the range of individuals' degrees is at least $K$.
    \end{restatable}
    Interestingly, these are not the parts of \cref{fig:inequality} where inequality is worst: the Gini coefficient seems to vary between $0$ and $0.05$, whereas in other parts of the parameter space it is $0.10$.
    This is because, when $p$ is close to 1 and $q = 1-p$, individuals regardless of degree tend to have utilities close to 1 (since $1-q$ is close to zero, so they have a low chance of failing to receive an exogenous opportunity).
    By contrast, where inequality is higher, equilibrium utilities tend to be close to zero and the difference between having $\dmax$ and $\dmin$ connections influences the Gini coefficient more.
    In fact, we can calculate a lower bound on the worst-case Gini coefficient over any $p, q, \gamma$ near $p = 0$ and show that it is about 0.1.

    \ifproofsinbody
    
    \proofof{\cref{prop:multiplicityofdreg}} First, notice that the utilities of individuals (and therefore equilibrium conditions) in separate components do not depend on each other. 
    So if we have two equilibrium networks, we can combine them into the same network (with two components) and the resulting network is an equilibrium.
    Our plan will be to construct equilibria this way by combining two regular graphs, one with low degree and one with high degree.
    
    Throughout the proof we will just consider the case that $\gamma / q = (1-p)/2$. (Although this condition is not necessary.) Notice that the result is implied by the following claim.
    For all $K > 0$, there exists some $p$ such that for 
    \begin{align*}
        \cD \defeq {\setcomp{d \in \N }{\frac{p}{d+1} \paren{1 - \frac{p}{d}}^d \leq \frac{1-p}{2} \leq \frac{p}{d} \paren{1 - \frac{p}{d}}^{d-1}} } 
    \end{align*}
    it holds $\max \cD - \min \cD \geq K$,
    since the set is exactly those $d$ satisfying the equilibrium conditions.
    First, we will claim that for small enough $\varepsilon$, it holds $1 \in \cD$.
    Notice $1 \in \cD$, if 
    \begin{align*}
        \frac{p}{2}(1-p) \leq \frac{1 - p}{2} \leq p
    \end{align*}
    which is true if $p \geq 1/3$.
    Next, we want to show that for some $p \geq 1/3$,
    \begin{align}
        \frac{1}{d+1} \paren{1 - \frac{p}{d}}^d \leq \frac{1-p}{2p} \leq \frac{1}{d} \paren{1 - \frac{p}{d}}^{d-1}, \label{eq:larged}
    \end{align}
    for $d \geq K$ which will imply the result.
    But, notice from \cref{lem:monotoneincreasing} and \cref{prop:decreasing} that for all $d$, \cref{eq:larged} is implied by
    \begin{align*}
        \frac{e^{-p}}{d+1}  \leq \frac{1-p}{2p} \leq \frac{e^{-p}}{d} 
    \end{align*}
    so it is sufficient to set $p$ such that
    \begin{align*}
        \frac{2p}{1-p} e^{-p} \in [d, d+1]
    \end{align*}
    which is achieved for any $d$ by setting $p$ close enough to 1.
    \qed
    \fi

    \begin{restatable}{proposition}{worstinequality} \label{prop:worstinequality}
        It holds
        \begin{align*}
            \sup_{n, q, p, \gamma \in [0, 1]} \mathrm{\gini}(n, q, p, \gamma) \geq 5 - 2\sqrt{6} \approx 0.1;
        \end{align*}
        and for the sequence of equilibria achieving the bound, approaching $\sqrt{6} - 2 \approx 45\%$ of the population derives $3 - \sqrt{6} \approx 55\%$ of the utility.
    \end{restatable}

    In other words, even if individuals are equally likely to receive opportunities from outside sources, inequality due to the network itself in equilibrium can be such that 45\% of the population derives 55\% of the expected utility.
    For comparison, in the United States, the top 40\% of earners receive about 70\% of the income according to recent data \cite{WB:2024}.
    
\section{Platform interventions} \label{sec:interventions}

    In this section, we consider how a platform might try to increase social welfare for its users. 
    We analyze light-touch interventions, in the sense that they do not explicitly constrain which connections form or dictate how opportunities are routed, since these kinds of interventions are the ones that platforms like LinkedIn are likely to implement.
    We consider two interventions: altering the \textit{friction} in forming connections and giving individuals access to more \textit{information} about others' existing opportunities.
    We prove mostly negative results about these interventions: 
    simple heuristics do not necessarily increase efficiency, and platforms would need a fine-grained understanding of the status quo parameter settings and how interventions might affect them in order to ensure that they increase social welfare.
    %
    
    \subsection{Varying friction.} \label{sub:friction}

        \begin{figure}
            \centering
            \includegraphics[width=0.5\textwidth]{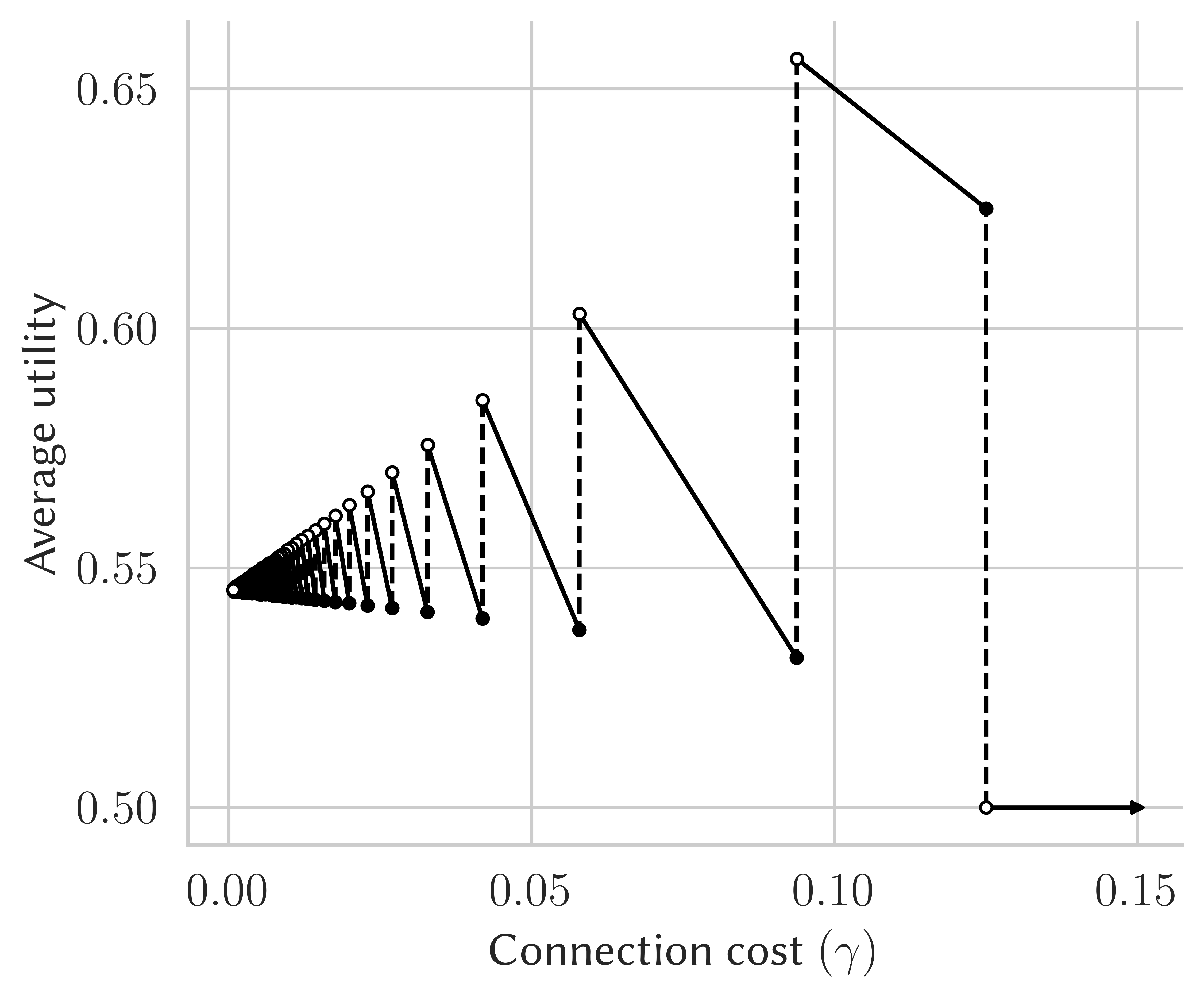}
            \caption{Average utility in a worst-case regular equilibrium as a function of $\gamma$ for $q = p = 0.5$, given in \cref{prop:mechdes}. Each discontinuity in the graph represents a point at which the set of regular equilibria changes.}
            \label{fig:varyingfriction}
        \end{figure}

        A natural means for platforms to change the way that individuals form connections is by tweaking the user experience to make it harder or easier to connect to others.
        For example, a platform could increase or decrease the size or frequency with which suggestions to form connections (the ``People You May Know'' panel on LinkedIn) appear on a user's feed.
        It might also reduce friction, as LinkedIn does, by sending push notifications with connection suggestions, or otherwise make it easier or harder for people to connect with and keep up with others.
        
        One might hypothesize that by increasing connection costs, platforms could induce sparser graphs and increase efficiency in the transfer of opportunities.
        Unfortunately, this does not necessarily work, and the benefits of inducing sparser graphs may be outweighed by the costs themselves.
        On the other hand, \textit{reducing} connection costs does not necessarily lead to better outcomes either, since lower connection costs may lead to equilibria in which everyone forms even more connections, leading to increased congestion in opportunity transfer.
        We formalize this in \cref{prop:mechdes} below.
        The first part of the proposition says that, with respect to a particular equilibrium, lowering $\gamma$ always increases social welfare, as long as it continues to be an equilibrium.
        The second part says that, for all exogenous opportunity distributions, if $\gamma$ is at the boundary between different sets of regular equilibria (i.e., if decreasing $\gamma$ allows for regular equilibria of larger degrees), setting $\gamma$ higher yields strictly greater worst-case social welfare at equilibrium than if it were lower, for any small pairs of increases and decreases to $\gamma$.
        In the second part of the result, we restrict our attention to $d$-regular equilibria, since we can compute equilibrium social welfare exactly in these cases.
        Because of 1-degree homophily, this gives us nearly tight upper bounds on worst-case equilibria.
        %
        %
        Let $\cE_{\mathrm{\eqbm}}^{\mathrm{\text{reg.}}}(q, p, \gamma)$ be the set of regular equilibria for the \indiscgm~with parameters $q, p, \gamma$.
    
        \begin{restatable}{proposition}{mechdes}\label{prop:mechdes}
            For any fixed edge set $E$ that is an equilibrium in the \indiscgm~for all connection costs $\gamma$ in the interval $[\underline{\gamma}, \overline{\gamma}]$, decreasing $\gamma$ within the interval is a strict Pareto improvement.
            On the other hand, for all $q, p > 0$ there exists $\gamma > 0$ and $\overline \varepsilon > 0$ such that for all $\varepsilon, \varepsilon' \in [0, \overline \varepsilon]$,  
            \begin{align*}
                \min_{E \in \cE_{\mathrm{\eqbm}}^{\mathrm{reg.}}(n, q, p, \gamma - \varepsilon)} u_{[n]}(E; q, p, \gamma - \varepsilon) < \min_{E \in \cE_{\mathrm{\eqbm}}^{\mathrm{reg.}} (n, q, p, \gamma + \varepsilon')} u_{[n]}(E; q, p, \gamma + \varepsilon').
            \end{align*}
        \end{restatable}
        
        An example of how the worst-case regular equilibrium changes as a function of $\gamma$ is given in \cref{fig:varyingfriction}.
        As $\gamma$ sweeps downwards, it passes through a series of discontinuities.
        Each discontinuity in the plot represents a point at which the set of regular equilibria changes for $q = p = 0.5$: for example, around 0.125, the worst-case equilibrium switches from degree 0 to degree 1; around 0.096, the worst-case equilibrium switches from degree 1 to degree 2, and so on.
        Within each equilibrium, lowering $\gamma$ increases utility, but for an interval around discontinuities, it is better for $\gamma$ to be above the discontinuity.
        
        \cref{prop:mechdes} indicates a platform cannot naively increase or decrease $\gamma$ and expect to have consistent effects on worst-case equilibrium social welfare, since the relationship between $\gamma$ and social welfare is non-monotonic.
        To have the intended effects on social welfare, a platform would need to know $q$, $p$ and status quo $\gamma$, as well as have fine-grained control over how $\gamma$ changes as a result of some change to the, e.g., user experience.
        \ifproofsinbody
        
        \proofof{\cref{prop:mechdes}} The first part of the proposition follows directly from the functional form individual utilities.
        In any equilibrium $E$, the utility of individual $i$ is
        \begin{align*}
            1 - q \prod_{j \in N_i(E)} \paren{1 - \frac{p}{d_j}} - \gamma d_i
        \end{align*}
        which is decreasing in $\gamma$. 
        Of course, this result depends on $E$ continuing to be an equilibrium as $\gamma$ changes.
        For the second part of the result, consider any $d \in \N$ and $0 < \gamma \leq q p (1-p)$ such that
            \begin{align*}
                \gamma = \frac{qp}{d} \paren{1 - \frac{p}{d}}^{d-1}.
            \end{align*}
        Notice that for 
        \begin{align*}
            \gamma > \frac{qp}{d} \paren{1 - \frac{p}{d}}^{d-1}
        \end{align*}
        the largest $k$-regular equilibrium is at most $d-1$.
        Then, the following sequence proves the result:
        \begin{align*}
            &1 - q \paren{1 - \frac{p}{d}}^d - (\gamma - \varepsilon') d < 1 - q \paren{1 - \frac{p}{d-1}}^{d-1} - (\gamma + \varepsilon) (d - 1) \\
            \iff &(\varepsilon + \varepsilon') d + \varepsilon < \gamma + q \paren{\paren{1 - \frac{p}{d}}^d - \paren{1 - \frac{p}{d-1}}^{d-1}} \\
            \impliedby & \overline{\varepsilon} < \frac{1}{2d+1} \paren{\gamma + q \paren{\paren{1 - \frac{p}{d}}^d - \paren{1 - \frac{p}{d-1}}^{d-1}}} \\
            \impliedby & \overline{\varepsilon} < \frac{\gamma}{2d+1} 
        \end{align*}
        \qed
        \fi
    \subsection{Broadcasting bandwidths.} \label{sec:broadcastbandwidths}

        If varying friction requires a fine-grained understanding of the status quo and the potential impacts of user experience interventions, might the platform improve outcomes by providing information about individuals' needs for opportunities?
        %
        %
        Indeed, LinkedIn provides a feature like this called ``\#OpenToWork''\footnote{\href{https://www.linkedin.com/help/linkedin/answer/a507508}{https://www.linkedin.com/help/linkedin/answer/a507508}}
        where individuals can add text in a frame around their profile image with the hashtag and are prioritized in some search results.
        
        To address whether this kind of intervention can increase efficiency in our context, we analyze an alternate opportunity transmission model, in which individuals are aware of their social contacts' realized exogenous opportunities and \textit{only send opportunities to others who received no exogenous opportunities}.
        We will assume individuals who have extra opportunities pass them to a social contact who {did not receive an exogenous opportunity} uniformly at random if there is one.
        However, there could still be wasted opportunities even under this opportunity transmission model if, for example, two people pass their extra opportunities to the same person.
        We call this the \game~(\gm).
        We provide a complete equilibrium and efficiency analysis in \Cref{sec:informed}, and here we explore the comparison between equilibria under the basic opportunity transfer model and the informed model.
        We will use the superscript ``\infd''~(e.g., $\informedy_i(E)$) for notation referring to the informed opportunity transfer model.
    
    The analysis of the informed opportunity transfer model is more complicated, but morally similar to our analysis in the previous sections. This is because, unlike in the \indiscgm, there is no simple explicit form for the utility of an individual as a function of their neighbors' degrees in the \gm:
    there may be dependencies between the probability that, for a given individual $i$, one of $i$'s social contacts (who may be connected to each other or share other connections in common) will transfer $i$ an opportunity.
    To see this, note
    \begin{align}
        u^{\mathrm{\infd}}_i(E; q, p, \gamma) = 1 - q \cdot \probgiv{\bigcap_{j \in N_i(E)} \curly{R_{ji}^\mathrm{\infd} = 0}}{X_i = 0} - \gamma d_i. \label{eq:uinformed}
    \end{align}
    In the \indiscgm, the events $\{ R_{ji} = 0 \}$ are independent of each other, so we can turn the intersection of these events into a product.
    However, in the \gm, this is not possible.
    For any $j, k \in N_i(E)$ such that $j$ and $k$ are connected, conditioning on the fact that $R_{ji}^\mathrm{\infd} = 0$ raises the probability that $R_{jk}^\mathrm{\infd} = 1$ (relative to the unconditioned probability).
    This, in turn, lowers the conditional probability that $X_k = 2$ (relative to the unconditioned probability).
    And, thus, the probability that $R_{ki}^\mathrm{\infd} = 0$, conditioning on $R_{ji}^\mathrm{\infd} = 0$, is higher than the unconditioned probability.
    A similar argument can be made if $j$ and $k$ are not directly connected by $N_j(E) \cap N_k(E)$ includes other individuals besides $i$.
    
        Intuitively, one would think that informed opportunity transmission should only improve efficiency, relative to the uninformed model. 
        (After all, no one ever passes an opportunity to someone who already received one from an outside source.)
        In one sense, this is true: for a fixed network, social welfare is greater under informed opportunity transmission. 
        We state this formally next.
        \begin{restatable}{proposition}{voifixed} \label{prop:voifixed}
            For all $p, q > 0, \gamma \geq 0,  n \in \N$ and fixed $E \subseteq [n]^2$, 
            \begin{align*}
                u_{[n]}(E; q, p, \gamma) \leq u_{[n]}^{\mathrm{\infd}}(E; q, p, \gamma).
            \end{align*}
            Further, for all $i \in [n]$ such that there exists some $j \in N_i(E)$ where $d_j \geq 2$,
            \begin{align*}
                u_{i}(E; q, p, \gamma) < u_{i}^{\mathrm{\infd}}(E; q, p, \gamma).
            \end{align*}
        \end{restatable}
        The result says that social welfare for a fixed edge set is no less in the informed opportunity transfer model than in the basic version of the model.
        It further says that, for any individual who has a neighbor of degree at least 2, the informed model is a strict Pareto improvement, yielding strictly greater utility.
        However, when equilibrium effects are taken into account, the net effect of informed opportunity transfer can actually be negative.
        Information can also increase the marginal benefits of connecting with others, leading them to form more connections. As a result, equilibrium social welfare can \textit{decrease} in the \gm~relative to the \indiscgm.
        We state this formally in \cref{prop:voieq}.
        It says that, when $\gamma = 0$, the informed model always yields greater equilibrium social welfare, but when $\gamma > 0$ it might yield either higher or lower equilibrium social welfare.

        \ifproofsinbody
        
        \proofof{\cref{prop:voifixed}} 
        Notice both parts of the result follows from showing the expected utility of each individual in the \gm~is greater than in the \indiscgm~over the edge set $E$.  
        We will first show that, over each draw of the random variables $\{ X_i \}_{i \in [n]}$, the expected utility of each individual in the \gm~is no less than in the \indiscgm.
        For this subsection, define $\informedy_i(E)$ to be $i$'s outcome under the \gm~opportunity transition model and $Y_i(E)$ as before for \indiscgm.
        For fixed $\{ x_j \}_{j \in [n]} \in \{ 0, 1, 2\}^n$ and $i \in [n]$, condition on $X_j = x_j$ for all $j \in [n]$ and notice
        \begin{align*}
            &\exgiv{Y_i(E)}{X_j = x_j, \forall j \in [n]} \leq \exgiv{\informedy_i(E)}{X_j = x_j, \forall j \in [n]} \\
            \iff &1 - \indic{x_i = 0} \prod_{j \in N_i(E) } \paren{1 - \frac{\indic{x_j = 2}}{d_j}} - \gamma d_j \\
            &\leq 1 - \indic{x_i = 0} \prod_{j \in N_i(E)} \paren{1 - \frac{\indic{x_j = 2}}{\max \curly{ 1 , \sum_{\ell \in N_j(E)} \indic{x_\ell = 0}}}}  - \gamma d_i \\
            \impliedby & \prod_{j \in N_i(E) } \paren{1 - \frac{ \indic{x_j = 2}}{d_j}} \geq \prod_{j \in N_i(E) } \paren{1 - \frac{\indic{x_j = 2}}{\max \curly{ 1 , \sum_{\ell \in N_j(E)} \indic{x_\ell = 0}}}} .
        \end{align*}
        The last line is implied by the fact that, for all $j \in [n]$,
        \begin{align}
            & {1 - \frac{ \indic{x_j = 2}}{d_j}} \geq {1 - \frac{\indic{x_j = 2}}{\max \curly{ 1 , \sum_{\ell \in N_j(E)} \indic{x_\ell = 0}}}} \nonumber \\
            \iff & {\frac{ \indic{x_j = 2}}{d_j}} \leq {\frac{\indic{x_j = 2}}{\max \curly{ 1 , \sum_{\ell \in N_j(E)} \indic{x_\ell = 0}}}} \nonumber\\
            \impliedby & {{d_j}} \geq {{\max \curly{ 1 , \sum_{\ell \in N_j(E)} \indic{x_\ell = 0}}}}. \label{eq:degreegex0}
        \end{align}
        Finally, to see that the expected utility of each individual in the \indiscgm, unconditional on realizations of $\{ X_i \}_{i \in [n]}$, is strictly greater than in the \gm, it is sufficient to show that for at least one realization $\{ x_j \}_{j \in [n]}$, the inequalities above are strict.
        But this is true for any $\{ x_j \}_{j \in [n]}$ and pair $(i,j) \in E$ where $x_j = 2$ and $x_i = 0$, and $d_j \geq 2$ and the inequality in \cref{eq:degreegex0} is strict.
        \qed
        \fi
        
        \begin{restatable}{proposition}{voieq} \label{prop:voieq}
        \phantom{~}
            \begin{enumerate}
                \item For all $q, p > 0$, $\gamma = 0$, $E \in \cE_{\mathrm{\eqbm}}(n, q, p, \gamma)$, and $E^{\mathrm{\infd}} \in \cE_{\mathrm{\eqbm}}^{\mathrm{\infd}}(n, q, p, \gamma)$ it holds
                \begin{align*}
                     u_{[n]}(E; q, p, \gamma) <  u_{[n]}^{\mathrm{\infd}}(E^{\mathrm{\infd}}; q, p, \gamma).
                \end{align*}
                \item There exists $q, p, \gamma > 0$, $E \in \cE_{\mathrm{\eqbm}}(n, q, p, \gamma)$ and $E^{\mathrm{\infd}} \in \cE_{\mathrm{\eqbm}}^{\mathrm{\infd}}(n, q, p, \gamma)$ such that
                \begin{align*}
                     u_{[n]}(E; q, p, \gamma) <  u_{[n]}^{\mathrm{\infd}}(E^{\mathrm{\infd}}; q, p, \gamma).
                \end{align*}
                \item There exists $q, p, \gamma > 0$, $E \in \cE_{\mathrm{\eqbm}}(n, q, p, \gamma)$ and $E^{\mathrm{\infd}} \in \cE_{\mathrm{\eqbm}}^{\mathrm{\infd}}(n, q, p, \gamma)$ such that
                \begin{align*}
                     u_{[n]}(E; q, p, \gamma) >  u_{[n]}^{\mathrm{\infd}}(E^{\mathrm{\infd}}; q, p, \gamma).
                \end{align*}
            \end{enumerate}
        \end{restatable}

        The first part of the result is about frictionless network formation, in which case, in both the informed and uninformed model, the unique equilibrium is the complete graph.
        Then \cref{prop:voifixed} tells us that social welfare is greater under informed opportunity transfer.
        The second and third parts of the result are about costly network formation, where equilibrium networks may be different across opportunity transfer models.
        Some equilibria are feasible in both the \indiscgm~and the \gm, in which case the equilibrium utility of the \gm~is greater.
        In other cases, for the same parameters the \gm\ admits equilibria with individuals of larger degrees, in which case the benefits of information can be outweighed by costs of higher degrees of individuals.
        \ifconferencesubmission
        \else 
        A limitation of the latter two parts of this result is that we are comparing arbitrary equilibria across the informed and uninformed variants of the game: a stronger result might compare worst-case equilibria.
        \fi

        As in the previous section, the results in this section show that the decision to implement an intervention depends on the part of the parameter space in which the platform sits. 
        Depending on $q, p$ and $\gamma$, it may or may not be a welfare-increasing decision to implement an intervention informing individuals of others' realized exogenous opportunities.
        \ifproofsinbody
        
        \proofof{\cref{prop:voieq}}  For the first item, observe 
        \begin{align*}
            & \sum_{i = 1}^n {u_i(E)} <  \sum_{i = 1}^n {u^{\infd}_i(E^{\infd})} \\
            \iff &1 - e^{-p} < 1 - e^{-\frac{p}{q}} \\
            \iff& -\frac{p}{q} < -p \\
            \iff& q < 1 .
        \end{align*}
        For the second item, let $\cE_{\mathrm{\eqbm}}^{\text{reg.}}$ be the set of equilibria edge sets that are regular and suppose $0 < p < 1/2$, $q = 1 - p$ and $\gamma = qp(1-p/2)/2$.
        It is easy to verify that the largest $d$ such that a $d$-regular network is an equilibrium for the \gm~and \indiscgm~is 2.
        Thus, the following sequence proves the result:
        \begin{align*}
            &1 - q\paren{1 - \frac{p}{1-p} \frac{1 - p^2}{2}}^2 - 2 \gamma > 1 - q \paren{1 - \frac{p}{2}}^2 - 2 \gamma \\
            \iff &\paren{1 - \frac{p}{2}}^2 > \paren{1 - \frac{p (1+p)}{2}}^2 \\
            \iff &{{p}} < {{p (1+p)}} 
        \end{align*}
        which holds for all $0 < p < 1/2$.

        For the third item, suppose $0 < p < 1/2$, $q = 1 - p$ and $\gamma = (2+p)(1-p)^2(1+p)/4$. It is easy to verify that the largest $d$ such that a $d$-regular network is an equilibrium for the \gm~is 2 and for the \indiscgm~it is 1.
        Thus, the following sequence proves the result:
        \begin{align*}
            &1 - q \paren{1 - \frac{p}{1-p} \frac{1 - p^2}{2}}^2 - 2 \gamma < 1 - q \paren{1 - p} - \gamma \\
            \iff &1 - p - \paren{1 - \frac{p(1 + p)}{2}}^2 < \frac{\gamma}{q} \\
            \iff &1 - p - \frac{1}{4}\paren{2 - p - p^2}^2 < \frac{\gamma}{q} \\
            \iff &1 - p - \frac{1}{4}(2+p)^2(1-p)^2 < \frac{(2+p)(1-p)^2(1+p)}{4(1-p)} \\
            \iff &1 - \frac{1}{4}(2+p)^2(1-p) < \frac{(2+p)(1+p)}{4} \\
            \iff &4  < (2+p)^2(1-p) + {(2+p)(1+p)} \\
            \iff &4  < (2+p) (3-p^2) 
        \end{align*}
        which is true for all $0 < p < 1/2$.
        \qed
        \fi
        An interesting question for future work is the equilibrium effects of \textit{choices about whether to opt-in} to a feature like \#OpenToWork.
        This is not captured by our analysis: we just compare the case in which either no one or everyone is aware of others' exogenous opportunities.
        \ifconferencesubmission
        \else
        Other theory models of referrals have shown how a market for lemons can emerge as a result of the use of referrals \cite{montgomery1991social}, and it would be interesting to explore whether and how the option to broadcast one's own bandwidth might signal one's ability to find out-of-network opportunities.
        \fi

\section{Discussion and future directions} \label{sec:discussion}

    In this paper, we proposed a novel model of opportunity transfer in networks and explored how equilibria induced by strategic individuals forming professional connections may affect individuals' economic outcomes.
    Our results provide clean mathematical formalization of how professional networks trade off connectivity and congestion.
    We develop several observations about the structures of equilibrium and efficient graphs, compute the price of anarchy, quantify inequality, and explore platform interventions.
    Our approach, which focuses on the choices of strategic individuals in networks, suggests several possible directions for future work within the study of opportunity transfer in networks.
    
    The first possible direction is more general characterizations of the basic phenomena we describe in this work. 
    The games we analyze are an instance of what we call a \textit{mutual support game}.
    For a population $\{ 1, \dots, n\}$ suppose each individual $i$ optionally has a vector of characteristics $z_i \in \R^r$ for some $r \in \N$.
    For an edge set $E \subseteq [n]$ and degree sequence $\{ d_i\}_{i \in [n]}$, suppose $i$ derives a benefit $f_i(E)$ from its direct connections (i.e., invariant to any $z_j$ or $d_j$ for some $j$ such that $(i,j) \not\in E$), and pays a cost $g_i(E)$ (possibly depending on more than just direct connections).
    Suppose their utility is $u_i(E) = f_i(E) - g_i(E)$.
    In the \generalizedgame, $z_i$ gives individual $i$'s exogenous opportunity distribution, and $f_i$ and $g_i$ are defined analogously to the \indiscgm.
    In the co-author game of \citet{jackson2003strategic}, $f_i$ is the sum of utility derived from working on a project with each of $i$'s connections, and $g_i$ is a cost for each connection.
    In the contagious risk game of \citet{blume2013network}, $f_i$ is a constant times $d_i$ and $g_i$ is the cost paid if $i$ fails.
    Many more games of interest can be described this way.
    Our results in this paper show how status homophily emerges as a result of the fact that individuals can find others who bring them similar marginal utility.
    It would be interesting to explore this phenomenon for other mutual support games.

    It would also be interesting to explore platforms' role in directly matching individuals to opportunities.
    This approach might be motivated by two stylized facts that we do not capture in our work: first, platforms have significantly reduced the cost of finding and applying for opportunities, leading to large increases in applications in several contexts \cite{demuels2023youre, majouirk2023deadline}, and second, through their search functionality, platforms exert significant influence on which jobs individuals apply for.
    This would enable analysis of a kind of congestion (different from the kind we consider) resulting from the fact that the most desirable opportunities receive too many applications and a small fraction of applicants receive a disproportionate share of the offers.

    These are just a few of the directions that we imagine future studies of network formation in opportunity transfer might take: other extensions of our work might consider repeated games, models with exogenous opportunities that arrive over time, whether platform interventions may decrease \textit{inequality} (rather than efficiency as we study).
    Generally, we hope that our work inspires more informed discussions of platform power and the role of social networks in labor markets.
    
\bibliographystyle{ACM-Reference-Format}
\bibliography{main}

\appendix

\section{Equilibrium and efficiency analysis of the \game} \label{sec:informed}

    Here we provide a complete analysis of equilibrium and efficient networks and the price of anarchy in the informed opportunity transfer model.
    
    \subsection{Frictionless network formation.}
    
     \paragraph{Equilibrium networks.} When $\gamma = 0$ in the \gm, equilibrium networks are identical to those of the \indiscgm, and for the same reason: connecting with others always helps the two individuals involved in the connection.
        \begin{restatable}{proposition}{informedselfish}\label{prop:informedselfish}
            In the \game, if $q, p > 0$, for all $E \subseteq [n]^2$ and $(i,j) \not \in E$,
            \begin{align*}
                u_i^{\mathrm{\infd}}(E \cup \{ (i,j) \}) > u_i^{\mathrm{\infd}}(E)
            \end{align*}
        \end{restatable}

        \ifproofsinbody
        
        \proofof{\cref{prop:informedselfish}} 
        The proof of \cref{prop:indiscselfish} can be straightforwardly adapted to the \gm.
        Notice
        \begin{align*}
            u_i^{\mathrm{\infd}}(E) &= 1 - q \probgiv{\cap_{\ell \in N_i(E)} \{R_{\ell i} = 0\}}{X_i = 0}, \;\text{and} \\
            u_i^{\mathrm{\infd}}(E \cup \{ (i, j) \}) &= 1 - q \probgiv{\cap_{\ell \in N_i(E) \cup \{ j \}} \{R_{\ell i} = 0\}}{X_i = 0}.
        \end{align*}
        But since
        \begin{align*}
            \probgiv{ R_{j i} = 0}{X_i = 0, \cap_{\ell \in N_i(E) } \{R_{\ell i} = 0\}} < 1,
        \end{align*}
        the latter is greater than the former.
        \qed
        \fi
        
        \cref{prop:indiscselfish} and \cref{prop:informedselfish} together show that equilibrium networks in \indiscgm~and \gm~ are identical to each other, unique and the complete graph. 
        However, we note that social welfare in each equilibrium is different. Social welfare in \gm~equilibrium networks is higher. This is because of the different opportunity transfer model: informed individuals reduce congestion. This can yield large benefits when $q$ is close to zero. As $q \to 0$, average utility in \indiscgm~goes to 1 at a rate of $\approx q$. On the other hand, average utility in \gm~goes to 1 at a rate of $\approx q e^{-p/q}$.

        \paragraph{Efficient networks.}
            In the \gm, socially optimal networks may be realized by denser networks than those of the \indiscgm.
            This is a result of the benefits of informed agents: when individuals know that others have not received an exogenous opportunity before they pass it to them, they can also form more connections before the negative externalities of connections outweigh their benefits to the connected individuals. 
        
            \begin{restatable}{proposition}{informedoptimal} \label{prop:informedoptimal}
                In the \game, if $q, p > 0$, a network is efficient if it is a 3- and 4-cycle-free $d$-regular graph where $d$ is given by a solution to
                \begin{align}
                    \argmax{0 \leq d \leq n, d \in \N} 1 - q\paren{1 - \frac{p}{q} \cdot \frac{1 - (1-q)^d}{d}}^d \label{eq:argmax}
                \end{align}
                and, for all $q, p > 0$, efficient networks are a strict Pareto improvement over equilibrium networks.
            \end{restatable}
        
            The result says that, in the \gm, $d$-regular 3- and 4-cycle-free graphs are efficient for $d$ sweeping over the natural numbers depending on $p$ and $q$ for $n$ large enough.
            Regularity comes from the fact that benefits to edges are most efficiently used when distributed evenly across all edges.
            The 3- and 4-cycle-free property comes from the fact that, in the \gm, outcomes of nodes are positively correlated, leading to inefficiency in opportunity transfer when an individual connects to two nodes who are connected or share some other connection.
            Notice that \cref{eq:argmax} can be evaluated in $O(n)$ time using enumeration.

        \ifproofsinbody
        
            \proofof{\cref{prop:informedoptimal}} 
                We will prove the result by showing 
                \begin{align}
                    \max_{E \subset [n]^2} \;\ex{\frac{1}{n}\sum_{i \in [n]} Y_i(E)} &\leq \max_{d \in \R: \; 1 < d \leq n} 1 - q\g{d}^d \label[ineq]{ineq:ub} 
                \end{align}
                and then by demonstrating that a 3- and 4-cycle-free $d$ regular graph achieves the stated maximum if $n$ admits a 3- and 4-cycle-free $d$ regular graph.
                To prove {\cref{ineq:ub}}, first, we will prove that for any network $G$ and individual $i$ in $G$, 
                it holds 
                \begin{align}
                    u_i^{\mathrm{\infd}}(E) &\leq 1 - q\prod_{j \in N_i(E)}  \paren{1 - \frac{p}{q} \cdot \frac{1 - p^{d_j}}{d_j}}. \label[ineq]{ineq:amgmyi}
                \end{align}
                
                The condition in \cref{ineq:amgmyi} is equivalent to proving
                \begin{align*}
                    \probgiv{\bigcap_{j \in N_i(E) } \{ R_{ji} = 0\}}{X_i = 0} \geq \prod_{j \in N_i(E)} \probgiv{  R_{ji} = 0 }{X_i = 0}. \yesnum \label[ineq]{ineq:intersectprod}
                \end{align*}
                
                To see this is true, we will use induction. Consider a sequence of sets $S_\ell \subseteq N_i(E) \defeq \{ k_1, \dots, k_{\abs{N_i(E)}}\}$ for $\ell = 1, \dots, \abs{N(i)}$ such that $S_{\ell} = \{ k_1, \dots, k_l\} $.
                For a base case, we condition on the empty set $S_0 = \varnothing$ and notice the tautological equality
                \begin{align*}
                    \probgiv{R_{k_1 i} = 0}{X_i = 0; \cap_{j \in S_{0}} \{ R_{ji} = 0\}} =  \probgiv{  R_{k_1 i} = 0 }{X_i = 0}.
                \end{align*}
                For the inductive step, assume that the property holds for the intersection of the events $\{ R_{ji} = 0 \}$ for $j \in S_{\ell -1}$. 
                Then we will prove
                \begin{align}
                    \probgiv{R_{k_\ell i} = 0}{X_i = 0; \bigcap_{j \in S_{\ell-1}} \{ R_{ji} = 0\}} \probgiv{\bigcap_{j \in S_{\ell-1}} \{ R_{ji} = 0\}}{X_i = 0} \geq \prod_{j \in N_i(E) } \probgiv{  R_{ji} = 0 }{X_i = 0}. \label{eq:rjideps}
                \end{align}
                Applying the inductive hypothesis, the inequality is implied by
                \begin{align}
                    \probgiv{R_{k_\ell i} = 0}{X_i = 0; \bigcap_{j \in S_{\ell - 1} }\{ R_{ji} = 0 \} } \geq \probgiv{  R_{k_\ell i} = 0 }{X_i = 0}. \label[ineq]{ineq:rkliineq2}
                \end{align}
                which can be seen by the fact that $R_{k_\ell i} $ can be written as
                \begin{align*}
                    R_{k_\ell i} = \indic{X_i = 0} \indic{X_{k_\ell} = 2} Z{\curly{\sum_{j \in N_{k_\ell}(E) \setminus \{ i \}}  \indic{X_j = 0}}} 
                \end{align*}
                where $Z\{m\}$ is a random variable equal to 1 with probability $1/(m+1)$ and zero otherwise.
                However, notice that for two sets $S, S'$, the variables 
                \begin{align*}
                    Z{\curly{\sum_{j \in S} \indic{X_j = 0}}}, Z{\curly{\sum_{j \in S'} \indic{X_j = 0}}},
                \end{align*}
                are positively correlated if $S \cap S' \neq \varnothing$ (since $\{ X_j \}_j$ are mutually independent and so the values of $Z{\{\cdot \}}$ depend on each other only insofar as they share some $X_j$ in common).
                Thus, $R_{k_\ell i}$ and $R_{ji}$ for $j \in S_{\ell - 1}$ are positively correlated, which proves \cref{ineq:rkliineq2}.

                Next, notice
                \begin{align*}
                    \probgiv{R_{ji} = 0}{X_i = 0} &= \prob{X_j < 2} + \prob{X_j = 2} \probgiv{R_{ji} = 0}{X_j = 2, X_i = 0} \tag{Law of total probability} \\
                    &=(1-p) + p\ex{1 - \frac{1}{\sum_{\ell \in N_j(E) \setminus \{ i \}} \indic{X_\ell = 0}} } \tag{Definitions of ${X_j}$, $R_{j,i}$} \\
                    &=  (1-p) + p\sum_{k = 0}^{d_j - 1}  \prob{\paren{\sum_{\ell \in N_E(j) \setminus \{ i \}} \indic{X_\ell = 0}} = k} \frac{k}{k+1} \tag{Law of total probability} \\
                    &= 1 - p\sum_{k = 0}^{d_j - 1} \binom{d_j - 1}{k}  q^k (1-q)^{d_j - 1 -k} \frac{1}{k+1} \tag{PMF of a binomial r.v.} \\
                    &=1 - p \sum_{k = 0}^{d_j - 1} \binom{d_j - 1}{k}  q^{ d_j - 1 - k} (1-q)^{k} \frac{1}{d_j - k} \tag{Change of index $k \to d_j - 1 - k$} \\
                    &=1 - \frac{p}{q} \sum_{k = 0}^{d_j - 1} \frac{(d_j - 1)!}{k! (d_j - k)!}  q^{ d_j - k} (1-q)^{k} \tag{Simplify} \\
                    &=1 - \frac{p}{q d_j} \sum_{k = 0}^{d_j - 1} \binom{d_j}{k}  q^{ d_j - k} (1-q)^{k} \tag{Simplify} \\
                    &=1 - \frac{p}{q} \cdot \frac{{1 - (1-q)^{d_j}}}{d_j}. \yesnum \label{eq:rji}
                \end{align*}
                where for any empty sum in the sequence (because $d_j = 1$), we default the value to 1. 
                The last line comes from {recognizing the sum as $1-(1-q)^{d_j}$ from the law of total probability and the binomial probabiliy mass function.}
                Plugging the expression in \cref{eq:rji} into \cref{ineq:intersectprod} yields the stated result.
                
                Next, we will show that, for any fixed average degree $d$, 
                \begin{align*}
                    \max_{E \subset [n]^2 \; : \; \frac{\abs{E}}{n} = d} \ex{\sum_{i \in [n]} Y_i(E)} &\leq 
                    \max_{E \subset [n]^2 \; : \; \frac{\abs{E}}{n} = d} \frac{1}{n}\sum_{i \in [n]}{1 - q\prod_{j \in N_i(E) } \g{d_j} } \\
                    &\leq {1 - q{\g{d}^d}}, \yesnum \label[ineq]{eq:desiredineq}
                \end{align*}
                The first line follows from applying linearity of expectation and plugging in the RHS of \cref{eq:rji} for ${u_i(E)}$ for all $i$.
                The second line is what we will prove next. For convenience, let 
                \begin{align*}
                    \h{d} \defeq \frac{p}{q} \cdot \frac{1-(1-q)^d}{d} .
                \end{align*}
                Then \cref{eq:desiredineq} can be equivalently written
                \begin{align}
                    1 - q\min_{E \subset [n]^2 \; : \; \frac{\abs{E}}{n} = d} \frac{1}{n}\sum_{i \in [n]}{\prod_{j \in N_i(E) } \paren{1- \h{d_j}} } \geq 1 - q{\paren{1 - \h{d}}^d}. \label{eq:minavgdegreesw}
                \end{align}

                The rest of the proof follows by plugging in $h(d)$ instead of $p/d$ for the proof of \cref{prop:obliviousoptimal}, starting at \cref{eq:amgm}. \qed
        \fi
        
        One might wonder under what circumstances $3$- and $4$-cycle-free $d$-regular graphs exist.
        A result due to \citet{erdos1963regulare} shows that for all $m$, there always exists some $n > m$ such that there exists a graph on $n$ nodes with this property.

        \begin{proposition} [\citet{erdos1963regulare}, adapted and simplified from English translation in \citet{chartrand2010graphs}] \label{thm:erdos1963}
            For every pair of integers $d \geq 3, g \geq 5$, there exists a $d$-regular graph of girth $g$ with at most $4(d-1)^{g-1}$ nodes.
        \end{proposition}
        If we set $g=5$ in the result, for any $d$, we can construct the $d$-regular graph with a number of nodes polynomial in $d$.
        Let $n_0 \leq 4(d-1)^{g-1}$ for $d \geq 4$ and $n_0 \leq d+1 $ for $d=1,2$.
        Then we can create a graph for infinitely many $n > n_0$ by creating $k$ disjoint copies of the $d$-regular girth-5 graph for $k=2,3 \dots$. 
        The proposition handles cases with $d \geq 3$, but it is simple to handle $d \in \{0, 1, 2\}$. For $d=0$, the property is satisfied by the empty graph. 
        For $d=1$, this property is satisfied by any perfect matching: i.e., any graph where each node has an edge to exactly one other node.
        For $d=2$, this property is satisfied by any cycle graph of length greater than or equal to 5.
        Next, we give an explicit form for the price of anarchy in the frictionless case.

        \begin{figure}[t]
            \centering
            \begin{subfigure}[b]{0.45\textwidth}
                \includegraphics[width=\textwidth]{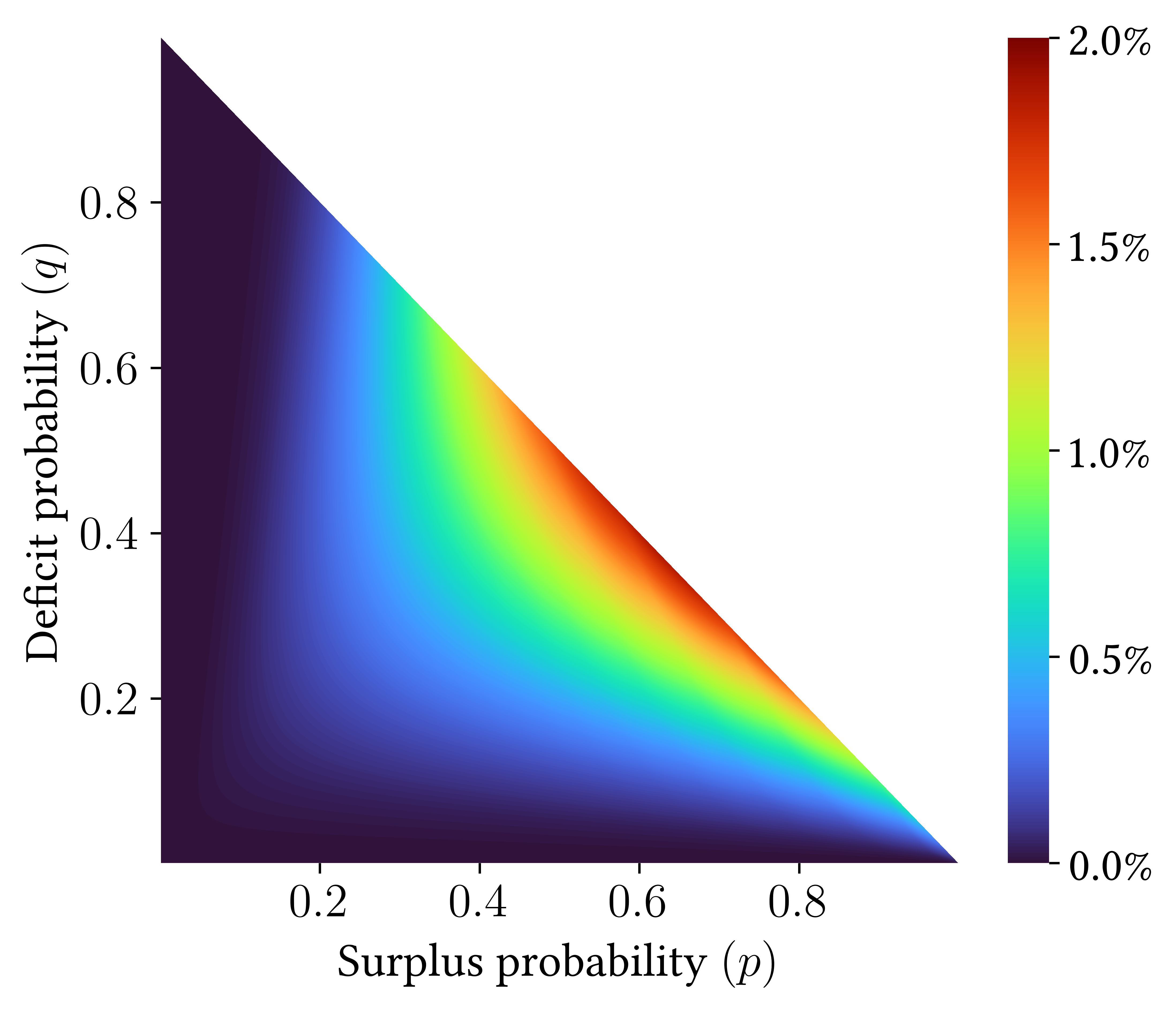}
                \caption{Informed frictionless network formation}
                \label{fig:informed}
            \end{subfigure}
            \hfill 
            \begin{subfigure}[b]{0.45\textwidth}
                \includegraphics[width=\textwidth]{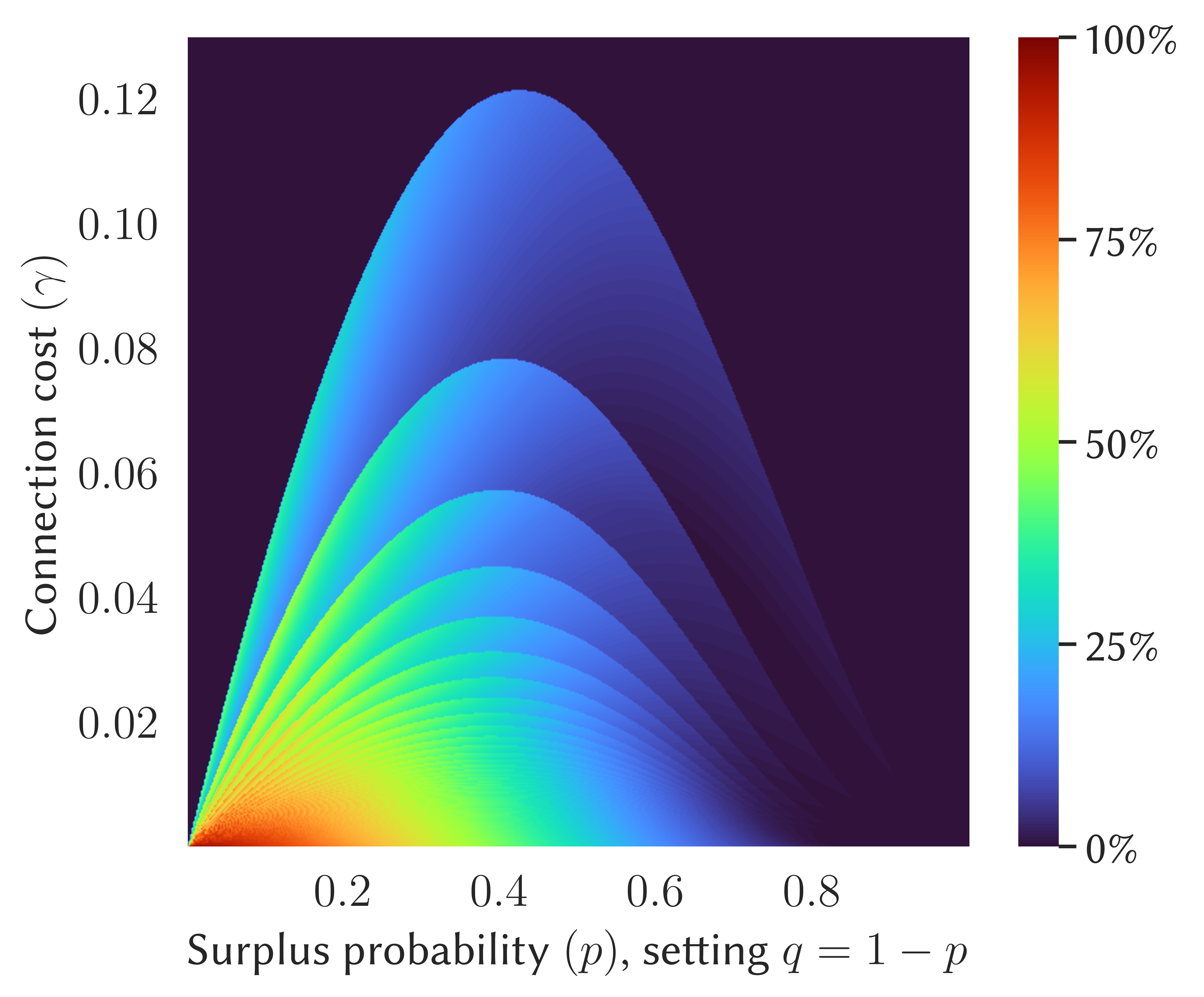}
                \caption{Informed costly network formation}
                \label{fig:informededgecosts}
            \end{subfigure}
            \caption{A visualization of the price of anarchy in professional networking games with edge costs. The percentage that the price of anarchy is above 1 is shown as the color. Left: We vary over $p,q$. Recall that $p + q \leq 1$, so the price of anarchy is plotted over the 2-dimensional simplex. Right: As in \cref{fig:comparison}, the percentage that the price of anarchy is above 1 is shown by color at given values of $p$, $\gamma$ on the x- and y-axis respectively.}
            \label{fig:informedcomparison}
        \end{figure}
        \begin{restatable}{theorem}{poainformed} \label{prop:poainformed}
            The price of anarchy in the \game~for $p, q > 0$ and $\gamma = 0$ is
            \begin{align*}
                \mathrm{\poa} (n, q, p, 0) = \frac{\max_{0 \leq d \leq n} 1 - q \paren{1 - \frac{p}{q} \cdot \frac{1-(1-q)^d}{d}}^d}{1 - q e^{-p/q}} \pm o(1).
            \end{align*}
        \end{restatable}
    
        The price of anarchy across different values of $p$ and $q$ is shown in \cref{fig:informed}.

        \ifproofsinbody
        
        \proofof{\cref{prop:poainformed}}
        The asymptotic social welfare for equilibrium networks is calculated similarly to before as
        \begin{align*}
         \frac{1}{n}\sum_{i=1}^n {u_i(\nashedgeset)}  &=  1 - q \paren{1 - \frac{p}{q} \cdot \frac{1 - (1-q)^{n-1}}{n-1}}^{n-1} \pm o(1)\\
         &= 1 - q e^{-p/q} \pm o(1).
        \end{align*}
        
        The asymptotic social welfare for efficient networks comes directly from \cref{prop:informedoptimal}, noticing that the largest 3- and 4-cycle-free regular subgraph for a population of size $n$ is within $o(1)$ of optimal.
        \qed 
        \fi
        
    \subsection{Costly network formation.}

        We can prove a similar structural result in the \gm~to what we did in \cref{prop:layeredgraph} for the \indiscgame.
        We overcome the difficulty of the dependence between opportunity transfers by observing that, at equilibrium, individuals will always prefer to connect with others who are shortest-path three or greater away from them.
        Thus, equilibrium networks are $C_3$- and $C_4$-free.
        Indeed, the fact that strategic individuals prefer to connect to others farther away in the network gives a quantitative justification for the value of long ties (i.e., local bridges) \cite{granovetter1973strength, burt2004structural}  which is an extensive topic of study in sociology and the social networks literature.
        We state the result formally next in \cref{prop:informedlayeredgraph}.
        Following that, we state a result about efficient networks in \cref{prop:informededgecostoptimal} and then give the price of anarchy in \cref{prop:informededgecostpoa}.

        \begin{restatable}{proposition}{informedlayeredgraph} \label{prop:informedlayeredgraph}
            For the \gm~on a population of size $n$ with connection costs $\gamma > 0$, in any \eqbm~edge set $E_{\mathrm{\eqbm}}$, there exists a set $S \subseteq [n]$ where $\abs{\cS} \geq n - (p/\gamma)^{2}(1 + (p/\gamma)^{3})$, where $\cS$ is 1-degree homophilous and where no individual $i \in \cS$ is incident to a 3- or 4-cycle in $E_{\mathrm{\eqbm}}$.
        \end{restatable}

        \ifproofsinbody
        
        \proofof{\cref{prop:informedlayeredgraph}} 
        \newcommand{\dist}{\mathtt{dist}}
        Let $\dist_E(i,j)$ be the shortest path from from individual $i$ to $j$ on an edge set $E$ where we let $\dist_E(i,j) = \infty$ if there is no path from $i$ to $j$ in $E$.
        Recall from \cref{prop:informeddegub} that the degree of any individual is upper bounded by a constant ($p/\gamma$) not depending on $n$.
        Let the maximum degree of an individual be denoted $\dmax$.
        Suppose for contradiction that there are two individuals $i,j$ such that $\dist_{E_{\mathrm{\eqbm}}}(i,j) \geq 4$, $d_i = d_j$, and where each $i$ and $j$ are connected to individuals $k$ and $\ell$ respectively where $d_k \geq d_i, d_\ell \geq d_j$ and at least one of the following is true:
        \begin{enumerate}
            \item the $(i,k)$ (resp. $(j,\ell)$) connection is incident to a 3- or 4-cycle in $E_{\mathrm{\eqbm}}$, or
            \item $d_k > d_i + 1$ (resp. $d_\ell > d_j + 1$).
        \end{enumerate}
        Then, since $i$ and $j$ would each rather defect from their existing connections to $k$ and $\ell$ to form a connection with each other, $E_{\mathrm{\eqbm}}$ cannot be an equilibrium, a contradiction.
        Thus, for each pair $i,j$ such that $d_i, d_j$ and where each $i$ and $j$ are connected to individuals of equal or greater degree with at least one of the conditions in the list above is true, it must hold that $\dist_{E_{\mathrm{\eqbm}}}(i,j) \leq 3$.
        But this implies that fewer than $\dmax^3 + 1$ nodes of degree $d_i$ can have this property, since at most $\dmax^3$ nodes of degree $d_i$ can be within distance 3 of individual $i$.
        Define $\cV$ to consist of each individual $i \in [n]$ such that for all $j \in N_i(E_{\mathrm{\eqbm}})$, it holds $d_j \leq d_i + 1$, and where $i$ is not incident to any 3- or 4-cycles.
        Notice that $\cV$ is no smaller than $n - (p/\gamma)(1 + (p/\gamma)^{3})$, since the number of distinct degrees of individuals in $\cV$ is no more than $p/\gamma$ by \cref{prop:informeddegub} and at most $\dmax^3 +1 \leq (p/\gamma)^3 + 1$ individuals of degree $d$ can be incident to 3- or 4-cycles or have a connection of degree greater than $d+1$.
        Let $\cS \defeq \{ i \in \cV \; : \; \forall j \in N_i(E_{\mathrm{\eqbm}}), j \in \cV\}$ consist of individuals in $\cV$ whose neighbors are each also in $\cV$.
        Notice for all $i \in \cS$ and $j \in N_i(E_{\mathrm{\eqbm}})$, it must also be that $d_i \leq d_j + 1$. 
        All that remains is to prove that $[n] \setminus \cS$ consists of no more than $(p/\gamma)^{2}(1 + (p/\gamma)^{3})$ individuals. 
        But since $[n] \setminus \cV$ consists of no more than $(p/\gamma)(1 + (p/\gamma)^{3})$ individuals and all individuals must have degree upper bounded by a $p/\gamma$, the number of individuals connected to individuals in $[n] \setminus \cV$ must be no more than the stated number.
        \qed
        \fi

            \begin{restatable}{proposition}{informededgecostoptimal} \label{prop:informededgecostoptimal}
                In the \game, if $q, p > 0$, a network is efficient if it is a 3- and 4-cycle-free $d$-regular graph where $d$ is given by a solution to
                \begin{align}
                    \argmax{0 \leq d \leq n, d \in \N} 1 - q\paren{1 - \frac{p}{q} \cdot \frac{1 - (1-q)^d}{d}}^d - \gamma d. 
                \end{align}
            \end{restatable}

        \ifproofsinbody
        
            \proofof{\cref{prop:informededgecostoptimal}} It is straight-forward to adapt the proof of \cref{prop:informedoptimal}. 
            All that is needed is to notice that for any fixed average degree $d$, the edge cost term in average utility is constant (and equal to $\gamma d$). 
            Thus, the minimization problem in \cref{eq:minavgdegreesw} is not influenced by the constant term and the result follows.
            \qed
        \fi
        
            \begin{restatable}{proposition}{informededgecostpoa} \label{prop:informededgecostpoa}
            The price of anarchy in the \game~for $p,q > 0$ and $0 < \gamma \leq qp$ is
                \begin{align*}
                    \mathrm{\poa} (n, q, p, \gamma) \equnder{\delta_1, \delta_2} \frac{\max_{0 \leq d^* \leq n} 1 - q \paren{1 - \frac{p}{q} \cdot \frac{1-(1-q)^{d^*}}{d^*}}^{d^*} - \gamma d^*}{1 - q \paren{1 - \frac{p}{q} \cdot \frac{1-(1-q)^{d + \delta_1 + \delta_2}}{d + \delta_1 + \delta_2}}^{d + \delta_1} - \gamma (d + \delta_1)} \pm o(1).
                \end{align*}
                where $d$ is the largest integer satisfying 
                \begin{align}
                     \frac{\gamma}{p} \leq {\frac{1 - (1-q)^d}{d}} \cdot \paren{1 - \frac{p}{q} \cdot \frac{1 - (1-q)^d}{d}}^{d-1}. \label{eq:informeddregd}
                \end{align}
                There is no price of anarchy otherwise.
            \end{restatable}
            
            \ifproofsinbody
            
            \proofof{\cref{prop:informededgecostpoa}} 
            We first calculate the social optimum utility. 
            From \cref{prop:informededgecostoptimal}, we have that efficient networks are achieved by 3- and 4-cycle-free $d$-regular graphs for $d$ a solution to a maximization problem over $0 \leq d \leq n$.
            To show that the asymptotic social welfare approaches that of those efficient networks, we simply need to show that for large enough $n$, such a network can be built so as to include an arbitrarily large fraction of nodes.
            Formally, for all $\varepsilon > 0$, we want to show that for some $m$ and all $n > m$, there exists a set $\cS$ such that $\abs{\cS} \geq (1-\varepsilon)n$ and all $i \in \cS$ are not incident to any 3- or 4-cycles and each is $d$ regular for the $d$ achieving the maximum in the optimization problem.
            Then, since all $[n] \setminus \cS$ have nonnegative utility, social welfare is at least $(1-\varepsilon)$ times the sum of utility of individuals in $\cS$.
            To define the set $\cS$, we can simply create disjoint copies of of such $d$-regular graphs of girth $5$ whose existence is implied by \cref{thm:erdos1963}.
            Since such graphs are made of no more than $4(d-1)^4$ nodes, at most that many nodes can be excluded from $\cS$ (otherwise we could fit another copy).
            This proves the asymptotic average socially optimal utility.

            We next deal with the simple case when $\gamma \geq qp.$
            If $\gamma > qp$, we claim the unique equilibrium network is the empty graph.
            For contradiction, suppose any node $i$ has at least one connection.
            Then their utility must be no more than
            \begin{align*}
                1 - q \paren{1 - {p}}^{d_i} - \gamma {d_i}
            \end{align*}
            if they are each of their connections' only connection.
            But this is the same upper bound on $i$'s utility that we gave in \cref{prop:poaedgecosts} for the \indiscgm~with $\gamma > 0$, so we know 
            that $i$'s utility would be higher if they had no connections, a contradiction of the equilibrium conditions.
            For the socially optimal utility, notice
            \begin{align*}
                \max_{0 \leq d^* \leq n} 1 - q \paren{1 - \frac{p}{q} \cdot \frac{1-(1-q)^{d^*}}{d^*}}^{d^*} - \gamma d^* \leq \max_{0 \leq d^* \leq n} 1 - q \paren{1 - {p}}^{d^*} - \gamma d^*
            \end{align*}
            where we just showed the right-hand side was achieved by $d^* = 0$.
            Thus, there is no price of anarchy when $\gamma > qp$.
            

            Finally, we handle the case when $\gamma \leq qp$.
            We have already shown that the social optimum utility is given in the numerator in the first paragraph of this proof.
            Thus, we just need to calculate the denominator.
            First, we will give an upper bound on worst-case average utility at an equilibrium using an equilibrium $d$-regular graph.
            Notice that a $d$-regular graph satisfies the equilibrium conditions
            \begin{align*}
                 \frac{\gamma}{p} \leq {\frac{1 - (1-q)^d}{d}} \cdot \paren{1 - \frac{p}{q} \cdot \frac{1 - (1-q)^d}{d}}^{d-1}
            \end{align*}
            which intuitively says that no individual would want to sever any of their existing connections, and
            \begin{align*}
                 \frac{\gamma}{p} \geq {\frac{1 - (1-q)^{d+1}}{d+1}} \cdot \paren{1 - \frac{p}{q} \cdot \frac{1 - (1-q)^d}{d}}^{d}.
            \end{align*}
            which intuitively says that no individual would want to create a connection to another node.
            As in the \indiscgm~with connection costs, we will notice that 
            \begin{align}
                \bigcup_{d=1}^\infty \sqparen{{\frac{1 - (1-q)^{d+1}}{d+1}} \cdot \paren{1 - \frac{p}{q} \cdot \frac{1 - (1-q)^d}{d}}^{d}, {\frac{1 - (1-q)^d}{d}} \cdot \paren{1 - \frac{p}{q} \cdot \frac{1 - (1-q)^d}{d}}^{d-1}} = (0, 1] \label{eq:cupintervals}
            \end{align}
            which shows that for any $\gamma / p$ there must exist some $d$ satisfying the equilibrium conditions.
            To see this, we will first show that each left endpoint of an interval is less than the successive right endpoint overlaps with the next one:
            \begin{align}
                {\frac{1 - (1-q)^{d+1}}{d+1}} \cdot \paren{1 - \frac{p}{q} \cdot \frac{1 - (1-q)^d}{d}}^{d} < {\frac{1 - (1-q)^{d+1}}{d+1}} \cdot \paren{1 - \frac{p}{q} \cdot \frac{1 - (1-q)^{d+1}}{d+1}}^{d} \label{prop:informedintervalsoverlap}
            \end{align}
            This is equivalent to showing
            \begin{align*}
                 &\frac{1 - (1-q)^{d+1}}{d+1} < \frac{1 - (1-q)^d}{d} \\
                 \iff &(dq + 1) (1-q)^d < 1.
            \end{align*}
            To prove the last line, we will argue the left-hand side is less than 1 for $d=1$ and the expression is monotonic decreasing on $d \geq 1$.
            For $d=1$, notice $(1 + dq)(1-q)^d = (1 + q)(1- q) = 1 -q^2 < 1$.
            For monotonicity, notice
            \begin{align*}
                \frac{\dif}{\dif d} (dq + 1) (1-q)^d = (1-q)^d ((dq + 1) \log (1-q) + q)
            \end{align*}
            and the first term on the right-hand side is positive, so we just need to show the left term is negative:
            \begin{align*}
                (dq + 1) \log (1-q) + q &= -(dq + 1) \sum_{k=1}^\infty \frac{q^k}{k} + q \\
                &= q - (dq + 1)q - (dq + 1) \sum_{k=2}^\infty \frac{q^k}{k} \\
                &= -dq^2  - (dq + 1) \sum_{k=2}^\infty \frac{q^k}{k}
            \end{align*}
            and since all the terms are negative, the whole expression is negative.
            This completes the proof of \cref{prop:informedintervalsoverlap}.
            Next, notice that the left endpoint of the intervals in \cref{eq:cupintervals} converge to zero. To see this, notice the terms are nonnegative, and
            \begin{align*}
                \lim_{d \to \infty}{\frac{1 - (1-q)^{d+1}}{d+1}} \cdot \paren{1 - \frac{p}{q} \cdot \frac{1 - (1-q)^d}{d}}^{d} \leq \lim_{d \to \infty} \frac{1}{d+1} = 0.
            \end{align*}
            Thus, for any value of $\gamma / p$, there exists some $d$ satisfying the equilibrium conditions.
            Finally, the largest $d$ satisfying the inequality in the statement of the result is the worst-case $d$-regular equilibrium, which is an upper bound on the worst-case equilibrium overall.
            That is, the worst-case $d$-regular equilibrium achieves utility
            \begin{align*}
                1 - q\paren{1 - \frac{p}{q}\frac{1 - (1-q)^d}{d}}^d - \gamma d
            \end{align*}
            for $d$ the largest integer satisfying \cref{eq:informeddregd}.
            For the rest of this proof, we will use $d$ (without subscripts) to refer to the $d$ of the worst-case $d$-regular equilibrium.

            Next, we give a lower bound on worst-case average utility at an equilibrium.
            We will use the fact that equilibrium networks include large 1-degree homophilous sets of individuals that are not incident to any 3- or 4-cycles.
            From here on, for a given equilibrium edge set $E_{\mathrm{\eqbm}}$, we will reason about the utility of individuals in a set $\cS$ that is 1-degree homophilous and 3- and 4-cycle-free where $\cS$ is composed of at least $n - (p/\gamma)^2(1 + (p/\gamma)^3)$ individuals as we know exists from \cref{prop:informedlayeredgraph}.
            Since utility of individuals exclude from $\cS$ is bounded below by 0 and above by 1, average utility will be arbitrarily close to that of individuals in $\cS$ for large enough $n$.

            Consider a node $i \in \cS$ with the largest degree of all nodes in $\cS$.
            Define $\dmax \defeq d_i$.
            Recall from the equilibrium conditions that for all $j \in N_i(E_{\mathrm{\eqbm}})$, 
            \begin{align*}
                \frac{\gamma}{p} &\leq \frac{1-(1-q)^{d_j}}{d_j} \prod_{\ell \in N_i(E_{\mathrm{\eqbm}}) \setminus \{ j \}} \paren{1 - \frac{p}{q} \cdot \frac{1 - (1-q)^{d_\ell}}{d_\ell}} \\
                &\leq \max_{u, v \in \{ \dmax-1, \dmax \}} \frac{1-(1-q)^{v}}{v}  \paren{1 - \frac{p}{q} \cdot \frac{1 - (1-q)^{u}}{u}}^{\dmax - 1} \\
                &\leq \frac{1-(1-q)^{\dmax - 1}}{\dmax - 1}  \paren{1 - \frac{p}{q} \cdot \frac{1 - (1-q)^{\dmax}}{\dmax}}^{\dmax - 1} 
            \end{align*}
            where the first line is the equilibrium condition for a 3- and 4-cycle-free node, the second line is a relaxation of the equilibrium condition using 1-degree homophily as before, and the third line comes from plugging in the maxima.
            As in the analogous proof of the \indiscgm, this gives us an upper bound on the degree of nodes in $\cS$ which is tighter than the simple bound we give in \cref{prop:informeddegub}.
            Next, we will argue that a lower bound on the utility of individuals in $\cS$ is that given by an individual of degree $\dmax$ connected exclusively to individuals of degree $\dmax + 1$.
            To see this, notice that 
            \begin{align}
                1 - q\paren{1 - \frac{p}{q}\frac{1 - (1-q)^{k+1}}{k+1}}^k - \gamma k \label{eq:informedlb}
            \end{align}
            is concave in $k$.
            Thus, the minimum must be achieved at $1$ or $\dmax$.
            We can show that the minimum is at $\dmax$ using the following sequence:
            \begin{align*}
                &1 - q \paren{1 - \frac{p}{q} \cdot \frac{1 - (1-q)^2}{2}} - \gamma > 1 - q\paren{1 - \frac{p}{q}\frac{1 - (1-q)^{\dmax+1}}{\dmax+1}}^{\dmax} - \gamma \dmax  \\
                \iff &\dmax \geq \frac{q}{\gamma} \paren{ \frac{p q (2-q)}{2} - 1 -\paren{1 - \frac{p}{q}\frac{1 - (1-q)^{\dmax+1}}{\dmax + 1}}^d  }+1 \\
                \impliedby &\dmax \geq \frac{q}{\gamma} \paren{ \frac{p q (2-q)}{2} - 1 - e^{-p/q} }+1 \\
                \impliedby &\dmax \geq 1.
            \end{align*}
            All that remains is to relate $\dmax$ to $d$ above:
            \begin{align*}
                d &= \max \curly{k \in \N  \; : \;\frac{\gamma}{p} \leq {\frac{1 - (1-q)^k}{k}} \cdot \paren{1 - \frac{p}{q} \cdot \frac{1 - (1-q)^k}{k}}^{k-1}} \\
                &= \max \curly{k \in \N  \; : \;\frac{\gamma}{p} \leq {\frac{1 - (1-q)^{k-1}}{k-1}} \cdot \paren{1 - \frac{p}{q} \cdot \frac{1 - (1-q)^{k-1}}{k-1}}^{k-2}} -1 \\
                &\geq \max \curly{k \in \N  \; : \;\frac{\gamma}{p} \leq {\frac{1 - (1-q)^{k-1}}{k-1}} \cdot \paren{1 - \frac{p}{q} \cdot \frac{1 - (1-q)^{k}}{k}}^{k-1}} -1 \\
                &\geq \dmax -1
            \end{align*}
            Plugging in $d+1$ for $k$ into \cref{eq:informedlb} completes the proof of the lower bound, which completes the proof of the result.
            \qed

            \fi

\ifconferencesubmission
\section{Further discussion: Equilibrium concept}

    Our equilibrium concept is equivalent to one in which an individual can sever multiple edges at once (rather than a single edge as we define it in the body of the paper). We prove this in the next result.

        \begin{restatable}{proposition}{multipleedgeeq} \label{prop:multipleedgeeq}
        The set of defection-free pairwise Nash equilibria is equivalent to one in which the second equilibrium condition is replaced with
        \begin{enumerate}
            \setcounter{enumi}{1}
            \item for all $i$ and all $S_i \subseteq N_i(E)$
            \begin{align}
                0 &\geq {u_i(E \setminus \{ (i,\ell) \; : \; \ell \in S_i \}) } - {u_i(E)}. \label{eq:nodelete}
            \end{align}
        \end{enumerate}
    \end{restatable}
    
    \proofof{\cref{prop:multipleedgeeq}} For simplicity, we will call a \eqbm as it was defined in the body of the paper as a \textit{single edge equilibrium} and the altered definition a \textit{multiple edge equilibrium}. To prove the result, first notice that any multiple edge equilibrium is also a single edge equilibrium. Next, suppose there is some individual $i$ with a nonempty set of edges that they could sever and increase their utility, so the edge set $E$ is not a multiple edge equilibrium. Let $S = \{ s_1, \dots, s_{|S|} \}$ be a set of edges user $i$ could remove and derive positive utility. Let $S_k =  \{ s_1, \dots, s_k \}$ and $S_0 = \varnothing$. Then
    $$
    u_i(E \setminus S) - u_i(E) = \sum_{\ell = 1}^{|S|} (u_i(E \setminus S_\ell) - u_i(E \setminus S_{\ell -1})),
    $$
    and 
    $$
    u_i(E \setminus s_\ell) - u_i(E) \geq u_i(E \setminus S_\ell) - u_i(E \setminus S_{\ell -1})
    $$
    so there must be some $u_i(E \setminus \{ (i, j) \}) - u_i(E)$ that is positive. Intuitively, a person would never increase utility from severing two edges jointly but not from severing either of the edges singly. \qed

\section{Further discussion: Existence of equilibria in the \indiscgame~with connection costs}

    In the \indiscgame~with connection costs, there always exists an equilibrium. In fact, if $n$ is even, we can show that there always exists a regular equilibrium! We state this next.

\else
\section{Technical lemmas}

    In this section, we provide technical lemmas that will be useful for proving many of our results in \cref{sec:deferredproofs}.
    First, we show that our equilibrium concept is equivalent to one in which an individual can sever multiple edges at once (rather than a single edge as we define it in the body of the paper).

        \begin{restatable}{proposition}{multipleedgeeq} \label{prop:multipleedgeeq}
        The set of defection-free pairwise Nash equilibria is equivalent to one in which the second equilibrium condition is replaced with
        \begin{enumerate}
            \setcounter{enumi}{1}
            \item for all $i$ and all $S_i \subseteq N_i(E)$
            \begin{align}
                0 &\geq {u_i(E \setminus \{ (i,\ell) \; : \; \ell \in S_i \}) } - {u_i(E)}. \label{eq:nodelete}
            \end{align}
        \end{enumerate}
    \end{restatable}
    
    \proofof{\cref{prop:multipleedgeeq}} For simplicity, we will call a \eqbm as it was defined in the body of the paper as a \textit{single edge equilibrium} and the altered definition a \textit{multiple edge equilibrium}. To prove the result, first notice that any multiple edge equilibrium is also a single edge equilibrium. Next, suppose there is some individual $i$ with a nonempty set of edges that they could sever and increase their utility, so the edge set $E$ is not a multiple edge equilibrium. Let $S = \{ s_1, \dots, s_{|S|} \}$ be a set of edges user $i$ could remove and derive positive utility. Let $S_k =  \{ s_1, \dots, s_k \}$ and $S_0 = \varnothing$. Then
    $$
    u_i(E \setminus S) - u_i(E) = \sum_{\ell = 1}^{|S|} (u_i(E \setminus S_\ell) - u_i(E \setminus S_{\ell -1})),
    $$
    and 
    $$
    u_i(E \setminus s_\ell) - u_i(E) \geq u_i(E \setminus S_\ell) - u_i(E \setminus S_{\ell -1})
    $$
    so there must be some $u_i(E \setminus \{ (i, j) \}) - u_i(E)$ that is positive. Intuitively, a person would never increase utility from severing two edges jointly but not from severing either of the edges singly. \qed

    \begin{restatable}{lemma}{edgecostseqdegree} \label{prop:edgecostseqdegree}
    In the \indiscgame~with connection costs $\gamma > 0$, for any \eqbm~edge set $E$ and all $i \in [n]$, it holds
        \begin{align*}
            d_i \leq  \frac{qp}{\gamma}.
        \end{align*}
    \end{restatable}
    \proofof{\cref{prop:edgecostseqdegree}} 
        Recall from the equilibrium conditions, for all $i,j$ such that $(i,j) \in E$ it holds
        \begin{align*}
            0 &\geq {u_j(E \setminus \{ (i,j) \} ) - u_j(E)} \\
            &= \gamma - {\frac{qp}{d_{i}}} \prod_{k \in N_j(E) \setminus \{ j \}}  \paren{1 - \frac{p}{d_k}} \\
            &\geq \gamma - {\frac{qp}{d_{i}}}.
        \end{align*}
    \qed

    \begin{restatable}{lemma}{informeddegub} \label{prop:informeddegub}
        In the \game~with connection costs $\gamma > 0$, for any \eqbm~edge set $E$ and all $i \in [n]$, it holds
        \begin{align*}
            d_i \leq \frac{p}{\gamma}.
        \end{align*}
    \end{restatable}

    \proofof{\cref{prop:informeddegub}} Recall from the equilibrium conditions, for all $i,j$ such that $(i,j) \in E$, it holds 
    \begin{align*}
            0 &\geq {u_j(E \setminus \{ (i,j) \} ) - u_j(E)} \\
            &= \gamma - q \paren{\probgiv{\cap_{\ell \in N_j(E) \setminus \{i \}} \{R_{\ell j} = 0\}}{X_j = 0} 
 - \probgiv{\cap_{\ell \in N_j(E)} \{R_{\ell j} = 0\}}{X_j = 0}} \\
            &\geq \gamma - q \paren{1 - \probgiv{R_{ij} = 0}{X_j = 0}} \probgiv{\cap_{\ell \in N_j(E)\setminus \{i \}} \{R_{\ell j} = 0\}}{X_j = 0} \\
            &\geq \gamma - q \paren{1 - \probgiv{R_{ij} = 0}{X_j = 0} } \\
            &= \gamma - q \cdot \paren{\frac{p}{q}\cdot \frac{1 - (1-q)^{d_i}}{d_i}} \\
            &\geq \gamma - { \frac{p}{d_i}}
    \end{align*}
    where the third line follows from the fact that 
    \begin{align*}
        \probgiv{\cap_{\ell \in N_j(E) } \{R_{ij} = 0\}}{X_j = 0} =  \probgiv{R_{ij} = 0}{X_j = 0}\probgiv{\cap_{\ell \in N_j(E) \setminus \{i \}} \{R_{ij} = 0\}}{X_i = 0, R_{ij} = 0} 
    \end{align*}
    and
    \begin{align*}
        \probgiv{\cap_{\ell \in N_j(E) \setminus \{i \}} \{R_{ij} = 0\}}{X_i = 0, R_{ij} = 0} \geq \probgiv{\cap_{\ell \in N_j(E) \setminus \{i \}} \{R_{ij} = 0\}}{X_j = 0}
    \end{align*}
    as we show in \cref{eq:rjideps}. \qed

    \begin{restatable}{lemma}{monotoneincreasing} \label{lem:monotoneincreasing}
        For all $\ptwo \in (0, 1]$, it holds $\paren{1 - {\ptwo}/{d}}^d$ is monotone increasing as a function of $d \in \N$.
    \end{restatable} 
    
    \proofof{\cref{lem:monotoneincreasing}}
        This can be accomplished by showing
        \begin{align*}
            \frac{\dif}{\dif d} \paren{1 - \frac{\ptwo}{d}}^d > 0
        \end{align*}
        for all $d \geq 2$, which implies that $(1-\ptwo/d)^d$ is increasing for all $d \geq 2$. 
        To complete the proof, we just need to compare utilities at $d=0, d=1, d=2$ and show that when $d=1$ the maximum is achieved.
        Notice
        \begin{align*}
            \frac{\dif}{\dif d} \paren{1 - \frac{\ptwo}{d}}^d = \paren{1 - \frac{\ptwo}{d}}^d\paren{\log \paren{1 - \frac{\ptwo}{d}} + \frac{\ptwo}{d-\ptwo}}.
        \end{align*}
        The first term in the RHS product is trivially positive since $d \geq 1, \ptwo \in (0,1)$.
        For the second term, we will use Taylor's Theorem to derive a lower bound. Notice
        \begin{align*}
            \frac{\ptwo}{d-\ptwo} + \log \paren{1 - \frac{\ptwo}{d}}  = \frac{\ptwo}{d-\ptwo} - \frac{\ptwo}{d} - \frac{\ptwo^2}{2 d^2 (1-\varepsilon)^2}
        \end{align*}
        for some $\varepsilon \in [0, \ptwo/d]$. Thus,
        \begin{align*}
            \frac{\ptwo}{d-\ptwo} - \frac{\ptwo}{d} - \frac{\ptwo^2}{2 d^2 (1-\varepsilon)^2} &= \frac{\ptwo^2}{d(d-\ptwo)} - \frac{\ptwo^2}{2 d^2 (1-\varepsilon)^2} \\
            &\geq \frac{\ptwo^2}{d(d-\ptwo)} - \frac{\ptwo^2}{2 d^2 \paren{1-\frac{\ptwo}{d}}^2}
        \end{align*}
        And we can eliminate $p^2$ and $(d-p)$ in both terms since it does not change the sign so that we have
        \begin{align*}
            \frac{1}{d} - \frac{d^2}{2 d^2 \paren{d-{\ptwo}}} &= \frac{1}{d} - \frac{1}{2 \paren{d-{\ptwo}}}
        \end{align*}
        Thus, all we need to show is
        \begin{align*}
            &\frac{1}{d} > \frac{1}{2 \paren{d-{\ptwo}}} \\
            \iff &2(d-\ptwo) > d \\
            \iff  &d> 2\ptwo  
        \end{align*}
        which is true for all $d\geq2$ and all $\ptwo \in (0,1)$.
        Finally, we just need to show that the value of the function is lesser at $d=1$ than at $d=2$.
        At $d=1$, it is $1-\ptwo$.
        At $d=2$, it is $(1-\ptwo/2)^2$.
        Since $1 - \ptwo < 1 - \ptwo + \ptwo^2/4$, the proof is completed.
    \qed

    \begin{restatable}{lemma}{decreasing} \label{prop:decreasing}
        For all $p \in (0, 1]$, it holds $(1-p/d)^{d-1}$ is monotone decreasing as a function of $d \in \N$.
    \end{restatable}

    \proofof{\cref{prop:decreasing}} 
    Notice
    \begin{align*}
        \frac{{\dif}}{\dif d} \paren{1 - \frac{p}{d}}^{d-1} &= \paren{1- \frac{p}{d}}^{d-1} \paren{\frac{(d-1)p}{\paren{1 - \frac{p}{d}}d^2} + \log\paren{1 - \frac{p}{d}} }
    \end{align*}
    Thus it is sufficient to show
    \begin{align*}
        0 &> \frac{(d-1)p}{\paren{1 - \frac{p}{d}}d^2} + \log\paren{1 - \frac{p}{d}} \\
        &= \frac{p}{d} \cdot \frac{(d-1)}{\paren{d - {p}} } - \frac{p}{d} - \sum_{\ell = 2}^\infty \frac{p^\ell}{\ell d^\ell } \\
    \end{align*}
    and we notice that since $(d-1)/(d-p) < 1$, the difference of the first two terms is negative and all of the rest of the terms are negative as well.
    \qed
    
    \begin{restatable}{lemma}{increasing} \label{prop:increasing}
        For all $p \in (0, 1]$, it holds $(1-p/(d-1))^{d}$ is monotone increasing as a function of $d \in \N_{\geq 2}$.
    \end{restatable}

    \proofof{\cref{prop:increasing}} Notice
    \begin{align*}
        \frac{\dif}{\dif d} \paren{1 - \frac{p}{d-1}}^d = \paren{1 - \frac{p}{d-1}}^d \paren{\frac{dp}{\paren{d - 1 - {p}} (d-1)} + \log \paren{1 - \frac{p}{d-1}}}
    \end{align*}
    Thus it is sufficient to show
    \begin{align*}
        0 &< \frac{dp}{\paren{d - 1 - {p}} (d-1)} + \log \paren{1 - \frac{p}{d-1}} 
    \end{align*}
    Notice, using Taylor's Theorem, that
    \begin{align*}
        \frac{dp}{\paren{d - 1 - {p}} (d-1)} + \log \paren{1 - \frac{p}{d-1}} &= \frac{dp}{\paren{d - 1 - {p}} (d-1)} - \frac{p}{d-1} - \frac{p^2}{2(d-1)^2 (1 - \varepsilon)^2} 
    \end{align*}
    for some $\varepsilon \in [0, p/(d-1)]$.
    Thus,
    \begin{align*}
        \frac{dp}{\paren{d - 1 - {p}} (d-1)} - \frac{p}{d-1} - \frac{p^2}{2(d-1)^2 (1 - \varepsilon)^2} &\geq \frac{p(1+p)}{(d - 1 - p) (d-1)} - \frac{p^2}{2 (d-1)^2 \paren{1 - \frac{p}{d-1}}^2} \\
        &= \frac{p(1+p)}{(d - 1 - p) (d-1)} - \frac{p^2}{2\paren{d- 1 - p}^2} \\
    \end{align*}
    
    Noticing that the $p$ and $d-1-p$ do not change the sign for $d \geq 2$, we can eliminate them, so the lemma holds if the following holds:
    \begin{align*}
        0 &< \frac{1+p}{d-1} - \frac{p}{2(d- 1 - p)} \\
        &= \frac{2(1+p) (d-1-p) - p(d-1)}{2(d-1)(d-1-p)}.
    \end{align*}
    The denominator is positive if $d \geq 2$ and the numerator is positive if
    \begin{align*}
        &0 < 2(1+p)(d-1-p) -p(d-1) \\
        \iff &0 < (2(1+p) - p)k - 2(1+p)^2 + p \\
        \iff &\frac{2(1+p)^2 - p}{2(1+p) - p} < d 
    \end{align*}
    which is true if $d \geq 2$.
    \qed

    \begin{restatable}{lemma}{concave} \label{prop:concave}
        For $d \geq 0$ and $p \in (0, 1)$,
        \begin{align*}
            0 &< \frac{{\dif}\,^2}{\dif d^2} \paren{1 - \frac{p}{d+1}}^d.
    \end{align*}
    \end{restatable}

    \proofof{\cref{prop:concave}} 
    Observe 
    \begin{align*}
        \frac{{\dif}\,^2}{\dif d^2} \paren{1 - \frac{p}{d+1}}^d &= \frac{p}{(1 + d)^2 \left(1 - \frac{p}{1 + d}\right)} \left(1 - \frac{p}{1 + d}\right)^d \left(2- \frac{2 d }{(1 + d) } -\frac{d p}{(1 + d)^2 \left(1 - \frac{p}{1 + d}\right)} \right) \\
        &+ \left(1 - \frac{p}{1 + d}\right)^d \left(\frac{d p}{(1 + d)^2 \left(1 - \frac{p}{1 + d}\right)} + \log\left(1 - \frac{p}{1 + d}\right)\right)^2
    \end{align*}
    and the second term is trivially positive.
    Thus, the following proves the result:
    \begin{align*}
        2 - \frac{2 d }{1 + d } -\frac{d p}{(1 + d)^2 \left(1 - \frac{p}{1 + d}\right)}  &= \frac{2}{1+d} -\frac{d p}{(1 + d)^2 \left(1 - \frac{p}{1 + d}\right)} \\
        &= \frac{2\left(1+d - {p}\right) - d p}{(1 + d)^2 \left(1 - \frac{p}{1 + d}\right)} \\
        &= \frac{2(1-p) + d(2-p)}{(1 + d)^2 \left(1 - \frac{p}{1 + d}\right)} \\
        &>0
    \end{align*}
    \qed

\ifproofsinbody
\else
\section{Deferred proofs} \label{sec:deferredproofs}

    We restate the proofs before the results for convenience.
    
    \indiscselfish*
            \proofof{\cref{prop:indiscselfish}} 
            Let $E$ be any edge set and suppose there exists some pair of individuals $(i,j) \not\in E$.
            %
            %
            Observe that
            \begin{align*}
                u_i(E) = 1 - q \prod_{\ell \in N_i(E)} \paren{1 - \frac{p}{d_\ell}}
            \end{align*}
            and 
            \begin{align*}
                u_i(E \cup \{ (i,j) \}) = 1 - q \paren{1 - \frac{p}{d_j+1}}\prod_{\ell \in N_i(E)} \paren{1 - \frac{p}{d_\ell}}
            \end{align*}
            which implies
            \begin{align*}
                u_i(E \cup \{ (i,j) \}) - u_i(E\} &= \frac{qp}{d_j+1} \prod_{\ell \in N_i(E)} \paren{1 - \frac{p}{d_\ell}} \\
                > 0.
            \end{align*}
        \qed

    \obliviousoptimal*

    \indiscpoa*
            \proofof{\cref{prop:indiscpoa}} 
            All that we need in addition to \cref{prop:indiscselfish} and \cref{prop:obliviousoptimal} is to calculate social welfare for each of the efficient and equilibrium networks.
            
            \cref{prop:indiscselfish} states that the complete graph is the unique (and therefore worst-case) equilibrium.
            Thus, the average utility is
            \begin{align*}
             \frac{1}{n}\sum_{i=1}^n {u_i(\nashedgeset)}  &= 1 - q \paren{1 - \frac{p}{n-1}}^{n-1} \\
             &= 1 - q e^{-p} \pm o(1).
            \end{align*}

            \cref{prop:obliviousoptimal} tells us that maximal matchings are efficient (up to a additive constant of $O(n^{-1}$).
            Thus, average utility is
            \begin{align*}
             \max_{E \in [n]^2} \;\; \frac{1}{n}\sum_{i=1}^n {u_i(E)}   = 1 - q (1 - {p}) \pm o(1).
            \end{align*}
            Dividing the social optimum by the unique equilibrium, we have
            \begin{align*}
                \frac{1 - q (1-p) \pm o(1)}{1 - q e^{-p} \pm o(1)} &= \frac{1 - q (1-p)}{1 - q e^{-p} \pm o(1)} \pm \frac{o(1)}{1 - q e^{-p} \pm o(1)} \\
                &= \frac{1 - q (1-p)}{1 - q e^{-p} \pm o(1)} \pm o(1) \\
                &= \paren{1 - q (1-p)} \sum_{k=1}^\infty \paren{qe^{-p} \pm o(1)}^k \pm o(1) \\
                &= \paren{1 - q (1-p)} \sum_{k=1}^\infty\paren{qe^{-p}}^k \pm o(1) \\
                &= \frac{1 - q (1-p) }{1 - q e^{-p} }\pm o(1)
            \end{align*}
        \qed

    \layeredgraph*

    \obliviousoptimaledgecosts*
    
        \proofof{\cref{prop:obliviousoptimaledgecosts}} 
        It is straightforward to generalize the proof of \cref{prop:obliviousoptimal}.
        All that is needed is to notice that 
            \begin{align*}
                \min_{d_1, \dots, d_n \in \N} \prod_{j \in [n]} \paren{1-\frac{p}{d_j}}^{d_j / n} - \gamma \sum_{j=1}^n d_j &= 
                \min_{d \in \N} \paren{1-\frac{p}{d}}^{d} - \gamma n d,
            \end{align*}
            which can be used to see that, if $\gamma \leq qp$, the optimal choice of $d$ is again 1 and otherwise achieved at $d = 0$, the empty network.
        \qed

    \poaedgecosts*

    \genhomophily*

    \inequality*

    \multiplicityofdreg*

    \worstinequality*
    \proofof{\cref{prop:worstinequality}} Let $q = 1-p$ and $\gamma = p(1-p)(1-p/2)/2$. 
We will show that for these parameters,
\begin{align*}
    \lim_{p \to 0} \mathrm{\gini}(p, q, \gamma) = 5 - 2 \sqrt{6}.
\end{align*}

First, notice that there exist $d$-regular equilibria for $d=1$ and $d=2$.
That is, in terms of the equilibrium conditions,
\begin{align*}
    \frac{1}{2}\paren{1 - p} \leq \frac{\gamma}{qp} \leq 1, \;\; \text{and} \\
    \frac{1}{3}\paren{1 - \frac{p}{2}}^2 \leq \frac{\gamma}{qp} \leq \frac{1}{2} \paren{1 - \frac{p}{2}}
\end{align*}
which, replacing $q, \gamma$ with their expression in terms of $p$ can be written
\begin{align*}
    \frac{1}{2}\paren{1 - p} \leq \frac{1}{2} \paren{1 - \frac{p}{2}} \leq 1, \;\; \text{and} \\
    \frac{1}{3}\paren{1 - \frac{p}{2}}^2 \leq \frac{1}{2} \paren{1 - \frac{p}{2}} \leq \frac{1}{2} \paren{1 - \frac{p}{2}},
\end{align*}
which are easily verified to hold.
Next, recall that for an individual with degree 1 connected to someone else of degree 1, their utility is
\begin{align*}
    1 - q(1-p) - \gamma.
\end{align*}
For an individual with degree 2 connected to two others of degree 2, their utility is
\begin{align*}
    1 - q \paren{1 - \frac{p}{2}}^2 - 2 \gamma.
\end{align*}
Let the first of these be denoted $\umax$ and the second $\umin$.
Plugging in the expressions for $q, \gamma$ in terms of $p$, we have
\begin{align*}
    \umax &= 1 - (1-p)^2 - \frac{1}{2}p(1-p)\paren{1 - \frac{p}{2}}, \\
    &= \frac{3}{2}p + \cO(p^2), \;\; \text{and} \\
    \umin &= 1 - (1-p) \paren{1 - \frac{p}{2}}^2 - p(1-p)\paren{1 - \frac{p}{2}},  \\
    &= p + \cO(p^2).
\end{align*}
Now, let 
\begin{align*}
    \lambda &\defeq \lim_{p \to 0} \frac{\sqrt{\umax \umin} - \umin}{\umax + \umin}, \\
    &= \sqrt{6} - 2.
\end{align*}
For large enough $n$, we can construct a network composed of a 1-regular part and a 2-regular part such that the 1-regular part is of size arbitrarily close to $\lambda n$.
Then, the Gini coefficient as $n \to \infty$ is 
\begin{align*}
    \lim_{p \to 0} \frac{\lambda (1- \lambda) (\umax - \umin)}{2 (\lambda \umax + (1-\lambda) \umin)} &= 5 - 2 \sqrt{6}.
\end{align*}
\qed

    \mechdes*
    
        \proofof{\cref{prop:mechdes}} The first part of the proposition follows directly from the functional form individual utilities.
        In any equilibrium $E$, the utility of individual $i$ is
        \begin{align*}
            1 - q \prod_{j \in N_i(E)} \paren{1 - \frac{p}{d_j}} - \gamma d_i
        \end{align*}
        which is decreasing in $\gamma$. 
        Of course, this result depends on $E$ continuing to be an equilibrium as $\gamma$ changes.
        For the second part of the result, consider any $d \in \N$ and $0 < \gamma \leq q p (1-p)$ such that
            \begin{align*}
                \gamma = \frac{qp}{d} \paren{1 - \frac{p}{d}}^{d-1}.
            \end{align*}
        Notice that for 
        \begin{align*}
            \gamma > \frac{qp}{d} \paren{1 - \frac{p}{d}}^{d-1}
        \end{align*}
        the largest $k$-regular equilibrium is at most $d-1$.
        Then, the following sequence proves the result:
        \begin{align*}
            &1 - q \paren{1 - \frac{p}{d}}^d - (\gamma - \varepsilon') d < 1 - q \paren{1 - \frac{p}{d-1}}^{d-1} - (\gamma + \varepsilon) (d - 1) \\
            \iff &(\varepsilon + \varepsilon') d + \varepsilon < \gamma + q \paren{\paren{1 - \frac{p}{d}}^d - \paren{1 - \frac{p}{d-1}}^{d-1}} \\
            \impliedby & \overline{\varepsilon} < \frac{1}{2d+1} \paren{\gamma + q \paren{\paren{1 - \frac{p}{d}}^d - \paren{1 - \frac{p}{d-1}}^{d-1}}} \\
            \impliedby & \overline{\varepsilon} < \frac{\gamma}{2d+1} 
        \end{align*}
        \qed

    \voifixed*

    \voieq*
    
        \proofof{\cref{prop:voieq}}  For the first item, observe 
        \begin{align*}
            & \sum_{i = 1}^n {u_i(E)} <  \sum_{i = 1}^n {u^{\infd}_i(E^{\infd})} \\
            \iff &1 - e^{-p} < 1 - e^{-\frac{p}{q}} \\
            \iff& -\frac{p}{q} < -p \\
            \iff& q < 1 .
        \end{align*}
        For the second item, let $\cE_{\mathrm{\eqbm}}^{\text{reg.}}$ be the set of equilibria edge sets that are regular and suppose $0 < p < 1/2$, $q = 1 - p$ and $\gamma = qp(1-p/2)/2$.
        It is easy to verify that the largest $d$ such that a $d$-regular network is an equilibrium for the \gm~and \indiscgm~is 2.
        Thus, the following sequence proves the result:
        \begin{align*}
            &1 - q\paren{1 - \frac{p}{1-p} \frac{1 - p^2}{2}}^2 - 2 \gamma > 1 - q \paren{1 - \frac{p}{2}}^2 - 2 \gamma \\
            \iff &\paren{1 - \frac{p}{2}}^2 > \paren{1 - \frac{p (1+p)}{2}}^2 \\
            \iff &{{p}} < {{p (1+p)}} 
        \end{align*}
        which holds for all $0 < p < 1/2$.

        For the third item, suppose $0 < p < 1/2$, $q = 1 - p$ and $\gamma = (2+p)(1-p)^2(1+p)/4$. It is easy to verify that the largest $d$ such that a $d$-regular network is an equilibrium for the \gm~is 2 and for the \indiscgm~it is 1.
        Thus, the following sequence proves the result:
        \begin{align*}
            &1 - q \paren{1 - \frac{p}{1-p} \frac{1 - p^2}{2}}^2 - 2 \gamma < 1 - q \paren{1 - p} - \gamma \\
            \iff &1 - p - \paren{1 - \frac{p(1 + p)}{2}}^2 < \frac{\gamma}{q} \\
            \iff &1 - p - \frac{1}{4}\paren{2 - p - p^2}^2 < \frac{\gamma}{q} \\
            \iff &1 - p - \frac{1}{4}(2+p)^2(1-p)^2 < \frac{(2+p)(1-p)^2(1+p)}{4(1-p)} \\
            \iff &1 - \frac{1}{4}(2+p)^2(1-p) < \frac{(2+p)(1+p)}{4} \\
            \iff &4  < (2+p)^2(1-p) + {(2+p)(1+p)} \\
            \iff &4  < (2+p) (3-p^2) 
        \end{align*}
        which is true for all $0 < p < 1/2$.
        \qed

    \informedselfish*
    
        \proofof{\cref{prop:informedselfish}} 
        The proof of \cref{prop:indiscselfish} can be straightforwardly adapted to the \gm.
        Notice
        \begin{align*}
            u_i^{\mathrm{\infd}}(E) &= 1 - q \probgiv{\cap_{\ell \in N_i(E)} \{R_{\ell i} = 0\}}{X_i = 0}, \;\text{and} \\
            u_i^{\mathrm{\infd}}(E \cup \{ (i, j) \}) &= 1 - q \probgiv{\cap_{\ell \in N_i(E) \cup \{ j \}} \{R_{\ell i} = 0\}}{X_i = 0}.
        \end{align*}
        But since
        \begin{align*}
            \probgiv{ R_{j i} = 0}{X_i = 0, \cap_{\ell \in N_i(E) } \{R_{\ell i} = 0\}} < 1,
        \end{align*}
        the latter is greater than the former.
        \qed

    \informedoptimal*

    \poainformed*
    
        \proofof{\cref{prop:poainformed}}
        The asymptotic social welfare for equilibrium networks is calculated similarly to before as
        \begin{align*}
         \frac{1}{n}\sum_{i=1}^n {u_i(\nashedgeset)}  &=  1 - q \paren{1 - \frac{p}{q} \cdot \frac{1 - (1-q)^{n-1}}{n-1}}^{n-1} \pm o(1)\\
         &= 1 - q e^{-p/q} \pm o(1).
        \end{align*}
        
        The asymptotic social welfare for efficient networks comes directly from \cref{prop:informedoptimal}, noticing that the largest 3- and 4-cycle-free regular subgraph for a population of size $n$ is within $o(1)$ of optimal.
        \qed 

    \informedlayeredgraph*

    \informededgecostoptimal*
    
            \proofof{\cref{prop:informededgecostoptimal}} It is straight-forward to adapt the proof of \cref{prop:informedoptimal}. 
            All that is needed is to notice that for any fixed average degree $d$, the edge cost term in average utility is constant (and equal to $\gamma d$). 
            Thus, the minimization problem in \cref{eq:minavgdegreesw} is not influenced by the constant term and the result follows.
            \qed

    \informededgecostpoa*

\fi
\fi

\end{document}